\documentclass{aastex631}
\usepackage{comment}
\usepackage{amsmath}
\usepackage{graphics}
\usepackage{color}
\usepackage{threeparttable}
\usepackage{rotating}
\usepackage{dcolumn}
\usepackage{multirow}
\newcolumntype{d}[1]{D{.}{.}{#1}}

\received{}
\revised{}
\accepted{\today}
\graphicspath{{./}{}}

\begin{document}
\title{CARMA-NRO Orion Survey: unbiased survey of dense cores and core mass functions in Orion A}

\author{Hideaki Takemura}
\affil{The Graduate University for Advanced Studies
(SOKENDAI), 2-21-1 Osawa, Mitaka, Tokyo 181-0015, Japan}
\affil{National Astronomical Observatory of Japan, 2-21-1 Osawa, Mitaka, Tokyo 181-8588, Japan}

\author{Fumitaka Nakamura}
\affil{The Graduate University for Advanced Studies
(SOKENDAI), 2-21-1 Osawa, Mitaka, Tokyo 181-0015, Japan}
\affil{National Astronomical Observatory of Japan, 2-21-1 Osawa, Mitaka, Tokyo 181-8588, Japan}
\affil{Department of Astronomy, The University of Tokyo, Hongo, Tokyo 113-0033, Japan}

\author{H\'ector G.~Arce}
\affil{Department of Astronomy, Yale University, New Haven, CT 06511, USA}

\author{Nicola Schneider}
\affil{I. Physik. Institut, University of Cologne, Z\"ulpicher Str. 77, 50937 Cologne, Germany}

\author{Volker Ossenkopf-Okada}
\affil{I. Physik. Institut, University of Cologne, Z\"ulpicher Str. 77, 50937 Cologne, Germany}

\author{Shuo Kong}
\affil{Steward Observatory, University of Arizona, Tucson, AZ 85719, USA}
\affil{Department of Astronomy, Yale University, New Haven, CT 06511, USA}

\author{Shun Ishii}
\affil{National Astronomical Observatory of Japan, 2-21-1 Osawa, Mitaka, Tokyo 181-8588, Japan}

\author{Kazuhito Dobashi}
\affil{Tokyo Gakugei University, Koganei, Tokyo, 184-8501, Japan}

\author{Tomomi Shimoikura}
\affil{Otsuma Women’s University, ‘Chiyoda-ku, Tokyo, 102-8357, Japan}

\author{Patricio Sanhueza}
\affil{The Graduate University for Advanced Studies
(SOKENDAI), 2-21-1 Osawa, Mitaka, Tokyo 181-0015, Japan}
\affil{National Astronomical Observatory of Japan, 2-21-1 Osawa, Mitaka, Tokyo 181-8588, Japan}

\author{Takashi Tsukagoshi}
\affil{National Astronomical Observatory of Japan, 2-21-1 Osawa, Mitaka, Tokyo 181-8588, Japan}

\author{Paolo Padoan}
\affil{Institut de Ci\' encies del Cosmos, Universitat de Barcelona, IEEC-UB, Mart\'i i Franqu\'es 1, E08028 Barcelona, Spain}
\affil{ICREA, Pg. Llu\'is Companys 23, E-08010 Barcelona, Spain}

\author{Ralf S.\ Klessen}
\affil{Universit\"{a}t Heidelberg, Zentrum f\"{u}r Astronomie, Albert-Ueberle-Str. 2, 69120 Heidelberg, Germany}
\affil{Universit\"{a}t Heidelberg, Interdisziplin\"{a}res Zentrum f\"{u}r Wissenschaftliches Rechnen, INF 205, 69120 Heidelberg, Germany}

\author{Paul. F. Goldsmith}
\affil{Jet Propulsion Laboratory, California Institute of Technology, 4800 Oak Grove Drive, Pasadena, CA 91109, USA}

\author{Blakesley Burkhart}
\affil{Department of Physics and Astronomy, Rutgers University, 136 Frelinghuysen Rd, Piscataway, NJ 08854, USA}
\affil{Center for Computational Astrophysics, Flatiron Institute, 162 Fifth Avenue, New York, NY 10010, USA}

\author{Dariusz C. Lis}
\affiliation{Jet Propulsion Laboratory, California Institute of Technology, 4800 Oak Grove Drive, Pasadena, CA 91109, USA}

\author{\'Alvaro S\'anchez-Monge}
\affil{I.~Physikalisches Institut, Universit\"at zu K\"oln, Z\"ulpicher Str. 77, D-50937 K\"oln, Germany}

\author{Yoshito Shimajiri}
\affil{National Astronomical Observatory of Japan, 2-21-1 Osawa, Mitaka, Tokyo 181-8588, Japan}

\author{Ryohei Kawabe}
\affil{The Graduate University for Advanced Studies
(SOKENDAI), 2-21-1 Osawa, Mitaka, Tokyo 181-0015, Japan}
\affil{National Astronomical Observatory of Japan, 2-21-1 Osawa, Mitaka, Tokyo 181-8588, Japan}


\begin{abstract}
The mass distribution of dense cores is a potential key to understand the process of star formation.
Applying dendrogram analysis to the CARMA-NRO Orion C$^{18}$O ($J$=1--0) data, we identify 2342 dense cores, about 22 \% of which have virial ratios smaller than 2, and can be classified as gravitationally bound cores.
The derived core mass function (CMF) for bound starless cores which are not associate with protostars has a slope similar to Salpeter's initial mass function (IMF) for the mass range above 1 $M_\odot$, with a peak at $\sim$ 0.1 $M_\odot$.
We divide the cloud into four parts based on the declination, OMC-1/2/3, OMC-4/5,
L1641N/V380 Ori, and L1641C, and derive the CMFs in these regions.
We find that starless cores with masses greater than 10 $M_\odot$ exist only in OMC-1/2/3,
whereas the CMFs in OMC-4/5, L1641N, and L1641C are truncated at around 5--10 $M_\odot$.
From the number ratio of bound starless cores and Class II objects in each subregion, the lifetime of bound starless cores is estimated to be 5--30 free-fall times, consistent with previous studies for other regions.
In addition, we discuss core growth by mass accretion from the surrounding cloud material to explain the coincidence of peak masses between IMFs and CMFs.
The mass accretion rate required for doubling the core mass within a core lifetime is larger than that of Bondi-Hoyle accretion by a factor of order 2.
This implies that more dynamical accretion processes are required to grow cores.
\end{abstract}

\keywords{Star formation (1569); Interstellar medium (847); Molecular clouds(1072); Protostars (1302)}

\section{Introduction}
\label{sec:intro}

How and when stellar masses are determined is an important unresolved problem in studies of star formation \citep{maclow04,mckee07}. Previous studies of low-mass star-forming regions suggest that CMFs are similar in shape to stellar IMFs, indicating that there is a one-to-one relation between dense cores and stars \citep{motte98,motte18,ikeda07,alves07,dib10,maruta10}. This can be interpreted as implying that the final stellar mass is more of less determined at the core formation and evolution stages.
Then, uncovering the properties of core mass functions (CMFs) is expected to provide a clue for understanding the above problem. The observed CMFs so far are characterized by two main parameters: a turnover mass and a power-law slope at the high-mass end ($\geq 1 M_\odot$).
The turnover masses reported so far are in the range of $0.1-1 M_\odot$. This variation may be interpreted as due to the following two effects, (1) the turnover mass depends on the cloud environment, (2) it is due to the observational biases such as the angular resolution and sensitivity of the observations \citep[e.g.,][Paper I hereafter]{ikeda07,takemura21a}, or both. Although previous studies mainly focus on the first effect, our recent CMF analysis of the finest angular resolution, large-scale maps (see below) indicates that the turnover mass previously reported strongly depends on the observed angular resolution (see also \citet{ikeda07}). In addition, the turnover mass derived from dust continuum maps tends to be overestimated due to overlap effects (Paper I).
On the other hand, the slope is more or less similar to those of the stellar IMFs, although recently CMFs with shallower slopes have been reported toward high-mass star-forming regions \citep{zhang15,sanchez-monge17,motte18,cheng18,liu18,kong19,sanhueza19,sadaghiani20}.

One of the most well studied star-forming regions is Orion A and this is the nearest giant molecular cloud (GMC).
Many dense core molecular line data surveys in Orion A have been performed prior to this study.
For example, \citet{ikeda07} observed $1.5\arcdeg \times0.5\arcdeg$ area in the H$^{13}$CO$^+$ ($J$=1--0) line using NRO45-m. The spatial resolution is 21\arcsec\ which corresponds to 0.05 pc. They identified 236 dense cores with \textsf{clumpfind} algorithm \citep{williams94} and mass detection limit of 1.6 $M_\odot$.
The mean and standard deviation of core radius and mass are 0.14 $\pm$ 0.03 pc and 12 $\pm$ 12 $M_\odot$, respectively.
The CMF has a best-fit power-law index of 2.3 $\pm$ 0.1 above 9.3 $\pm$ 1.5 $M_\odot$.
However, the authors claimed that low-mass cores are affected by confusion with large cores and several small cores are misidentified. They concluded that confusion-corrected CMF has no turnover.
A dense core survey of $20\arcmin \times 20\arcmin$ area of the central region in Orion A called OMC-1 in the C$^{18}$O ($J$=1--0) line was carried out using the NRO45-m telescope \citep{ikeda09}. The map resolution is $26\arcsec.4$. Applying \textsf{clumpfind} algorithm, they identified 65 dense cores and there are 57 H$^{13}$CO$^+$ cores in the same area \citep{ikeda07}.
The mean and standard deviation of core radius and mass are 0.18 $\pm$ 0.03 pc and 7.2 $\pm$ 4.5 $M_\odot$, respectively. The best-fit power-law index of the high-mass end (above 5 $M_\odot$) of observed CMF is 2.3 $\pm$ 0.3 but the confusion-corrected CMF has no turnover as well.
\citet{shimajiri15} conducted a dense core survey using the \textit{AzTEC} 1.1 mm continuum map and NRO45-m C$^{18}$O ($J$=1--0) data which have angular resolutions of 36\arcsec\ and $26\arcsec.4$, respectively. The authors identified 619 dust cores and 235 C$^{18}$O cores using \textsf{clumpfind} method.
The mean and standard deviation of the radius and mass of dust cores are 0.09 $\pm$ 0.03 pc and 5.52 $\pm$ 9.55 $M_\odot$. For C$^{18}$O cores, these values are 0.23 $\pm$ 0.04 pc and 12.4 $\pm$ 10.4 $M_\odot$.
As for high-resolution molecular line observations, \citet{hacar18} carried out ALMA observations with N$_2$H$^+$ ($J$=1--0) in 3\arcsec\ angular resolution.
They identified many velocity-coherent elongated structures called fibers in the main filament. These structures seem to influence the core formation and evolution.

In \citet{takemura21a}, we studied the CMF in a small area ($15\arcmin \times 15\arcmin$) of the Orion Nebular Cluster (ONC) region in Orion A, using the high angular resolution C$^{18}$O ($J$= 1--0) dataset taken with the Combined Array for Research in Millimeter-wave Astronomy (CARMA) interferometer and the Nobeyama Radio Observatory (NRO) 45-m telescope of Orion A \citep{kong18,nakamura19}.
In the CARMA-NRO Orion Survey, we produced wide-field maps of $^{12}$CO ($J$=1--0), $^{13}$CO ($J$=1--0), and C$^{18}$O ($J$=1--0) toward the Orion A cloud which cover $\sim$ 2 deg$^2$, by combining the CARMA interferometric data and the NRO 45-m single-dish telescope data.
The resulting angular resolution is $\simeq$ 8\arcsec, corresponding to $\sim$ 3300 au at a distance of 414 pc \citep{menten07}. The velocity resolution is $\simeq$ 0.1 km s$^{-1}$.
The spatial resolution is more than three times better than that resolution used in the previous survey in wide area of Orion A \citep[][hereafter Paper II]{takemura21b}, which allows us to investigate much more effectively sub-solar mass structures in Orion A.
We used astrodendro ver. 0.2.0 \citep{rosolowsky08}\footnote{https://dendrograms.readthedocs.io/en/stable/} to identify 692 dense cores in the position-position-velocity (PPV) space.
Using the \textit{Herschel} H$_2$ column density map and
the C$^{18}$O data cube, we estimated the core masses in the PPV space by removing the mass of the ambient gas from the two-dimensional H$_2$ column density map.
The MF of gravitationally bound starless cores is characterized by a turnover mass of 0.1 $M_\odot$ and a power-law slope of 2.12 $\pm$ 0.29 which is similar to Salpeter's IMF \citep{salpeter55} which has a power-law index of $\alpha=2.35$ when it is written as $\mathrm{d}N/\mathrm{d}M \propto M^{-\alpha}$.
We then compared the CMF to the stellar IMF in the same area \citep{dario12}, and found that the slope is similar to the Salpeter value, and the turnover masses of the CMF and IMF almost coincide with each other.
If stars form from individual cores, almost the same amount of gas mass in the cores should go into stars formed, so that the IMF can be reproduced by the CMF.

The coincidence of the turnover mass is a different result from the previous studies of the nearby star-forming regions,
in which the turnover masses are reported to be larger than those of the IMFs. For example, \citet{alves07}
revealed that the turnover mass of the CMF in the Pipe Nebular is larger than that of the ONC's IMF.
From the JCMT/SCUBA 850$\mu$m image in the Orion molecular cloud, \citet{nutter07} reported the turnover mass of $\sim 1\ M_\odot$.
Previous observations of CMFs in nearby star-forming regions \citep[e.g.,][]{ikeda09, maruta10, konyves15} seem to be consistent with the Pipe Nebular and James Clerk Maxwell Telescope (JCMT) Orion A CMFs.
These turnover masses of the CMFs are about 10 times larger than that of the ONC region, $\sim 0.1\ M_\odot$, which could be caused by coarser spatial resolutions (see Paper I).
The interpretation of this relation between the IMF and observed CMF in the ONC region is that the IMF can be reproduced directly from the observed CMFs, although the inefficient protostellar outflow feedback appears to contradict theoretical results \citep{matzner00,machida12}.
Another interpretation is that the prestellar cores or protostars gain additional mass from the surroundings through mass accretion if the protostellar outflow feedback blows away a significant amount of core material.

In this paper, we extend Paper I's analysis to the entire map of the CARMA-NRO Orion Survey to understand how CMFs depend on the cloud environment.
The map covers a wider area of Orion A than NRO-45m data of \citet{nakamura19} used in Paper II.
We present the core catalog of Orion A and reveal the physical properties of the cores.
A difference from the analysis in Paper I's is that we use the updated H$_2$ column density map
with 18\arcsec \ angular resolution which is a factor of two better than the previous map of \citet{kong18}. We also use the YSO catalog of the Vienna survey in Orion (VISION) with the European Southern Observatory's (ESO) Visible and Infrared Survey Telescope for Astronomy (VISTA) \citep{meingast16} which is more complete compared to the catalog of \textit{Herschel} Orion protostar survey (HOPS) \citep{furlan16}.
In addition, we use the updated CARMA-NRO Orion maps which are extended up to Declination $\simeq -7^\circ$ 15\arcmin.
Other procedures such as the core identification scheme and core mass estimate are the same as those of Paper I.
As we discuss in the following section, we adopt 390 pc as a distance in this paper, although we used 414 pc \citep{menten07} in previous papers (Paper I, Paper II).
The difference makes the core masses smaller by 6 \%.
It is worth noting that the Orion A filamentary cloud is tilted relative to the plane of sky, according to a recent Gaia analysis \citep{grobschedl18}. However, this effect seems to be small at least in our observed area, and the 390 pc is a reasonable representative distance of the entire area we have observed. Therefore, in this paper, we adopt this distance to evaluate the core physical quantities.

The paper is organized as follows.
In Section \ref{sec:obs}, we briefly describe the data used in this study.
In Section \ref{sec:result}, we present how we identify cores and derive their physical properties.
Then, we discuss in Section \ref{subsec:core_property} the core properties.
In Section \ref{sec:cmf}, we present the core mass function for different regions in Orion A and discuss how the CMF depends on the cloud environment.
Finally, we summarize our results in Section \ref{sec:discussion}.

\section{Observations and data}
\label{sec:obs}

\subsection{C$^{18}$O ($J$=1--0) data}

\begin{figure}
 \begin{center}
  \includegraphics[height=20cm]{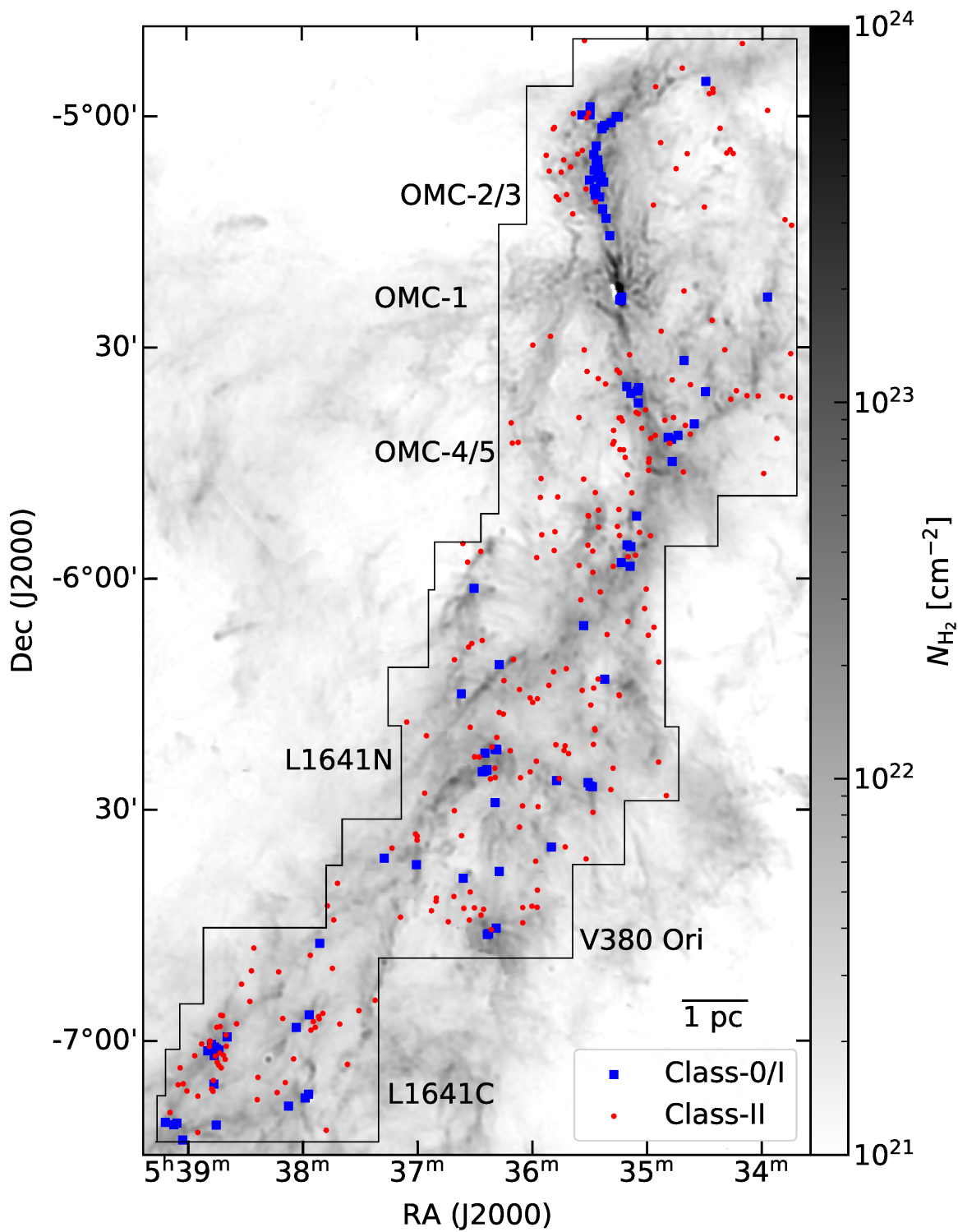}
 \end{center}
\caption{The C$^{18}$O ($J$=1--0) observation area is drawn with the black solid lines on the H$_2$ column density map.
Our map covers a wide area of 1 $\times$ 2 square degrees which is from OMC-1/2/3/4 to L1641N, V380 Ori, and L1641C.}
\label{fig:obsarea}
\end{figure}

We use the C$^{18}$O ($J$=1--0, 109.782182 GHz) data which cover the $1^\circ \times 2^\circ$ area of the Orion A molecular cloud.
In Figure \ref{fig:obsarea}, we show the observed area. The map contains the OMC-1/2/3/4/5, L1641-N, V380 Ori, and L1641C regions.
The map used in the present paper is slightly larger than that of Paper I and \citet{kong18} which do not include the L1641C region.
We obtained the data by combining the CARMA interferometer and the NRO 45-m single-dish telescope data as described in the dense core survey in the ONC region (Paper I).
Details of the procedure are described in detail in \citet{kong18}.
The angular and velocity resolutions are the same as in the previous map, corresponding to an angular resolution of 8\arcsec$\ $ and a velocity resolution of 0.1 km s$^{-1}$.
The average noise level in the C$^{18}$O map is 0.68 K in units of $T_{\rm MB}$, almost the same as in Paper I.
We used this value as 1$\sigma$ in core identification described in Section \ref{subsec:core_id}.

\begin{figure}
 \begin{center}
 \includegraphics[height=20cm]{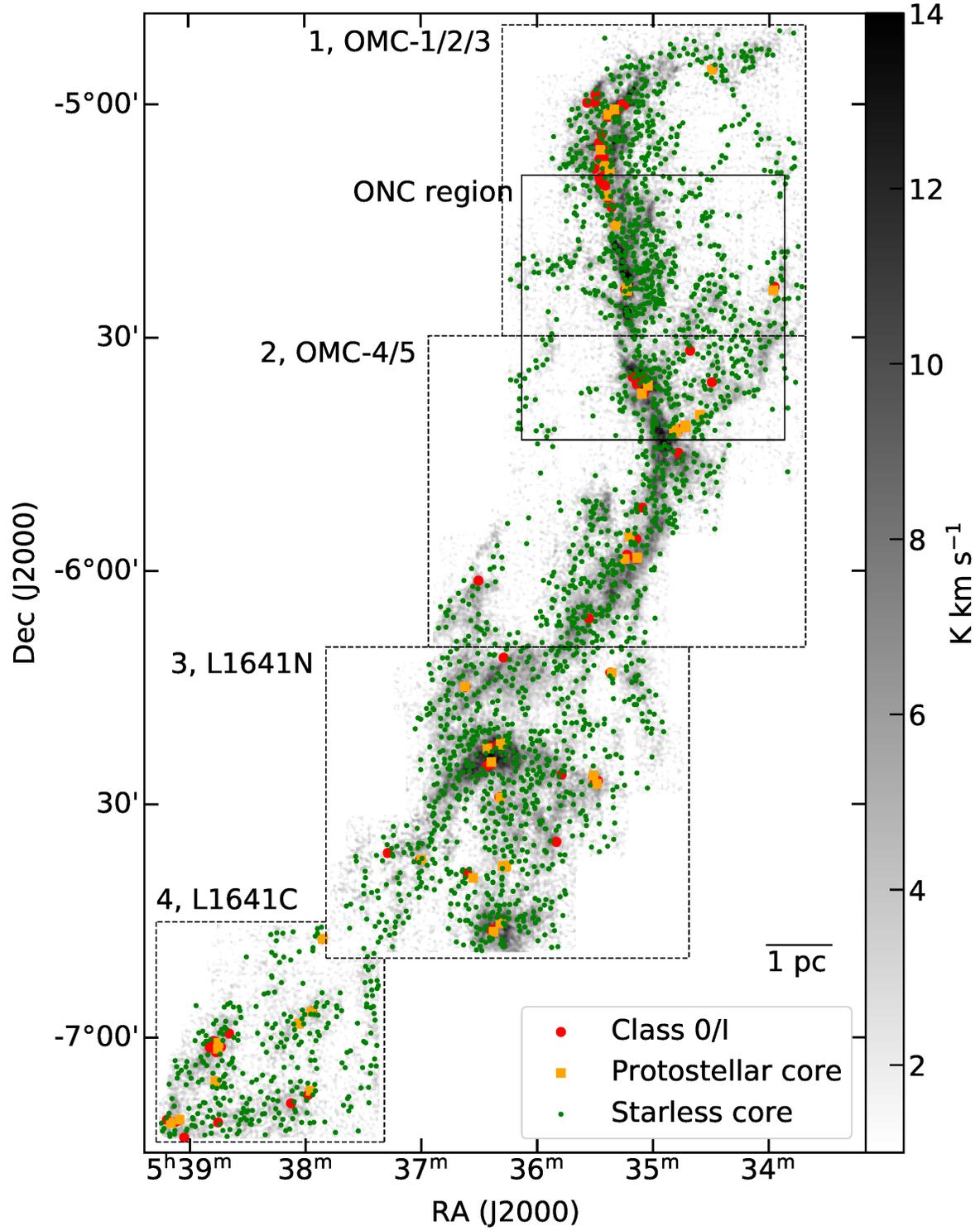}
 \end{center}
 \caption{The spatial distribution of starless and protostellar cores in Orion A.
 The dashed black lines represent the area of four subregions: OMC-1/2/3, OMC-4/5, L1641N, and L1641C. The solid black square is the ONC region which is analysed by Paper I.}
 \label{fig:core}
\end{figure}

In Paper I, we used a part of the map containing the most crowded and dense region, corresponding to the OMC-1 and the ONC region.
In this paper, we divide Orion A into four areas according to only their declination as follows.
\begin{itemize}
    \item[(a)]
    OMC-1/2/3 ($-5^\circ 30\arcmin \ \lesssim \delta$)
    \item[(b)]
    OMC-4/5 ($-6^\circ 10\arcmin \ \lesssim \delta \lesssim -5^\circ 30\arcmin$)
    \item[(c)]
    L1641N/V380 Ori ($-6^\circ 50\arcmin \ \lesssim \delta \lesssim -6^\circ 10\arcmin$),
    \item[(d)]
    L1641C ($\delta \lesssim -6^\circ 45\arcmin$)
\end{itemize}
The dashed rectangles in Figure \ref{fig:core} show the accurate definitions of the areas.
The ONC region studied in the previous paper (Paper I) is shown with a black solid square.
Our observed region contains areas with different cloud environments.
In OMC-1/2/3, the main cloud follows a single dense filamentary ridge, while outside the ridge, several fainter filamentary structures can be seen.
In L1641N/V380 Ori, the dense filamentary ridge is not seen but OMC-4/5 contains several small filaments.
In addition, OMC-1/2/3, especially OMC-1, is influenced by the UV radiation and stellar winds from massive stars such as $\theta^1$ Ori	C in the Trapezium star cluster \citep{pabst19}.
It is noted that for massive stars of this region, \citet{fukui18} showed that this area has experienced triggered O-type star formation caused by a cloud-cloud collision (see also \citet{lim21}).
In contrast, L1641N and L1641C look more quiescent. Although protostars or protostellar outflows are detected in L1641N \citep{ali04,nakamura12, tanabe19,feddersen20} and L1641C \citep{fukui86, chen93}, feedback from massive stars in these regions is not observed.
In the following, we refer to these areas to discuss their physical properties.

\subsection{H$_2$ column densities}

In this paper, we estimate the core masses using the H$_2$ column density map, following Paper I.
However, the H$_2$ column density map we use in the present paper is different from that of Paper I, as follows.

The H$_2$ column density map was determined from observations of the \textit{Herschel} fluxes and is shown in Figure \ref{fig:obsarea}. Orion A was observed in 2010 October within the Gould Belt keyprogram \citep{andre10}. Imaging was performed in parallel mode, i.e., using PACS at 70 and 160 $\mu$m and SPIRE at 250,
350, and 500 $\mu$m with a fast scanning speed of 60\arcsec/s. The dust column density map was obtained by fitting spectral energy distributions (SEDs) to the
\textit{Herschel} 160, 250, 350, and 500 $\mu$m flux data, assuming a grey body optically thin dust emission.
The 250, 350, and 500 $\mu$m \textit{Herschel} maps were taken from the \textit{Herschel} archive advanced data products that provide 'HiRes' flux maps.
These are constructed using a deconvolution method based on the Richardson-Lucy algorithm \citep{richardson72,lucy74}, applied to standard pipeline processed extended-source calibrated maps. An uncertainty of 20\% for the PACS 160 $\mu$m and 10\% for SPIRE bands was considered.
The gas surface density and dust temperature are derived from the \textit{Herschel} maps assuming a dust opacity of $\kappa_\lambda=0.1\times\left(\lambda/300\ \mu \mathrm{m}\right)^\beta\ [\mathrm{cm^2\ g^{-1}}]$ with $\beta=-2$.
The detailed knowledge of the SPIRE Photometer beam allows for iterative image restoration to produce the maps with a factor of two better angular resolution \citep{xu14} than the original maps having 36\arcsec\ angular resolution.
The column density of H$_2$ is calculated with a mean molecular weight per hydrogen molecule of $\mu=2.8$.
The core masses calculated with this new \textit{Herschel} dust column density map are essentially the same as those determined using the lower-resolution image. The angular resolution of 18\arcsec\ of the H$_2$ map is still about twice that of our C$^{18}$O ($J$=1--0) map with 8\arcsec\ resolution. However, for a better comparison, we regridded the H$_2$ map to match that of the C$^{18}$O ($J$=1--0) data using the CASA task, \textit{imregrid}.

\subsection{Catalog of young stellar objects (YSOs)}
\label{subsec:YSO}

We utilize the VISION protostar catalog to identify all the young stellar objects in the Orion A molecular cloud.
To estimate the core lifetime in Section \ref{sec:correlations},
we extracted the Class II objects from this catalog, where we
chose the objects with the spectral indices of $-0.3> \alpha > -1.6$ as the Class II objects.
Our observed area contains 40 Class 0, 61 Class I, and 264 Class II objects.
The positions of the Class 0/I and II objects are denoted by the blue and red points, respectively, in Figure \ref{fig:obsarea}.
The Class 0/I objects are concentrated in the dense filamentary region, whereas the Class II objects are distributed more extensively over the entire observed area. The northern portion (OMC-1/2/3/4/5) contains about half of the Class 0/I objects.
We note that an area around OMC-1 ($\sim 11 \arcmin \times 11 \arcmin$) is excluded from the analysis in the survey because the background emission in this region on a very small scale is too large to allow source detection.
A detailed description can be found in section 3.7.4. of \citet{meingast16}.
In addition, we probably undercount protostars and protostellar cores in OMC-1/2/3 and give a detailed description in Section \ref{subsec:core_id}.

\section{Dense cores in Orion A}
\label{sec:result}

\subsection{Core Identification}
\label{subsec:core_id}

In our previous papers, we confirmed that the C$^{18}$O integrated intensity emission is proportional to
the H$_2$ column density in the range of a few $\times 10^{22}- $ a few $\times 10^{23}$ cm$^{-2}$ (Paper I, Paper II).
Therefore, we consider that the C$^{18}$O emission reasonably traces the structure in Orion A in a dynamic range of its column density.
We identified the dense cores in the position-position-velocity (PPV) space using the C$^{18}$O ($J$=1--0) data cube.
We applied astrodendro to the CARMA-NRO C$^{18}$O ($J$=1--0) data cube.
The algorithm identifies the hierarchical structure of the data and the hierarchies are named as {\it leaf}, {\it branch}, and {\it trunk} similar to the structure of a tree.
A {\it leaf} corresponds to a peak and two {\it leaves} are combined into one {\it branch}.
A {\it branch} can also be made by a combination of one {\it leaf} and one {\it branch} or two {\it branches}.
A {\it trunk} is the bottom of the hierarchy and the next higher hierarchy contains two {\it branches} or a {\it leaf} and a {\it branch}.
We define a {\it leaf} (the smallest structure identified by astrodendro) as a core.
Then, we estimate the masses of the cores using the \textit{Herschel} H$_2$ column density map, but we remove the contribution of the ambient gas distributed outside the cores in the position-position-velocity (PPV) space.
This procedure is the same as that employed in Papers I and II.

In the actual identification, the three input parameters are set to
min\_delta = 2$\sigma$, min\_value = 2$\sigma$, and min\_npix = 60 ($\approx$ 1 beam $\times$ 3 channels), following the suggestions of \citet{rosolowsky08}.
Each parameter gives the fineness of the identification, the minimum flux per pixel of the structure, and the minimum area or volume of the structure, respectively
\citep[see][for more detailed definitions and discussions]{rosolowsky08}.
Three additional selection criteria are imposed to minimize the effect of the spatially varying noise levels.
\begin{itemize}
    \item[(1)] the peak intensity of the {\it leaf} should be larger than $4\sigma$ at the corresponding spatial position.
    \item[(2)] more than three successive channels should contain more than 20 pixels ($\approx$ the map angular resolution) for each channel.
    \item[(3)] an identified core should not contain any pixels located at the boundaries of the observation area.
\end{itemize}
The definition of the parameters and the additional criteria are the same as those used in the dense core survey of the ONC region (Paper I).
In addition to the positions of the pixels in each core, the output of astrodendro includes core properties such as total flux, position angle, area, and velocity width.
Since the H$_2$ column density for one core in the OMC-1/2/3 area (core id 927) was not derived due to the saturation of the \textit{Herschel} observations, we remove this core from the sample so that our final sample size is 2341.

We then classified the cores into two groups, starless and protostellar cores, using the VISION catalog \citep{meingast16}.
If a core overlaps spatially with at least one object in the VISION catalog, we classified it as a protostellar core.
A core without overlapping objects is categorized as a starless core.
As a result, we identified 2295 starless cores and 46 protostellar cores.
We note that almost all the VISION class 0/I objects are associated with identified {\it leaves}, but about half of such {\it leaves} do not satisfy our additional condition (2).
Therefore, they are not classified as protostellar cores and we simply omit such small core candidates in this paper.
In addition, the VISION catalog seems not to include all protostars in Orion A as described in Section \ref{subsec:YSO}.
Due to the above reasons, we probably undercount the protostellar cores and overestimate the ratio of starless cores and protostellar cores.

We show the spatial distribution of starless and protostellar cores in Figure \ref{fig:core}.
While most of the cores are concentrated in the main ridge of the Orion A cloud, a significant number of cores are more widely distributed, mostly along the less dense filamentary structure, but also outside of
any filamentary structure. Interestingly, \citet{polychroni13} obtained a similar result; they found in their core distribution using only \textit{Herschel} data for the southern L1641 region (mostly not covered by our observations) that $\sim$30\% of the cores are not located in filaments.
Since we identify the cores in PPV space, approximately 55\% of identified cores overlap with more than one core along the line of sight by one pixel or more.
The results of core identification are summarized in Table \ref{tab:oriona_core_id}.
We describe the mass calculation method considering the overlap effect in Section \ref{subsec:derivation}.

\subsection{Derivation of the Core Physical Quantities}
\label{subsec:derivation}

We define the physical quantities of the identified cores as follows.
The positions and the line of sight velocity of a core are determined by the mean positions of the structure identified and the intensity-weighted first-moment velocity, respectively.
The core radius is defined as
\begin{equation}
    R_{\mathrm{core}}=\left(\frac{A}{\pi}\right)^{1/2} \ ,
\end{equation}
where $A$ is the area of the core projected onto the plane of the sky.
The aspect ratio of the core is calculated as the ratio of the major axes to minor axes.
The major and minor axes are computed from the intensity-weighted second moment in direction of maximum elongation and perpendicular to the major axis, respectively,
in the plane of the sky.
The position angle (P.A.) of the core is determined counter-clockwise from the west.
The FWHM velocity width, $dV_\mathrm{core}$ is obtained by multiplying
the intensity-weighted second moment of velocity by a factor of $2\sqrt{2\mathrm{ln 2}}$.

The core mass is evaluated with the same procedure described in Papers I and II.
In brief, we calculated the core mass using the \textit{Herschel} H$_2$ column density ($N_{\mathrm H_2}^{Herschel}$) and C$^{18}$O intensity-ratio of the {\it leaf} and the {\it trunk} ($I_{\rm leaf}/I_{\rm trunk}$) to calculate the mass, where $I_\mathrm{leaf}$ is the intensity of the core.
We assigned the H$_2$ column density to each core using the intensity ratio and calculated the core mass as
\begin{equation}
M_\mathrm{core}=\mu m_\mathrm{H}\sum N_{\mathrm H_2}^{Herschel} (i,\,j)\times \frac{I_{\mathrm{leaf}} (i,\,j)}{I_{\mathrm{trunk}} (i,\,j)} ,
\label{eq:core_mass}
\end{equation}
where $i$ and $j$ are the indices of the cell of interest on the R.A.--Dec. plane, respectively.
Using the Dendrogram's hierarchical structures such as leaves and trunks in three-dimensional space, we can estimate the masses of cores that overlap each other along the line of sight.
We note that we implicitly assumed the C$^{18}$O excitation temperatures $T_\mathrm{ex,C^{18}O}$ and fractional abundances $X_\mathrm{C^{18}O}$ have no variations along the line of sight.
Strictly speaking, we can identify cores and calculate core masses when $X_\mathrm{C^{18}O}T_\mathrm{ex,\ C^{18}O}\mathrm{exp}(5.27/T_\mathrm{ex,\ C^{18}O})$ (see Equation (1) of \citet{ikeda09}) is constant in the data even if $T_\mathrm{ex,\ C^{18}O}$ and $X_\mathrm{C^{18}O}$ vary within the cloud.
In addition, we assume a subthermal emission of C$^{18}$O since the cloud seems not to be dense enough in most region.
This somewhat affects the mass estimate in the ambient gas. In real situations, the fractional abundance of C$^{18}$O can be lower in denser regions due to depletion \citep[e.g.,][]{caselli99,christie12}, particularly in dense starless regions.
The selective dissociation by UV radiation also contributes to reducing the fractional abundance in less dense regions \citep[e.g.,][]{shimajiri14,lin16,ishii19,komesh20}.
The lower abundance of C$^{18}$O and depletion of C$^{18}$O in a dense core make a core mass larger compared to that given by Equation \ref{eq:core_mass}.
Dense starless gas is prone to have lower temperatures than ambient gas.
This can produce temperature variations along the line of sight.
If the temperature of the ambient gas is higher than that of a dense core, a core mass becomes smaller than that from Equation \ref{eq:core_mass} when the abundance of C$^{18}$O does not change along the line of sight.
The mass ratio of core and ambient gas is proportional to a ratio of $B(T_\mathrm{core})$ and $B(T_\mathrm{ambient})$ when the fractional abundance of C$^{18}$O to H$_2$ is constant along the line of sight.
Here, $B(T)$ is the Planck function at the frequency of C$^{18}$O ($J$=1--0) and $T_\mathrm{core}$ and $T_\mathrm{ambient}$ are temperature of core and ambient gas.
Since the dense starless regions are expected to have smaller fractional abundance and lower excitation temperatures, these two effects tend to cancel each other out.
When $T_\mathrm{leaf}$ = 20 K, $T_\mathrm{ambient}$ = 40 K (= 2$T_\mathrm{leaf}$), and $X_\mathrm{C^{18}O,trunk}/X_\mathrm{C^{18}O,leaf}$ = 2, we underestimate the core mass by $\sim$15 \%.
Thus, the effects of fractional abundance and excitation temperature variations along the line of sight are expected to be minor for our core mass calculation.
See Appendix 1 in Paper II for a more quantitative discussion.
The other source of uncertainty for the core masses is noise in the H$_2$ map.

\begin{figure}[htbp]
    \centering
    \includegraphics[height=20cm]{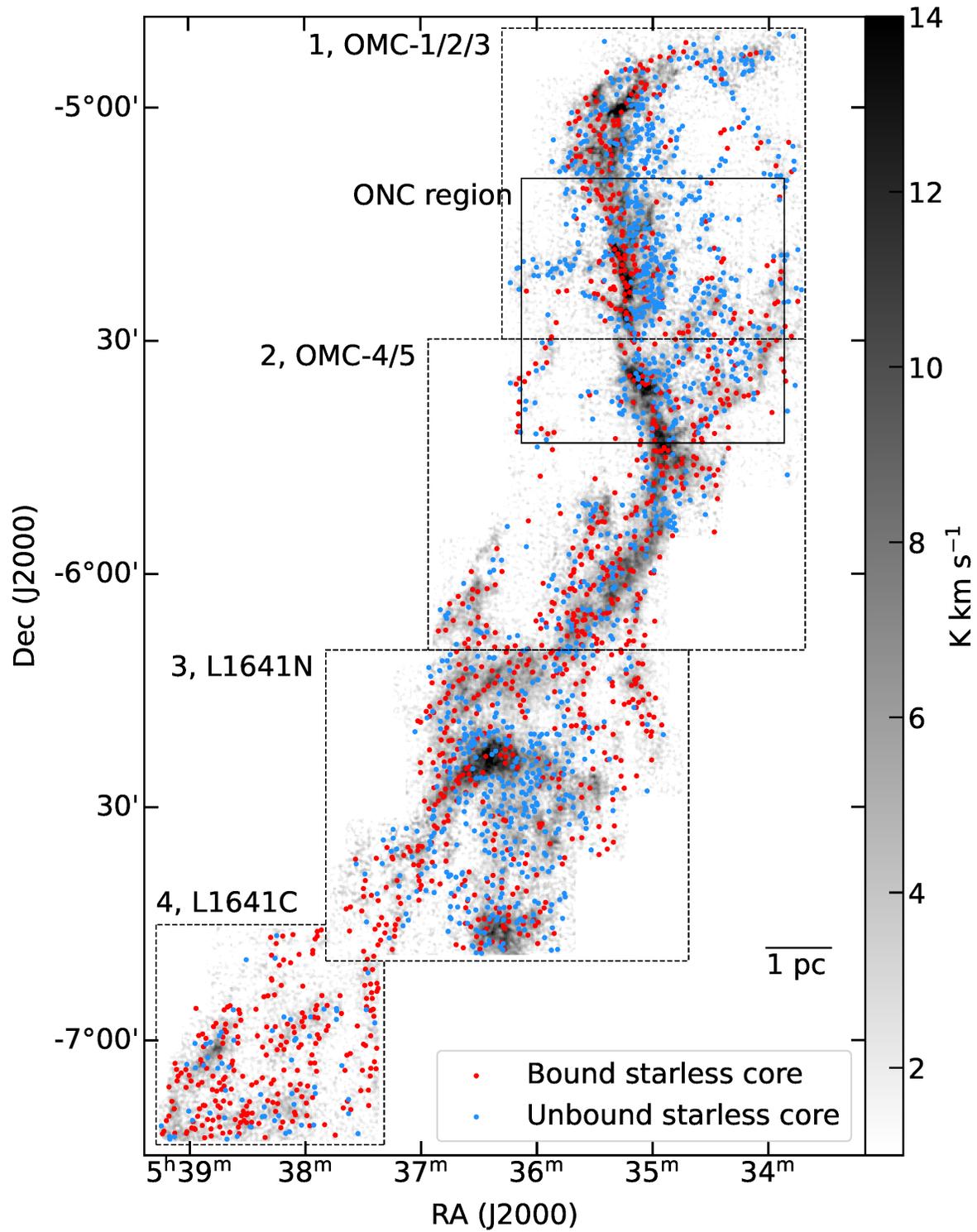}
    \caption{The spatial distribution of bounded starless cores (red dots) and unbounded starless cores (blue dots)
    overlaid onto the integrated intensity map of the C$^{18}$O ($J$=1--0) emission.}
    \label{fig:oriona_core_bound}
\end{figure}

\begin{figure}[htbp]
    \centering
    \includegraphics[width=15cm]{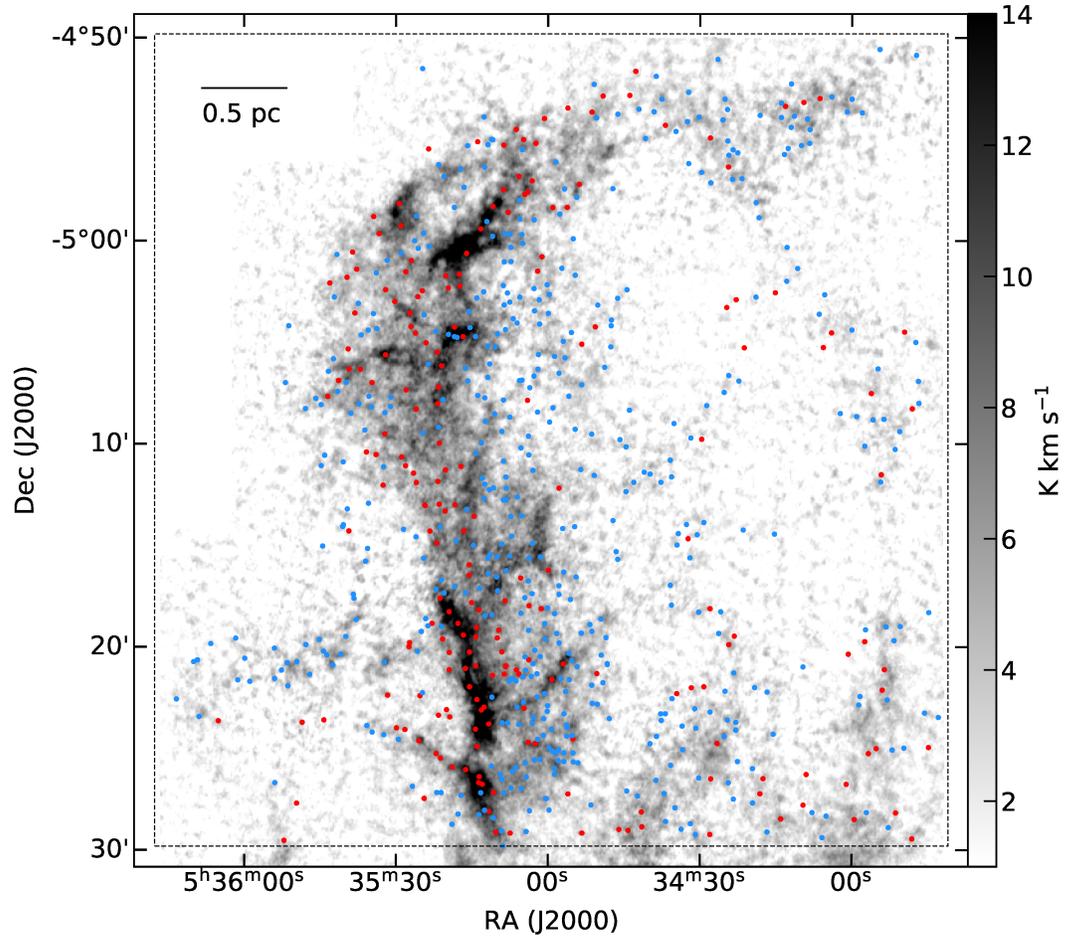}
    \caption{The enlarged view of OMC-1/2/3 area corresponds to the area 1 in Figure \ref{fig:oriona_core_bound}.}
    \label{fig:oriona_core_bound1}
\end{figure}

\begin{figure}[htbp]
    \centering
    \includegraphics[width=15cm]{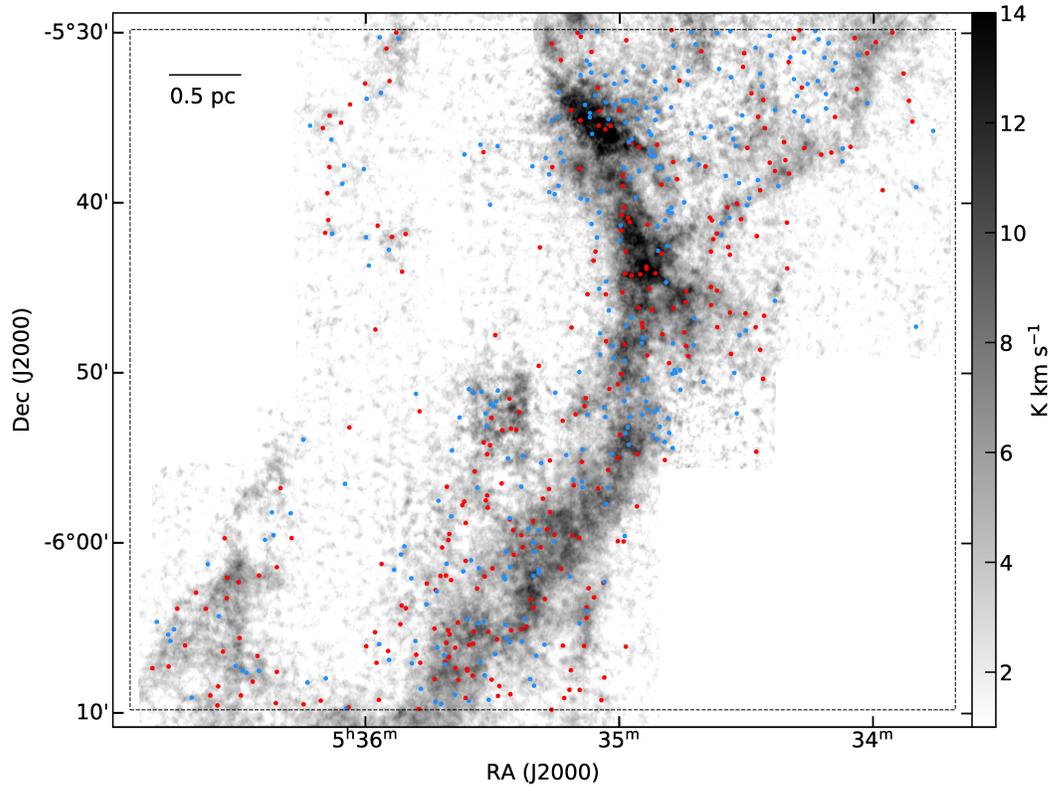}
    \caption{The same figure as Figure \ref{fig:oriona_core_bound1} for OMC-4/5 area which is the area 2 in Figure \ref{fig:oriona_core_bound}.}
    \label{fig:oriona_core_bound2}
\end{figure}

\begin{figure}[htbp]
    \centering
    \includegraphics[width=15cm]{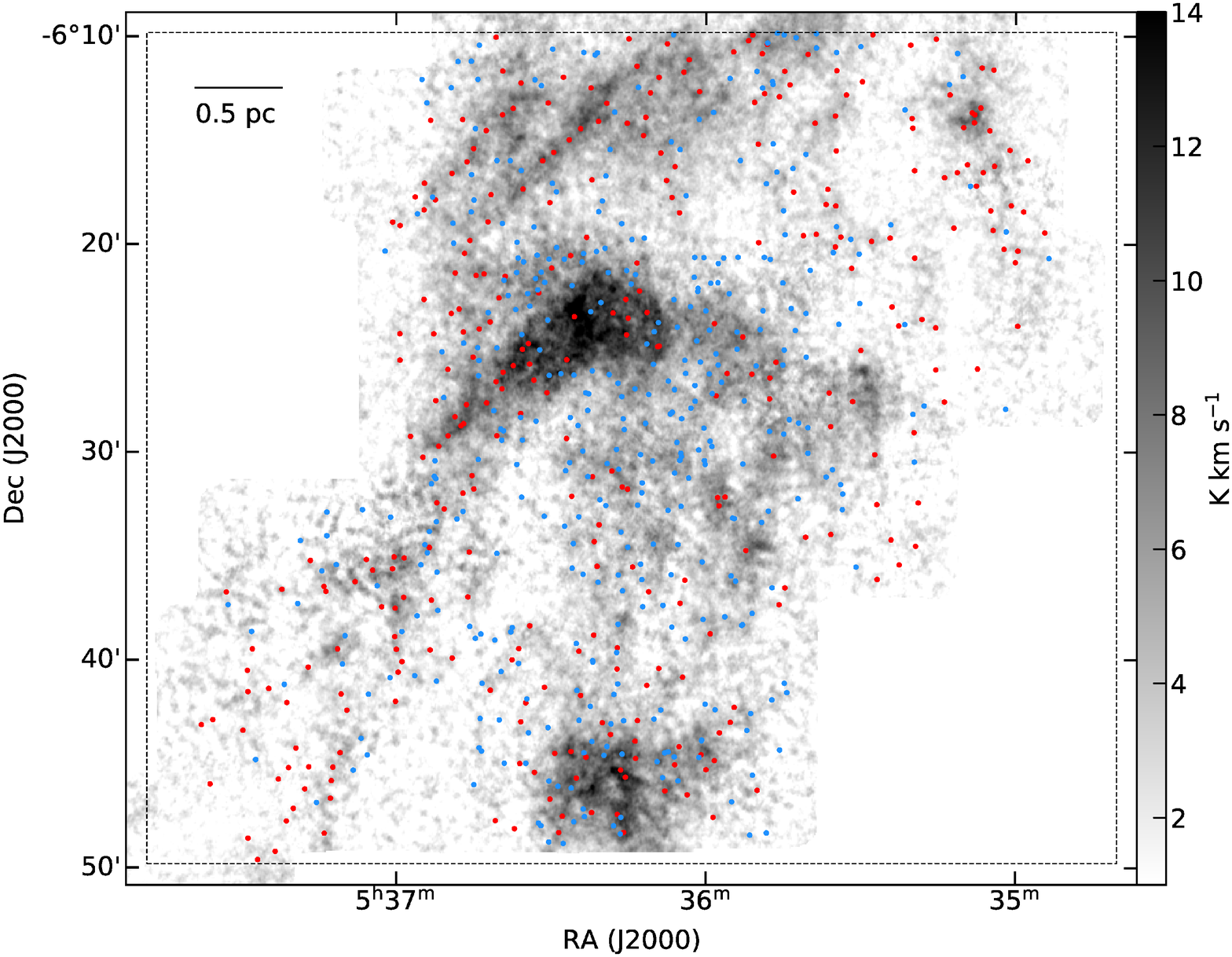}
    \caption{The same figure as Figure \ref{fig:oriona_core_bound1} for L1641N/V380Ori area which is the area 3 in Figure \ref{fig:oriona_core_bound}.}
    \label{fig:oriona_core_bound3}
\end{figure}

\begin{figure}[htbp]
    \centering
    \includegraphics[width=15cm]{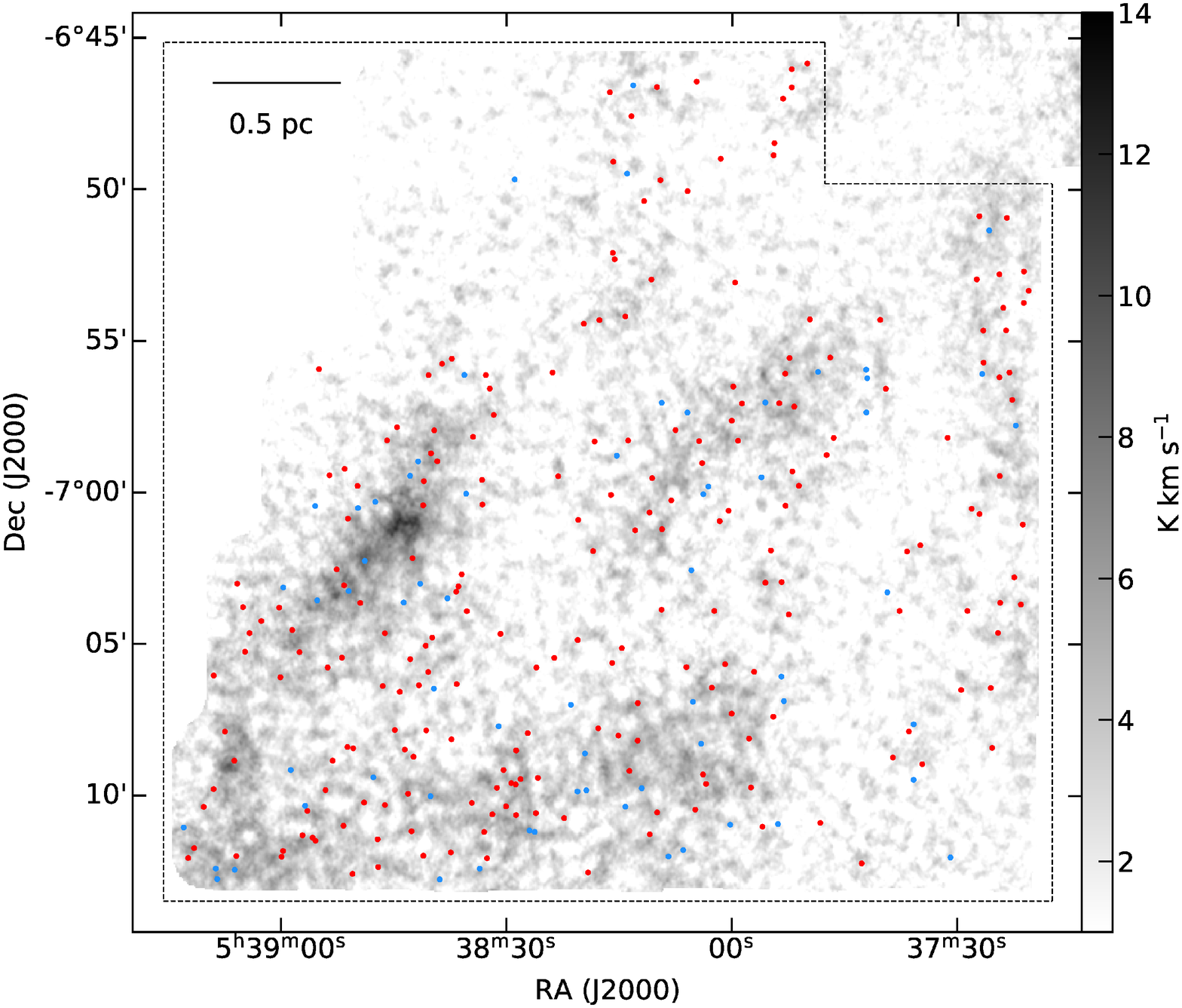}
    \caption{The same figure as Figure \ref{fig:oriona_core_bound1} for L1641C area which is the area 4 in Figure \ref{fig:oriona_core_bound}.}
    \label{fig:oriona_core_bound4}
\end{figure}

Similar to Paper I, the virial mass and virial ratio of a core are calculated as
\begin{equation}
    M_{\mathrm{vir}}=126\left(\frac{R_{\mathrm{core}}}{\mathrm{pc}}\right)
    \left(\frac{\Delta V_{\mathrm{core}}}{\mathrm{km\ s^{-1}}}\right)^{2} \ ,
\end{equation}
and
\begin{equation}
    \alpha_{\mathrm{vir}}=\frac{M_{\mathrm{vir}}}{M_{\mathrm{core}}} \ ,
\end{equation}
where the core is assumed to be a centrally condensed sphere without magnetic support and external pressure.
We define a gravitationally-bound core (hereafter referred to as a bound core) as a core having a virial ratio of smaller than 2 ($\alpha_{\mathrm{vir}} < 2$).
The cores with larger virial ratios are defined as a gravitationally-unbound
(hereafter referred to as an unbound core).
We classified 1045 starless cores as bound cores which comprise $\sim 46\%$ of the starless cores.
The spatial distributions of bound and unbound starless cores in Orion A are shown in Figure \ref{fig:oriona_core_bound} with red and blue dots.
Most of the bound cores are located in the inner regions of the main filamentary structure. The unbound cores are distributed over a much larger area.
In addition, more unbound cores are distributed in the western part of the main filament in the northern region.
In the southern region, the fraction of bound cores is larger.
We present the blow-ups of several regions in Figures \ref{fig:oriona_core_bound1}, \ref{fig:oriona_core_bound2}, \ref{fig:oriona_core_bound3}, and \ref{fig:oriona_core_bound4}.

We calculate the free-fall time of a core with a core density $\rho_\mathrm{core}$ as follows.
\begin{equation}
    t_{\rm ff} = \sqrt{\frac{3\pi}{32 G\rho_{\rm core}}}
\end{equation}
To derive the mean core density, we assumed a uniform sphere with a radius $R_\mathrm{core}$ and used a mean molecular weight of $\mu=2.8$.

Previous unbiased core surveys have roughly estimated
the core lifetime statistically using the number ratio of prestellar cores and YSOs \citep{beichman86,lee99,jessop00,das21}.
Here, we estimate the prestellar core lifetimes for four subregions, applying the same procedure.
We assume that all Class II YSOs have the same median lifetime of
$2\times 10^6$ yr \citep{evans09}.

\begin{equation}
    \tau_{\rm SF} = \frac{\rm number \ of \ prestellar \ cores}{\rm number \ of \ Class \ II \ objects} \times (2 \times 10^6 \ \rm yr)
\label{eq:lifetime}
\end{equation}
An important assumption is that the star formation rate is constant over the cloud lifetime, which is at best an approximation assumption \citep[e.g., ][]{federrath15, burkhart18, grudic19}.

We note that a smaller value of the virial ratio may be suitable as a boundary between bound and unbound cores when we consider magnetic support.
In Appendix \ref{appsec:virial_ratio}, we set $\alpha_{\mathrm{vir}} = 1$ as the boundary and conducted the analyses of core lifetimes and CMFs.

\subsection{Physical Properties of Cores}
\label{subsec:core_property}

\begin{figure}[htbp]
    \centering
    \includegraphics[width=15cm]{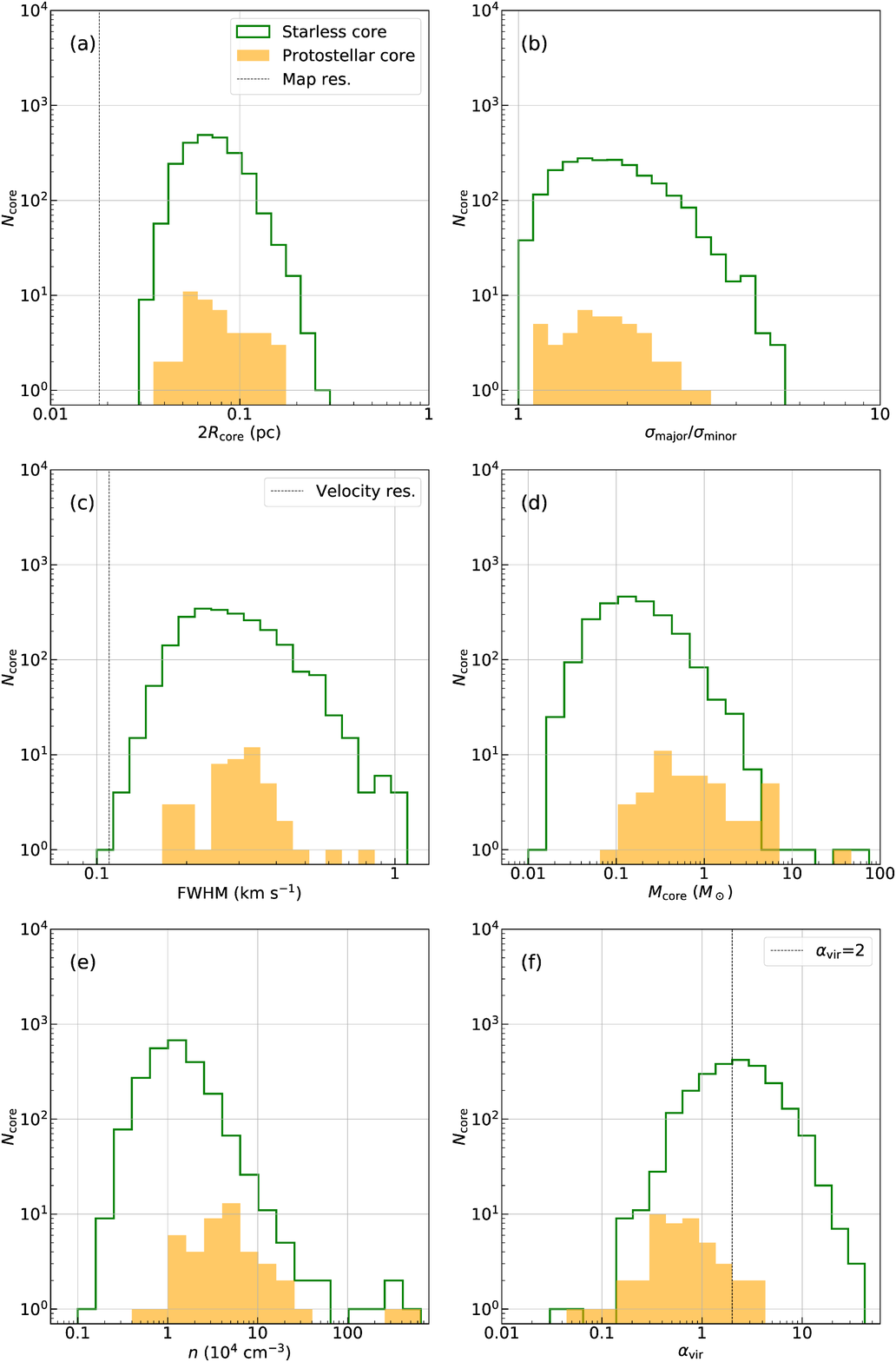}
    \caption{Histograms of properties of
    starless cores (solid green line) and protostellar cores (dashed orange line) in Orion A:
    (a) diameter, (b) aspect ratio, (c) velocity width in FWHM, (d) mass, (e) number density
    and (f) virial ratio.
    In panel (f) the vertical dotted line represents $\alpha_\mathrm{vir}=2$.}
    \label{fig:histo_oriona_sl_ps}
\end{figure}

In Orion A, more than 50 \% of starless cores are not bound by gravity.
The number of protostellar cores is much smaller than that of starless cores.
We summarize the minimum, maximum, mean, and standard deviation of physical properties of identified cores
in Table \ref{tab:oriona_property}.
The histograms of physical properties including the diameter, aspect ratio, FWHM velocity width, mass, number density,
and virial ratio of identified cores in Orion A are shown in Figure \ref{fig:histo_oriona_sl_ps}.
In addition, we conducted a two-sample Kolmogorov–Smirnov (KS) test of the physical properties of starless cores and protostellar cores as listed in Table \ref{tab:ks_test_oriona}. In the KS test, the null hypothesis that two samples have identical distributions is rejected with the significance level of 5\% when the p-value is less than 0.05.
For example, we can say that the diameters of bound starless cores and unbound starless cores have different distributions since the p-value is below 0.05.
However, it is not clear that the two samples have the same distributions even if the p-value of the KS test is larger than 0.05.
According to the results of KS test, there are no apparent differences in the diameter and aspect ratio between starless cores and protostellar cores.
In contrast, the distributions between starless cores and protostellar cores of velocity width, mass, and density are different.
Especially, it is clearly seen that the protostellar cores tend to have larger masses and densities than the starless cores.
As a result, the virial ratio tends to be smaller for protostellar cores.
The larger mass for protostellar cores is also observed in other star-forming regions,
e.g., $\rho$ Oph \citep{maruta10} and the Dragon infrared dark cloud also known as G28.37+0.07 or G28.34+0.06 \citep{kong21}.
\citet{kong21} argued that the larger mass in protostellar cores is evidence of the core growth.
The 91 \% of the protostellar cores are also gravitationally bound, whereas only 46 \% of the starless cores are bound.
Thus, the fraction of bound protostellar cores is greater than bound starless cores.

\begin{figure}[htbp]
    \centering
    \includegraphics[width=15cm]{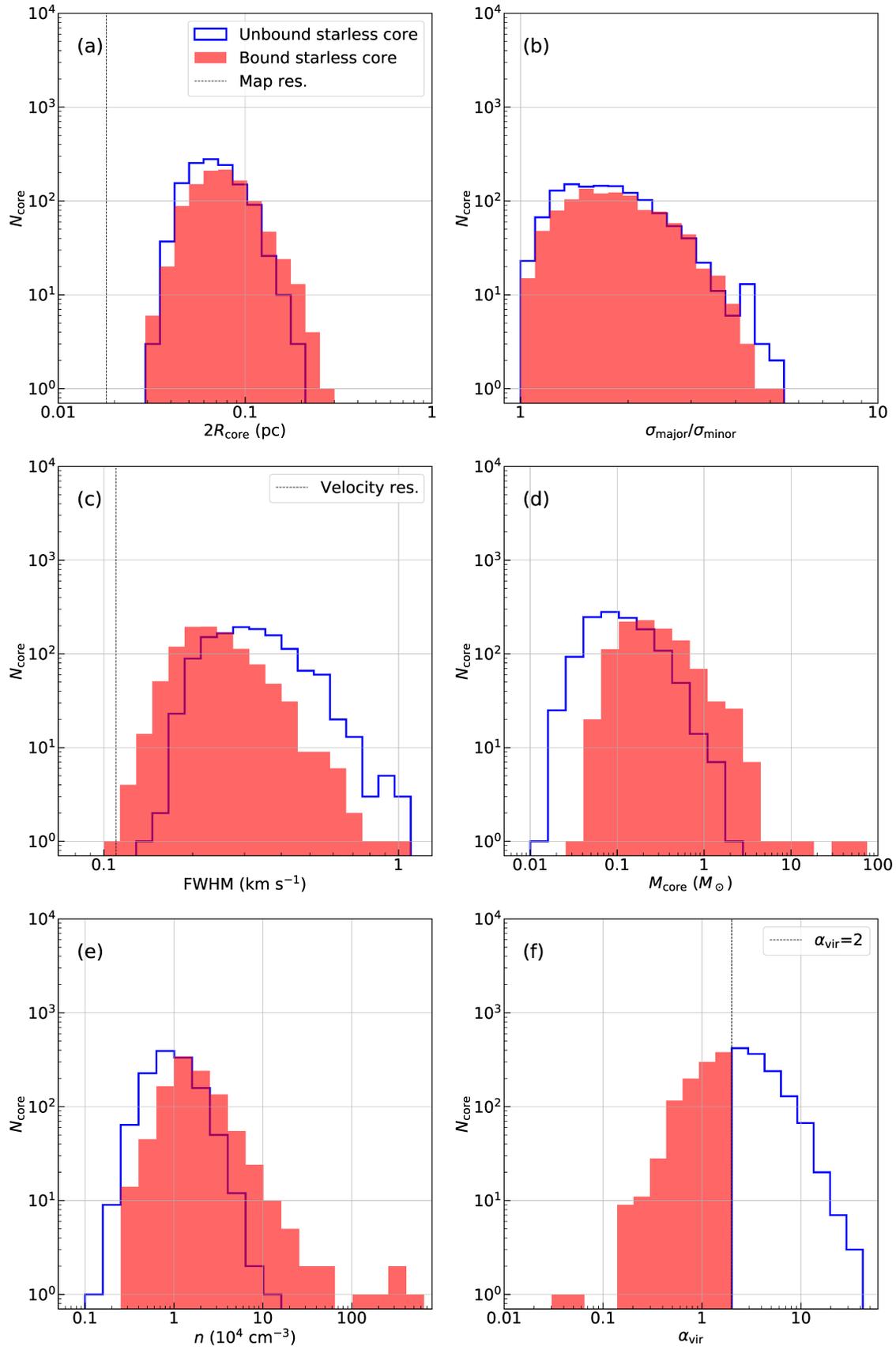}
    \caption{Histograms of properties of bounded starless cores (thick solid blue line),
    and unbounded starless cores
    (thick dashed cyan line) in Orion A:
    (a) diameter, (b) aspect ratio, (c) velocity width in FWHM, (d) mass, (e) number density, and (f) virial ratio.
    }
    \label{fig:histo_oriona_bound_unbound}
\end{figure}

Figure \ref{fig:histo_oriona_bound_unbound} shows the histograms of core physical properties for bound and unbound starless cores in Orion A.
The unbound cores tend to have slightly smaller diameters, similar aspect ratios, and larger velocity widths compared to the bound cores.
The mean masses and densities for bound cores are larger than those of unbound cores.
The differences between the distributions of bound and unbound cores are also confirmed with the results of KS test (see Table \ref{tab:ks_test_oriona}).
We can conclude that the bound and unbound cores have different distributions of diameters, velocity widths, mass, and density with a significance level of 1\%.

\begin{figure}[htbp]
    \centering
    \includegraphics[width=15cm]{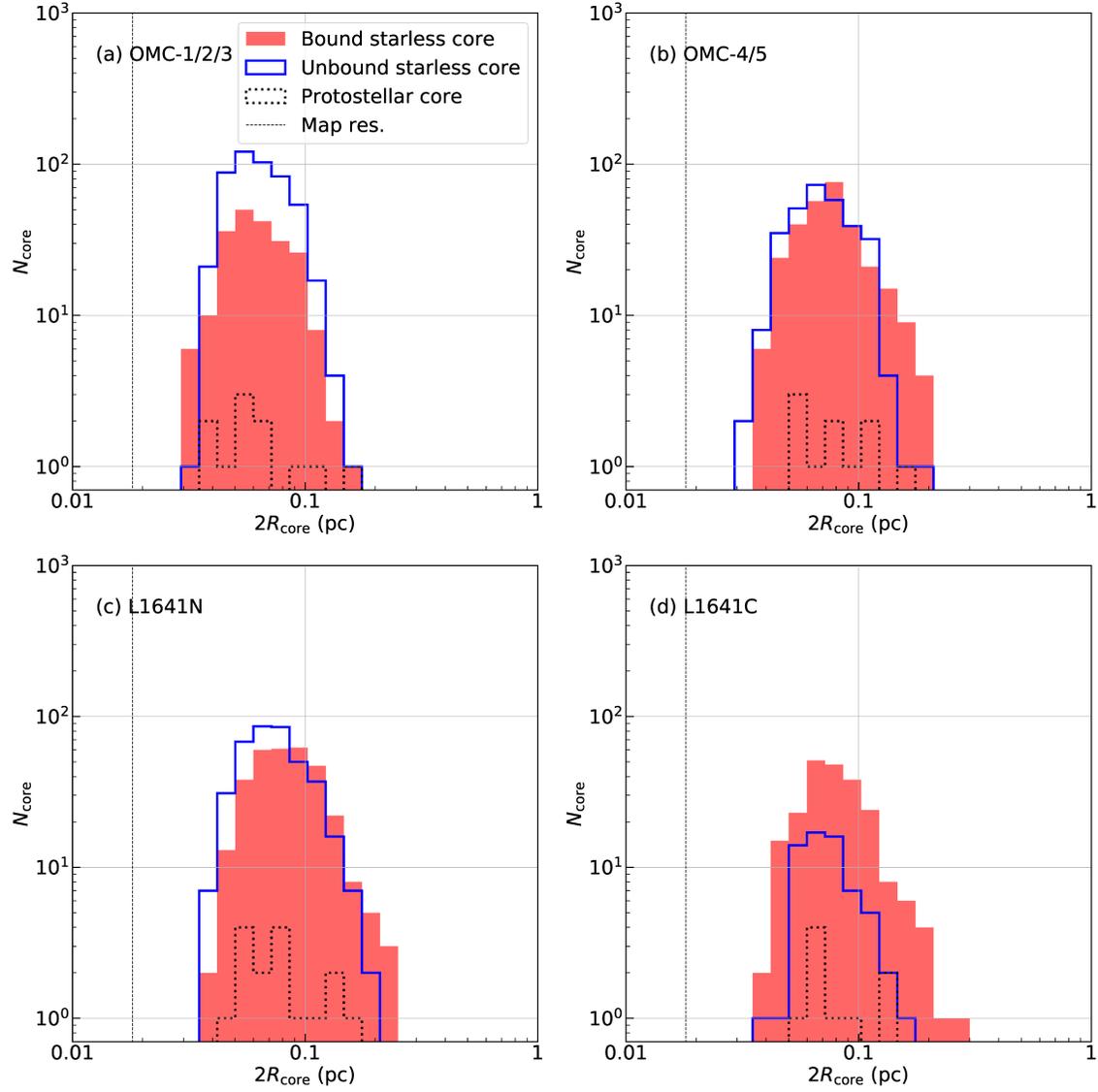}
    \caption{Histograms of core diameters for
    unbound starless cores (red), bound starless cores (blue), and protostellar cores (dashed line) in the four subregions:
    (a) OMC-1/2/3. (b) OMC-4/5, (c) L1641N/V380Ori, and (d) L1641C, respectively.}
    \label{fig:histo_size}
\end{figure}

\begin{figure}[htbp]
    \centering
    \includegraphics[width=15cm]{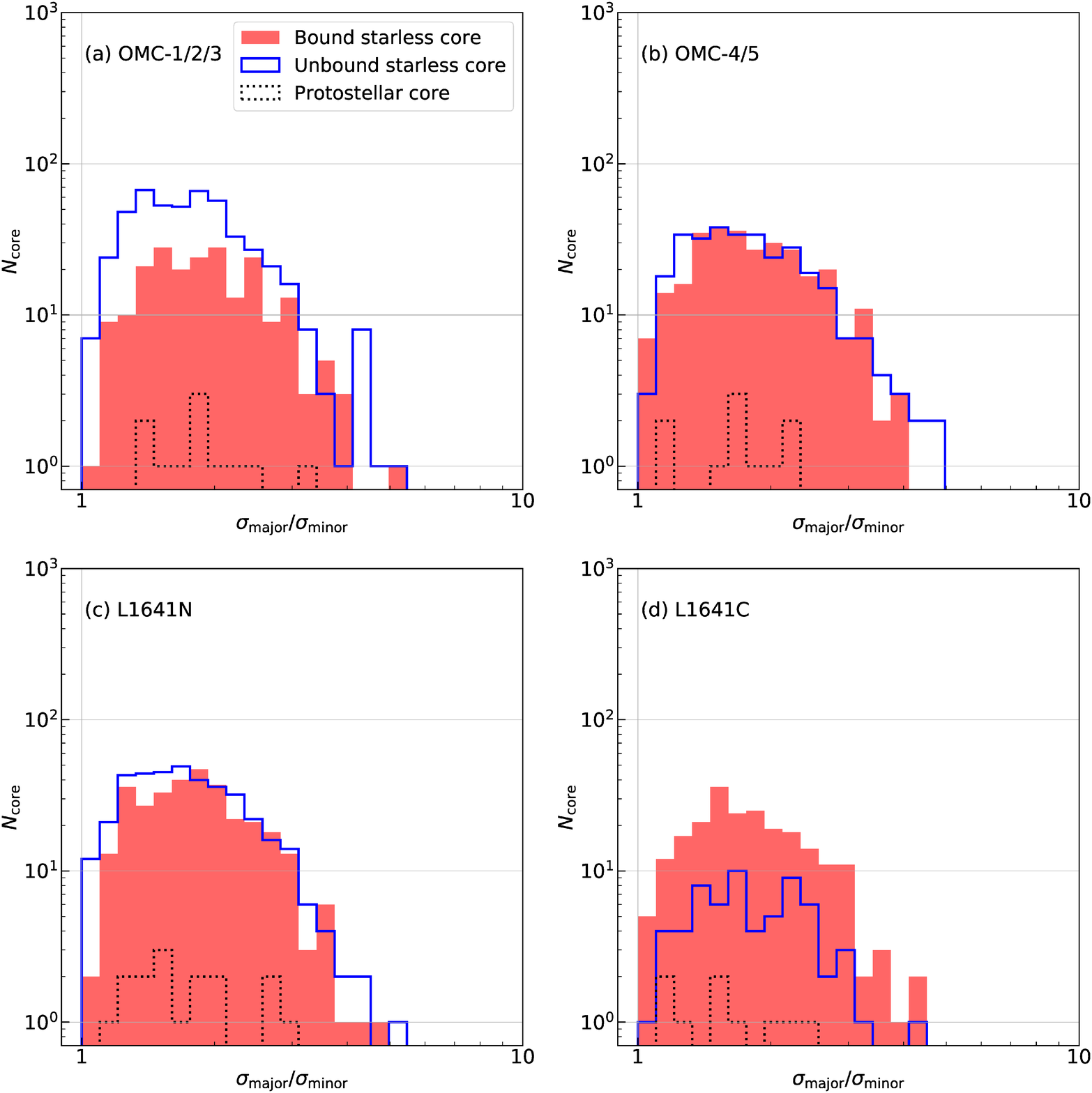}
    \caption{Histograms of core aspect ratios for
    unbound starless cores (red), bound starless cores (blue), and protostellar cores (dashed line) in the four subregions.
    Each panel is for the same area as Figure \ref{fig:histo_size}.}
    \label{fig:histo_aspect}
\end{figure}

\begin{figure}[htbp]
    \centering
    \includegraphics[width=15cm]{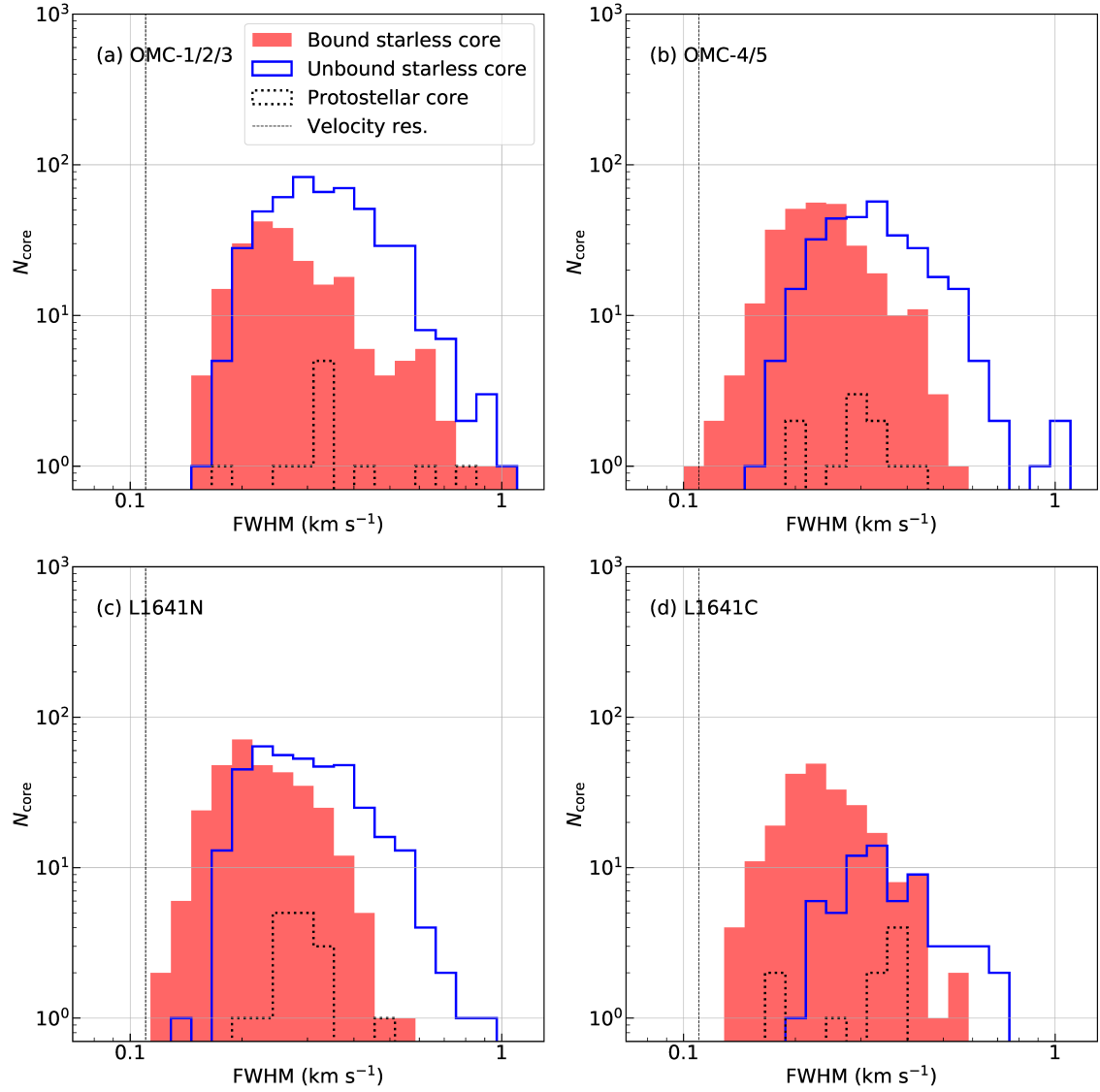}
    \caption{Histograms of FWHM velocity widths for
    unbound starless cores (red), bound starless cores (blue), and protostellar cores (dashed line) in the four subregions.
    Each panel is for the same area as Figure \ref{fig:histo_size}.}
    \label{fig:histo_fwhm}
\end{figure}

\begin{figure}[htbp]
    \centering
    \includegraphics[width=15cm]{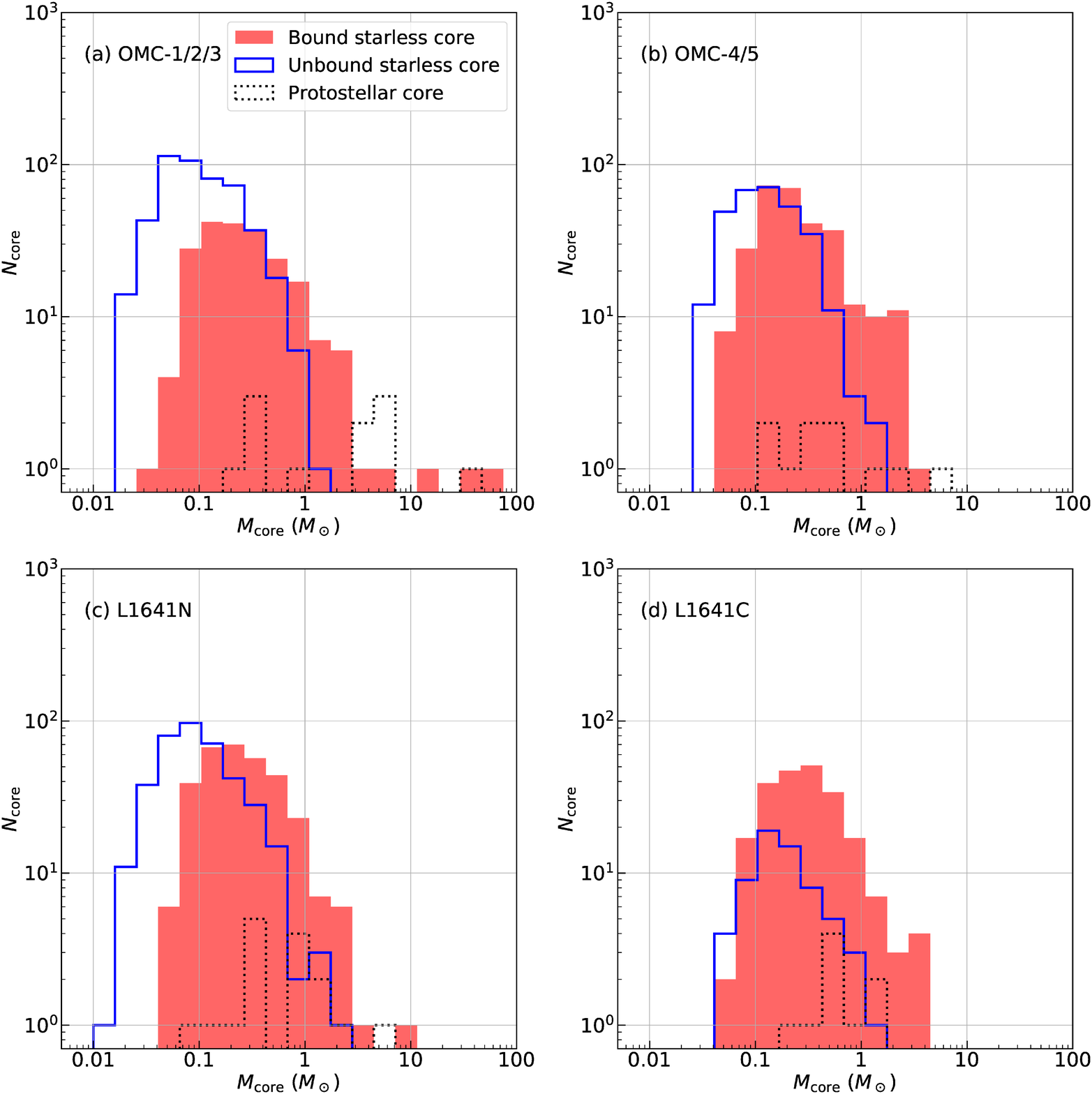}
    \caption{Histograms of core masses for
    unbound starless cores (red), bound starless cores (blue), and protostellar cores (dashed line) in the four subregions.
    Each panel is for the same area as Figure \ref{fig:histo_size}.}
    \label{fig:histo_mass}
\end{figure}

\begin{figure}[htbp]
    \centering
    \includegraphics[width=15cm]{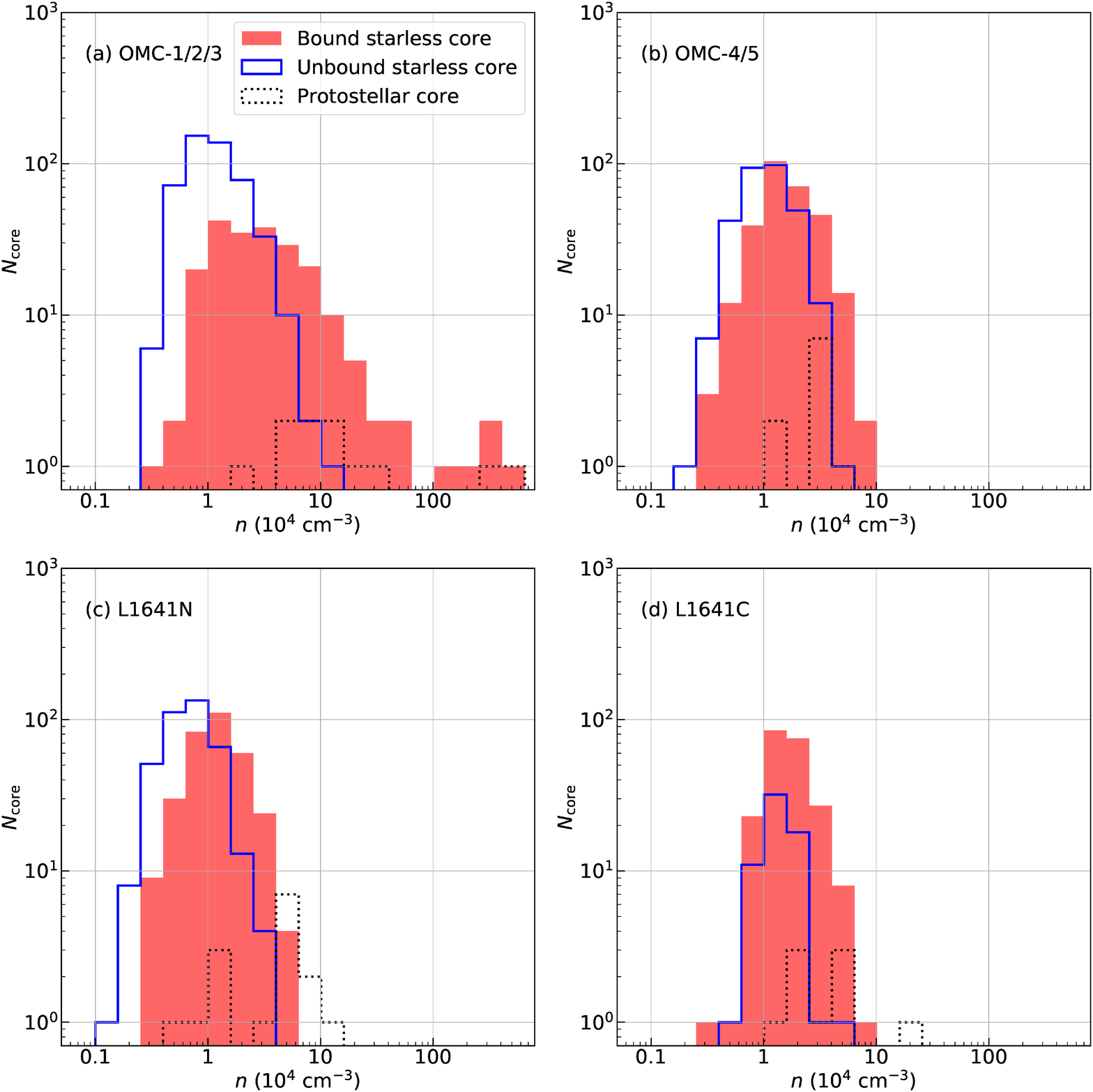}
    \caption{Histograms of densities for
    unbound starless cores (red), bound starless cores (blue), and protostellar cores (dashed line) for the four subregions.
    Each panel is for the same area as Figure \ref{fig:histo_size}.}
    \label{fig:histo_density}
\end{figure}

\begin{figure}[htbp]
    \centering
    \includegraphics[width=15cm]{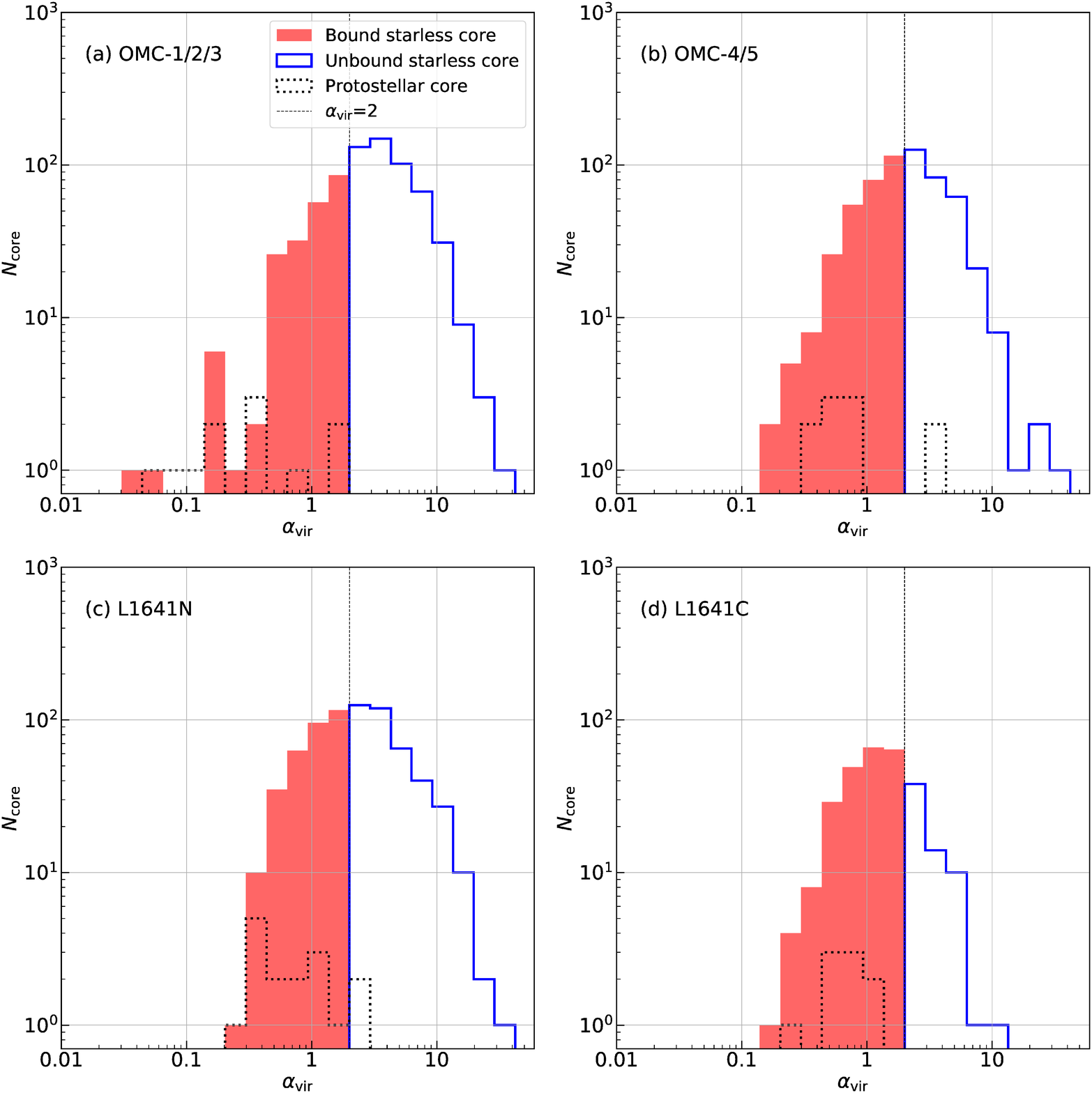}
    \caption{Histograms of virial ratios for
    unbound starless cores (red), bound starless cores (blue), and protostellar cores (dashed line) for the four subregions.
    Each panel is for the same area as Figure \ref{fig:histo_size}.}
    \label{fig:histo_virial}
\end{figure}

Here, we search for the differences and similarities in core physical properties in different cloud environments by comparing the core properties in the four subregions in Figures \ref{fig:core} and \ref{fig:oriona_core_bound}.
The minimum, maximum, mean, and standard deviation of core physical properties of each subregion are summarised in Table \ref{tab:omc123_property}, \ref{tab:omc45_property}, \ref{tab:l1641n_property}, and \ref{tab:l1641c_property}, respectively.
We also show the histograms of the physical properties in Figures \ref{fig:histo_size}, \ref{fig:histo_aspect}, \ref{fig:histo_fwhm}, \ref{fig:histo_mass}, \ref{fig:histo_density}, and \ref{fig:histo_virial}.
Figures \ref{fig:histo_size} and \ref{fig:histo_aspect} show the histograms of core diameter and aspect ratio for four individual regions, respectively.
The distributions of the core diameter look similar from area to area. The distributions of the core aspect ratios also resemble each other.
For quantitative comparison, We performed KS test of core physical properties in each subregion in Appendix \ref{appsec:ks_test}.

Figure \ref{fig:histo_fwhm} shows histograms of the FWHM velocity widths for the four areas.
The distributions of the velocity widths are somewhat different.
In the northernmost area, OMC-1/2/3, some bound starless cores have large velocity widths ($\gtrsim 0.5$ km s$^{-1}$), although the peaks of the velocity width distributions resemble each other ($\approx 0.2 $ km s$^{-1}$).
Such cores with large velocity widths are deficient in other areas.
As for the unbound cores, their velocity widths tend to be large in the entire Orion A region.
The velocity width distribution looks similar in OMC-4/5 and L1641N/V380 Ori.
In L1641C, the velocity widths of the bound cores tend to be small.
When we compare OMC-1/2/3 and the other subregions, the p-values are smaller than 0.01 and the differences are confirmed with the significance level of 5\%.

In case of aspect ratio, we cannot judge whether four subregions have different distributions in each core category or not based on the results of KS test except for bound cores in OMC-1/2/3 and L1641C.
The velocity widths of bound cores in OMC-1/2/3 clearly have a different distribution from them in OMC-4/5 and L1641N since their p-values are smaller than 0.01. The distribution of unbound cores in L1641N is also different from those in OMC-4/5 and L1641C.

Figure \ref{fig:histo_mass} presents histograms of the core mass in the four areas.
The fraction of massive starless cores ($\gtrsim$ 5 -- 10 $M_\odot$) is larger in OMC-1/2/3.
In contrast, no such massive cores reside in other areas except in L1641N where one core has a mass of about 10 $M_\odot$. In L1641C, cores having small masses ($\lesssim$ 0.04 $M_\odot$) are deficient compared to other areas.
Even though OMC-1/2/3 contains more massive starless cores, the p-values of KS test about a core mass between OMC-1/2/3 and the other subregions are not smaller than 0.05.
Then, whether OMC-1/2/3 has a distinguishable mass distribution of bound starless core from the other subregions cannot be judged from this study.

Figure \ref{fig:histo_density} presents histograms of the core densities in the four areas.
The starless cores having higher densities
($\gtrsim$ $10^5$ cm$^{-3}$) reside only in OMC-1/2/3. This arises from the fact that the starless cores in OMC-1/2/3 tend to be more massive with comparable sizes to other regions.
In OMC-1/2/3, the unbound cores have significantly smaller densities than bound cores.
The cores in OMC-4/5 and L1641N/V380 Ori have similar density distributions for both unbound and bound cores.
The range of the core density is narrower in L1641C.
Although differences of bound starless core mass distributions between OMC-1/2/3 and the other subregions are not confirmed, the p-values of their densities are smaller than 0.01. The differences seem to arise from the different core diameters among subregions.

Figure \ref{fig:histo_virial} shows the histograms of the virial ratios in the four areas.
The unbound cores look slightly more abundant in OMC-1/2/3.
In L1641C, the average virial ratio is somewhat smaller than in other regions.
However, The overall distributions are similar, having a peak at around 2--3.
The p-values of KS test among bound starless cores in OMC-1/2/3, OMC-4/5, and L1641N is not smaller than 0.05 and the differences are not confirmed. On the other hand, the p-values among L1641C and the other subregions are smaller than 0.05 and then L1641C statistically has different distributions of virial ratio from the other regions.

\subsection{Correlations among the Core Physical Quantities}
\label{sec:correlations}

\begin{figure}[htbp]
    \centering
    \includegraphics[width=12cm]{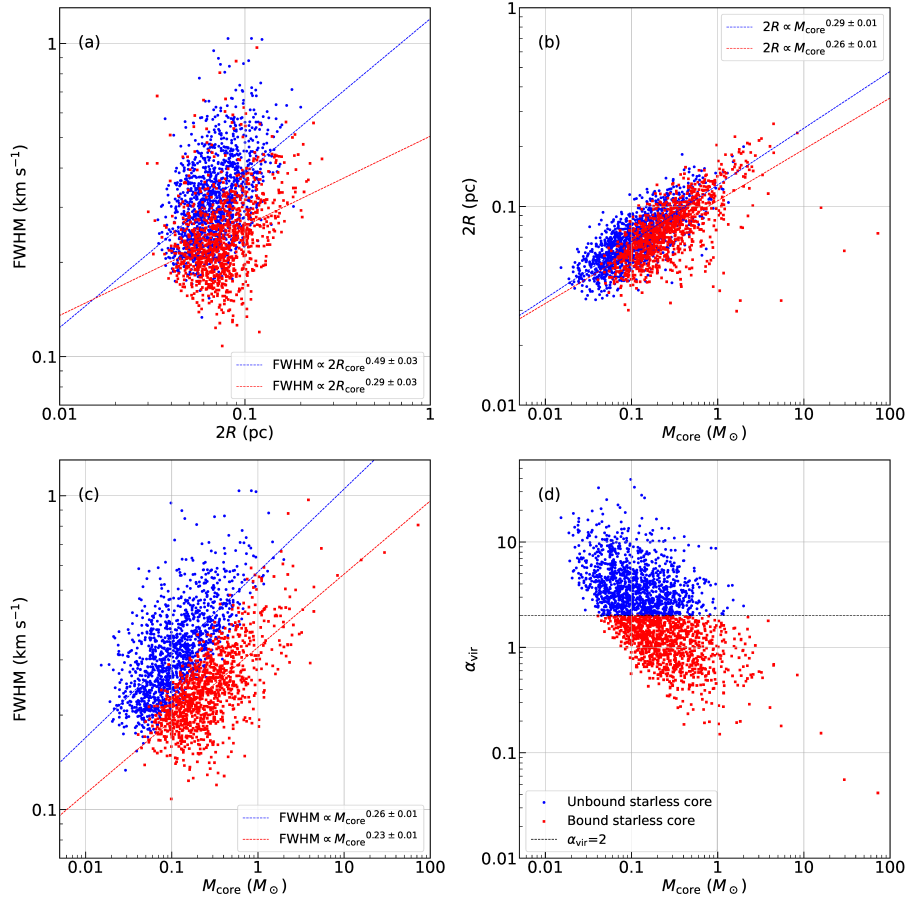}
    \caption{(a) The velocity width -- diameter relation,
    (b) diameter -- mass relation, (c) velocity width -- mass relation and
    (d) virial ratio -- mass relation for starless cores in Orion A.
    The best fit lines for bounded cores, unbounded cores, and all starless cores are shown in (a), (b), and (c).
    The red and blue dots are the bound and unbound cores, respectively.
    The dashed line in (d) at $\alpha_\mathrm{vir}=M_\mathrm{vir}/M_\mathrm{core}=2$ shows the boundary between bound and unbound cores.}
    \label{fig:oriona_correlation}
\end{figure}

Figure \ref{fig:oriona_correlation} shows the correlations of the physical properties of starless cores in Orion A: (a) velocity width -- diameter relation, (b) diameter -- mass relation, (c) velocity width -- mass relation, and (d) virial ratio -- mass relation.
In spite of a large dispersion in the velocity width -- diameter relation, bound starless cores have smaller velocity widths than unbound starless cores for a given diameter.
There are good correlations between diameters and masses and bound cores and unbound cores have similar best-fit power-law indices of $\sim$ 0.3, close to the power-law index of the uniform core density of 1/3.
We note that several bound cores are below the best-fit line of nearly uniform density which implies that they have high densities.
Good correlations are also seen between velocity FWHM and mass and both cores have similar best-fit power-law indices.
However, bound cores have smaller velocity widths than unbound cores when we compare cores having the same masses.
Lastly, the virial ratio -- mass relation shows that cores with larger masses tend to have smaller virial ratios.
The cores below the horizontal dashed line in Figure \ref{fig:oriona_correlation} (d) with virial ratios smaller than 2 are classified as gravitationally bound cores.
These trends are similar to those of cores in other star-forming regions \citep[e.g.,][]{bertoldi92,dobashi01,maruta10,sanchez-monge13,kauffmann13,tanaka13} under the assumption of no magnetic support.

As shown in the previous subsection, core properties depend on the cloud environments.
Particularly, the cores in OMC-1/2/3 have somewhat different properties than those in other areas.
Therefore, the correlations of core properties are expected to show similar environmental differences.
We thus investigate correlations among core physical properties by separating cores into four groups using the areas shown in Figure \ref{fig:core} and \ref{fig:oriona_core_bound}.

\begin{figure}[htbp]
    \centering
    \includegraphics[width=16cm]{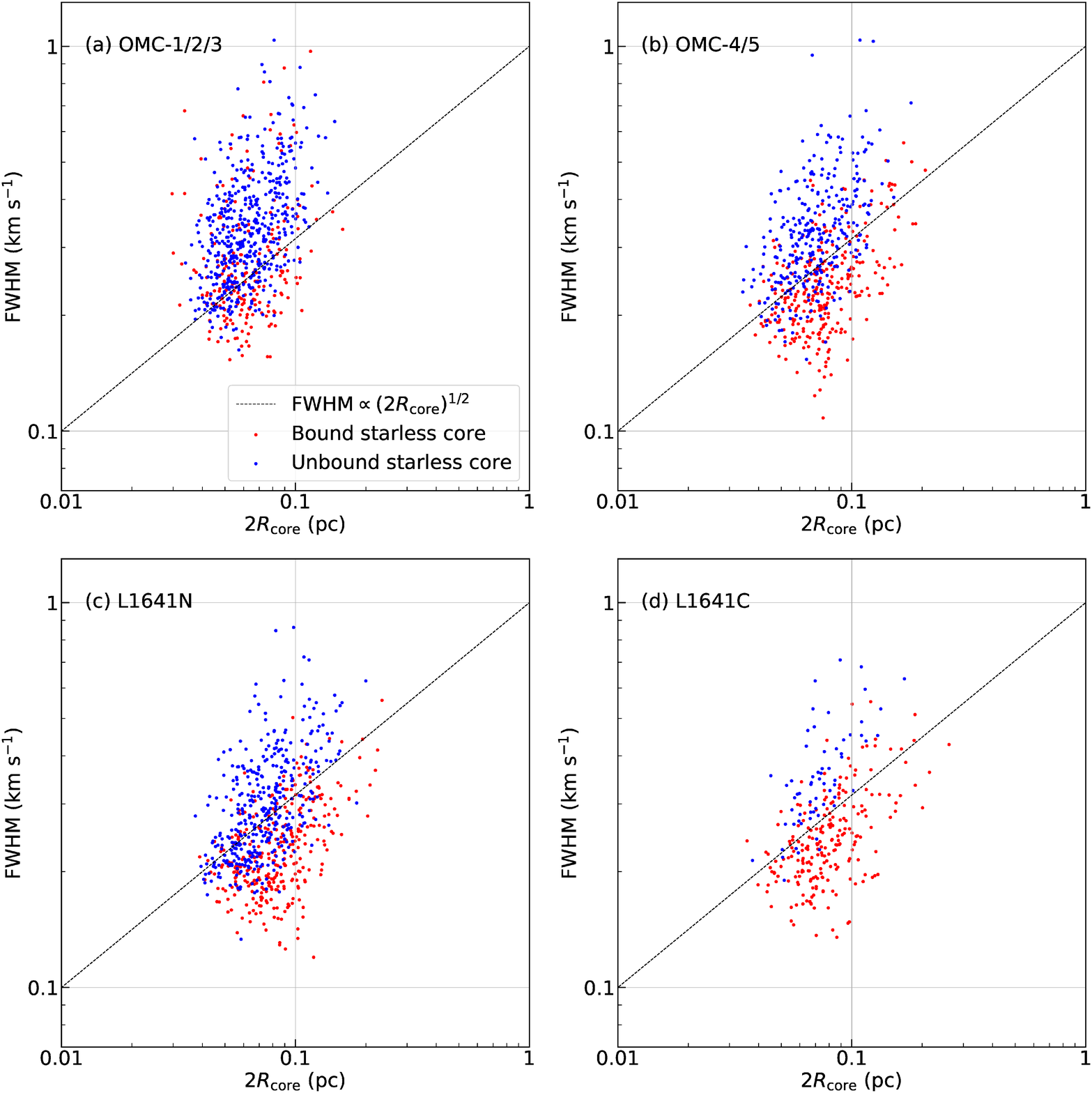}
    \caption{The velocity width -- diameter relation in four subregions.}
    \label{fig:oriona_correlation_v-dv}
\end{figure}

\begin{figure}[htbp]
    \centering
    \includegraphics[width=16cm]{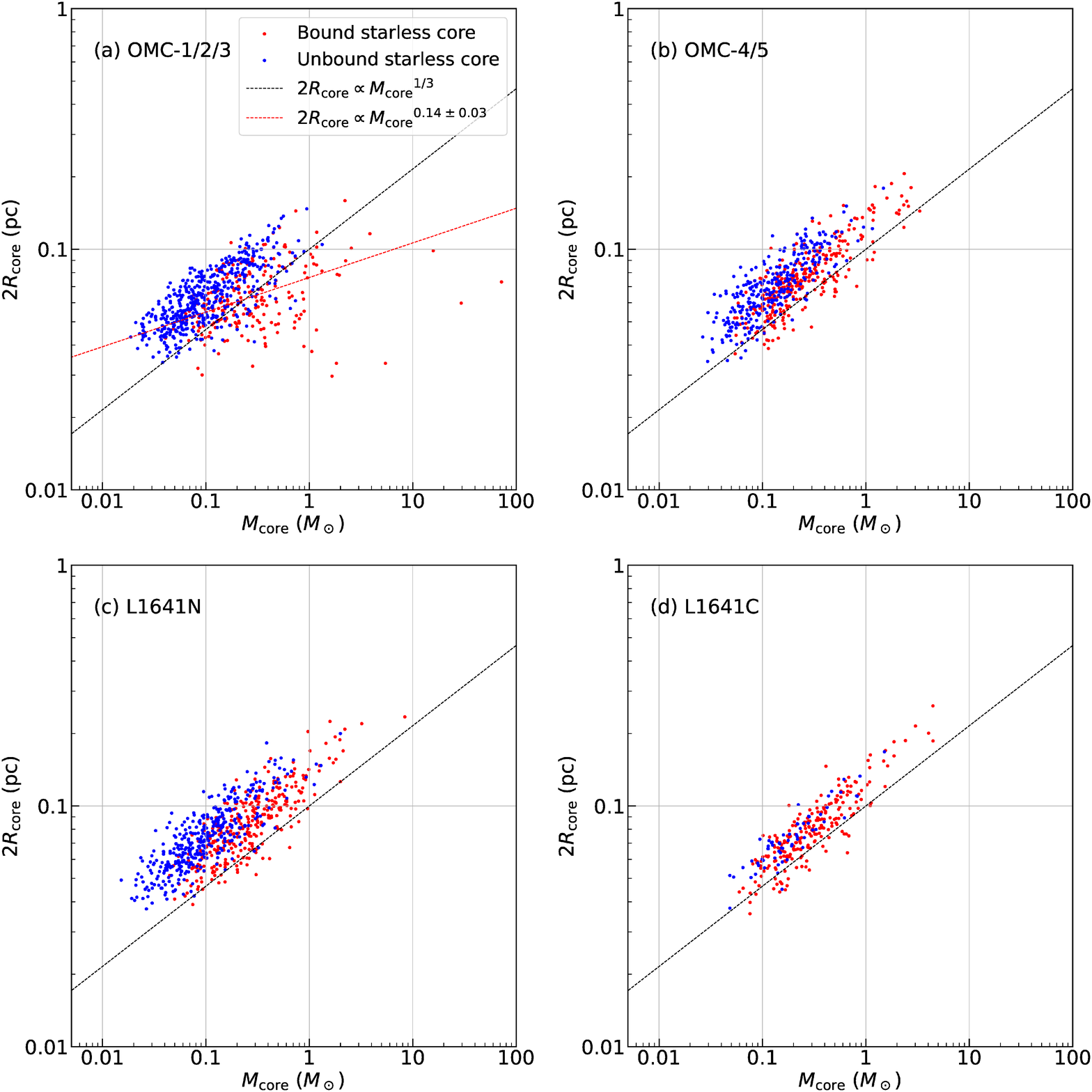}
    \caption{The core diameter -- mass relation in 4 areas.
    The dotted lines denote the relation of $M_{\rm core} \propto (2R_{\rm core})^3$.
    The red and blue dots are the bound ($\alpha_{\rm vir} \le 2$) and unbound ($\alpha_{\rm vir} > 2$) cores, respectively.
    The red dashed line in panel (a) is a best-fit function for low-mass ($<1 M_\odot$) bound starless cores.}
    \label{fig:oriona_correlation_m-r}
\end{figure}

\begin{figure}[htbp]
    \centering
    \includegraphics[width=16cm]{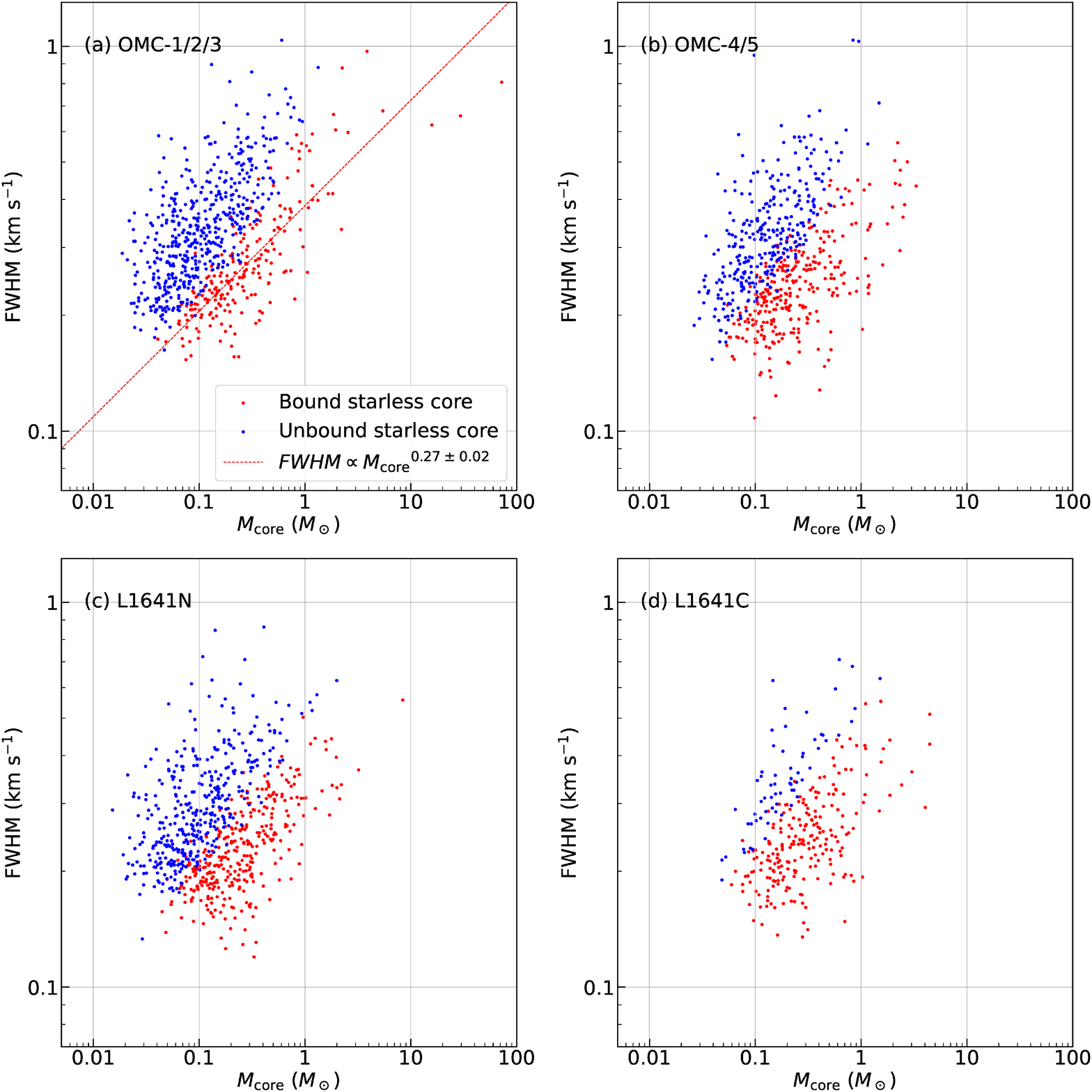}
    \caption{The velocity width -- mass relation in four subregions. The red dashed line in panel (a) represents a best-fit function for low-mass ($<1 M_\odot$) bound starless cores as well as Figure \ref{fig:oriona_correlation_m-r}.}
    \label{fig:oriona_correlation_m-dV}
\end{figure}

Figure \ref{fig:oriona_correlation_v-dv} shows the velocity width -- diameter relation with best-fit lines.
Bound cores in OMC-1/2/3 appear to be distributed in almost the same area as the unbound cores in panel (a). In other words, the bound and unbound cores have similar velocity widths. In contrast,
for other areas, bound cores typically have smaller velocity widths than unbound cores. In addition, in OMC-1/2/3, both bound and unbound cores appear to have larger velocity widths than those in other areas (the fraction of dots distributed above the dotted line is larger).
Figure \ref{fig:oriona_correlation_m-r} presents the core diameter -- mass relation for the four areas.
The core-diameter-mass relation basically follows the $R_\mathrm{core} \propto M_\mathrm{core} ^{1/3}$ dependence, which is shown as a dashed line in each panel.
In OMC-1/2/3, several massive bound cores appear to deviate from this relation.
For other areas, the correlations look similar to one another.
We note that the spatial resolution may not be sufficient to resolve an inner structure of a core since the mean diameter of a starless core
(0.076 pc, see Table \ref{tab:oriona_property}) is only 4.5 times of the map resolution.
Then, although cores with uniform density also have a power-law index of 1/3 in this relation, actual cores may have internal structures or density gradients.
The velocity width--mass relations and the virial ratio -- mass relations of the four areas are shown in Figures \ref{fig:oriona_correlation_m-dV} and \ref{fig:oriona_correlation_m-alpha}, respectively.
The massive cores ($>10\ M_\odot$) in OMC-1/2/3 have smaller radii and velocity widths than expected from the extrapolation of low-mass cores ($<1\ M_\odot$).
The best-fit slopes of diameter -- mass relation and velocity width--mass relation for bound cores with masses less than 1 $M_\odot$ are 2$R_\mathrm{core} \propto M_\mathrm{core} ^{0.14\pm0.03}$ and FWHM$\propto M_\mathrm{core} ^{0.27\pm0.02}$ as shown in Figure \ref{fig:oriona_correlation_m-r} (a) and Figure \ref{fig:oriona_correlation_m-dV} (a).
These cores have small virial ratios compared to the other bound cores.
Except for that, there is no clear difference in the correlations among the four areas.

\begin{figure}[htbp]
    \centering
    \includegraphics[width=16cm]{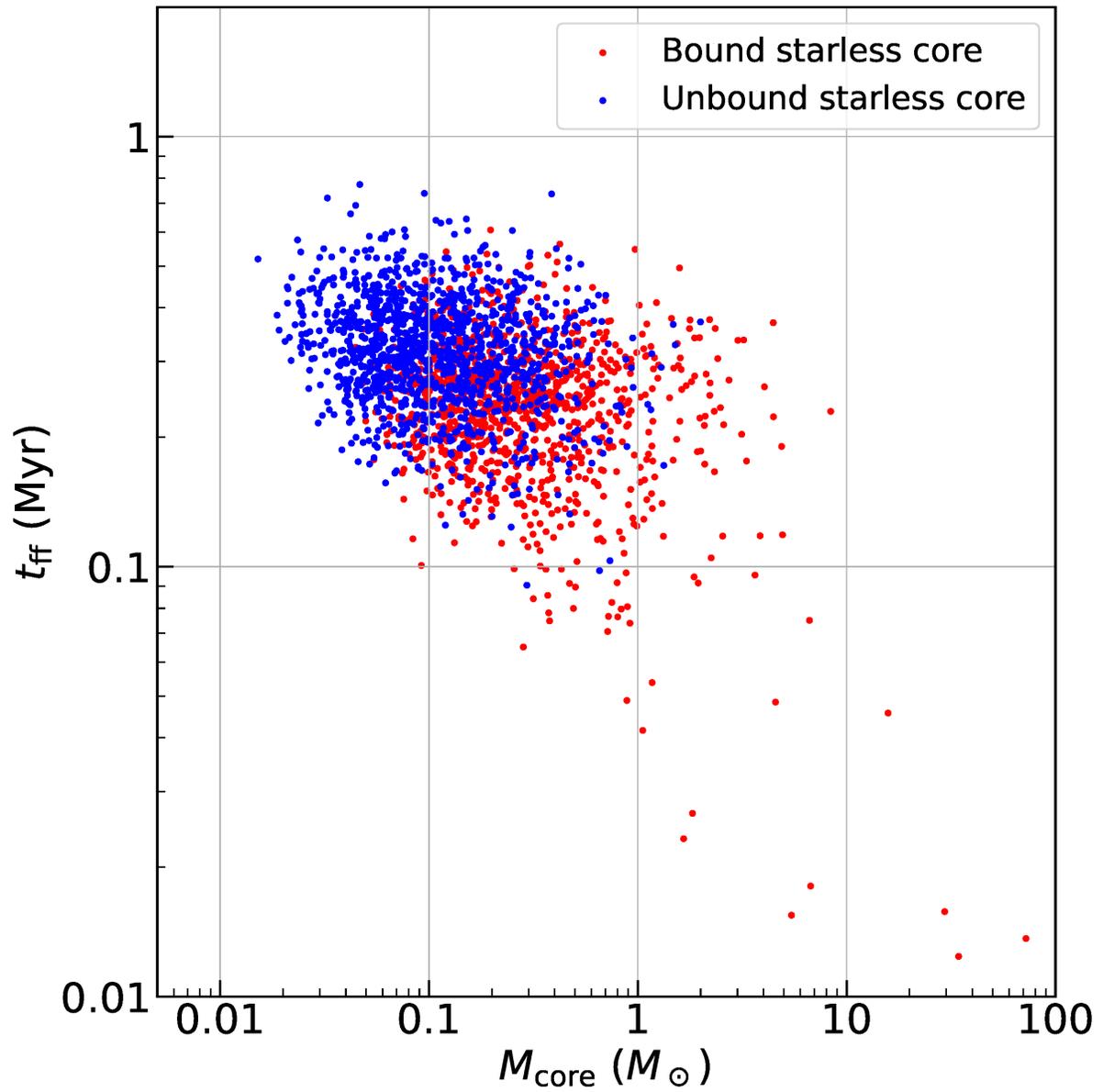}
    \caption{The free-fall times of starless cores as a function of core masses in Orion A.}
    \label{fig:freefall}
\end{figure}

\begin{figure}[htbp]
    \centering
    \includegraphics[width=16cm]{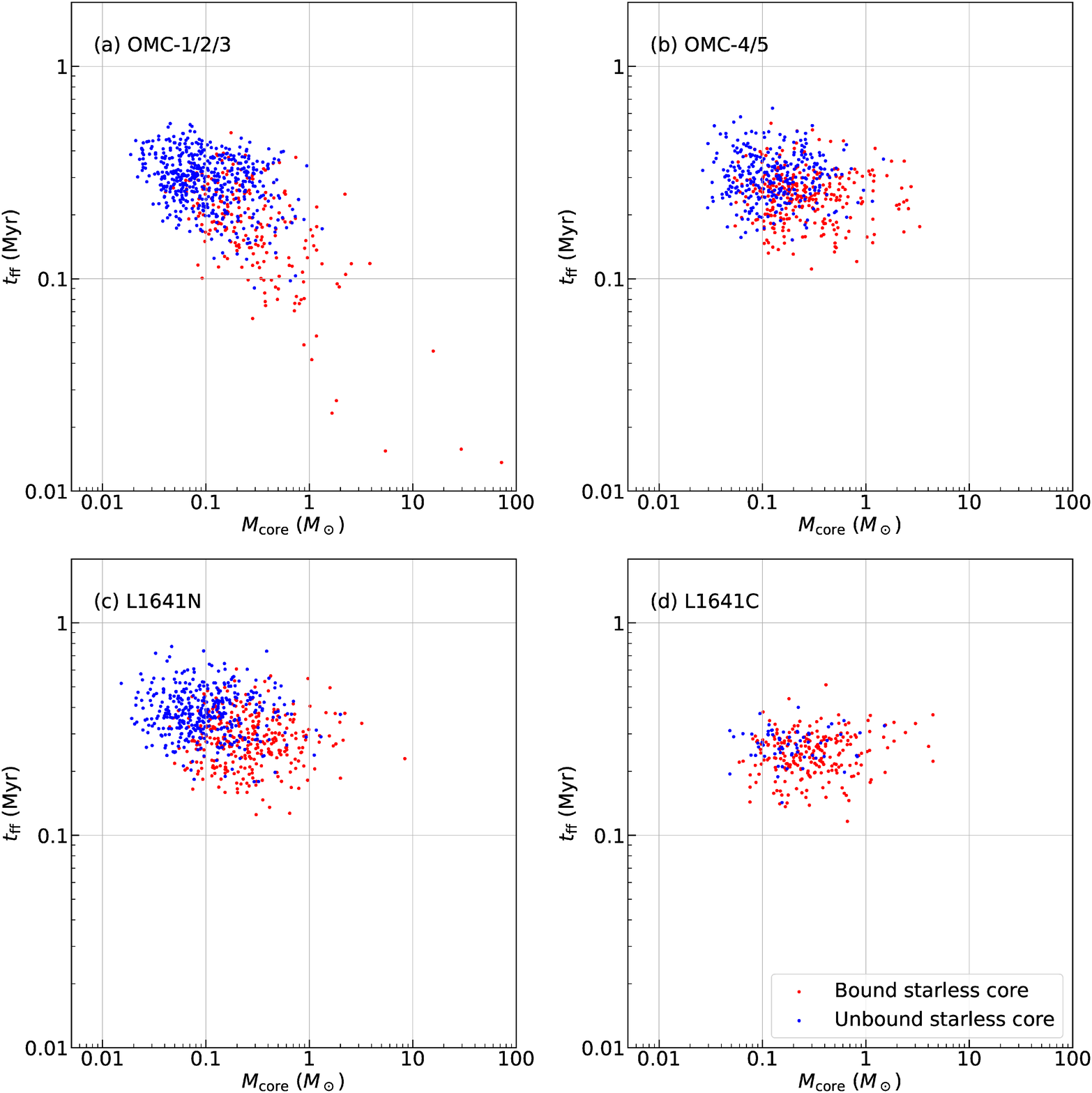}
    \caption{The same figure as Figure \ref{fig:freefall} but for four subregions in Orion A.
    Each panel is for the same area as Figure \ref{fig:histo_size}.}
    \label{fig:freefall_4blocks}
\end{figure}

Figure \ref{fig:freefall} shows the distribution of the core free-fall times as a function of core mass for starless cores in Orion A.
Several high-mass cores have shorter free-fall times than the others.
However, there is no clear correlation between the core mass and the core free-fall time except for several high-mass cores.
Thus, in the following discussion, we assume for simplicity that all bound starless cores have the same free-fall time.
The high-mass cores are more distributed in OMC-1/2/3 than in other areas as discussed in the previous sections.
In other areas, the dependence of the core free-fall time on the core mass is weak and the dispersion is large as shown in Figure \ref{fig:freefall_4blocks}.
The unbound cores, which have smaller masses, tend to have longer free-fall times.
In other words, the blue dots are more broadly distributed in the upper-left part of the panel.

Figure \ref{fig:lifetime_4blocks} shows the derived core lifetimes as a function of the mean density of cores for four subregions.
We assumed that all starless cores will form stars and calculated the lifetimes of not only bound starless cores but also starless cores and unbound cores.
The plots represent the lifetimes of cores with a density above each density calculated with Equation \ref{eq:lifetime} similar to \citet{konyves15}.
Here, we only show the core lifetime--density plots in subregions since the lifetime estimation strongly depends on the choices of the areas. For comparison, we also present the core lifetime--density relation using the data of the whole Orion A in Figure \ref{fig:lifetime}. In this plot, the core lifetimes approach the free-fall time for densities larger than $10^5$ cm$^{-3}$, whereas they are several to 10 free-fall times for smaller densities.
The typical core lifetime decreases with core density.
Most of the estimated core lifetimes for bound cores lie between 5 and 30 times the free-fall time which is shown as a shaded area in each panel in Figure \ref{fig:lifetime_4blocks}.
For densest cores with $10^6 $ cm$^{-3}$, which reside only in OMC-1/2/3, the core lifetime is close to $\sim$ 5 $t_{\rm ff}$.
We note that we may undercount Class II objects in OMC-1/2/3 as mentioned in \ref{subsec:YSO} and thus overestimate core lifetimes in this area.
If the real number of Class II objects is doubled, the core lifetime becomes a half, resulting in 2-3 $t_{\rm ff}$ in OMC-1/2/3.

These characteristics seem to be qualitatively consistent with earlier studies.
However, our core lifetimes are somewhat longer than previous estimates. For example,
\citet{konyves15} derived the core lifetime in Serpens south, which shows
a trend that the lifetime reaches one free-fall time at a high density of $10^5$ cm$^{-3}$.
\citet{das21} proposed that the core lifetime in Aquila Rift can be fitted well by the magnetically critical model.
Since our estimated core lifetime is longer than that of \citet{das21},
the Orion A cores might be more strongly magnetized.
In fact, strong magnetic fields are observed around OMC-1 from polarization observations \citet{hwang21} who reported that the mass-to-magnetic flux ratio is smaller than its critical value for the outer parts of OMC-1 ridge.
Interestingly, the core lifetimes for the dense unbound cores become slightly shorter than the free-fall times.
This is partly because there is a lacking unbound cores with high densities. This implies that the assumption that all unbound cores form stars may be wrong, and the majority of unbound cores may not form stars unless they first become gravitationally bound.

\subsection{Core Physical Quantities along the Declination axis}
\label{subsec:declination}

\begin{figure}[htbp]
    \centering
    \includegraphics[width=16cm]{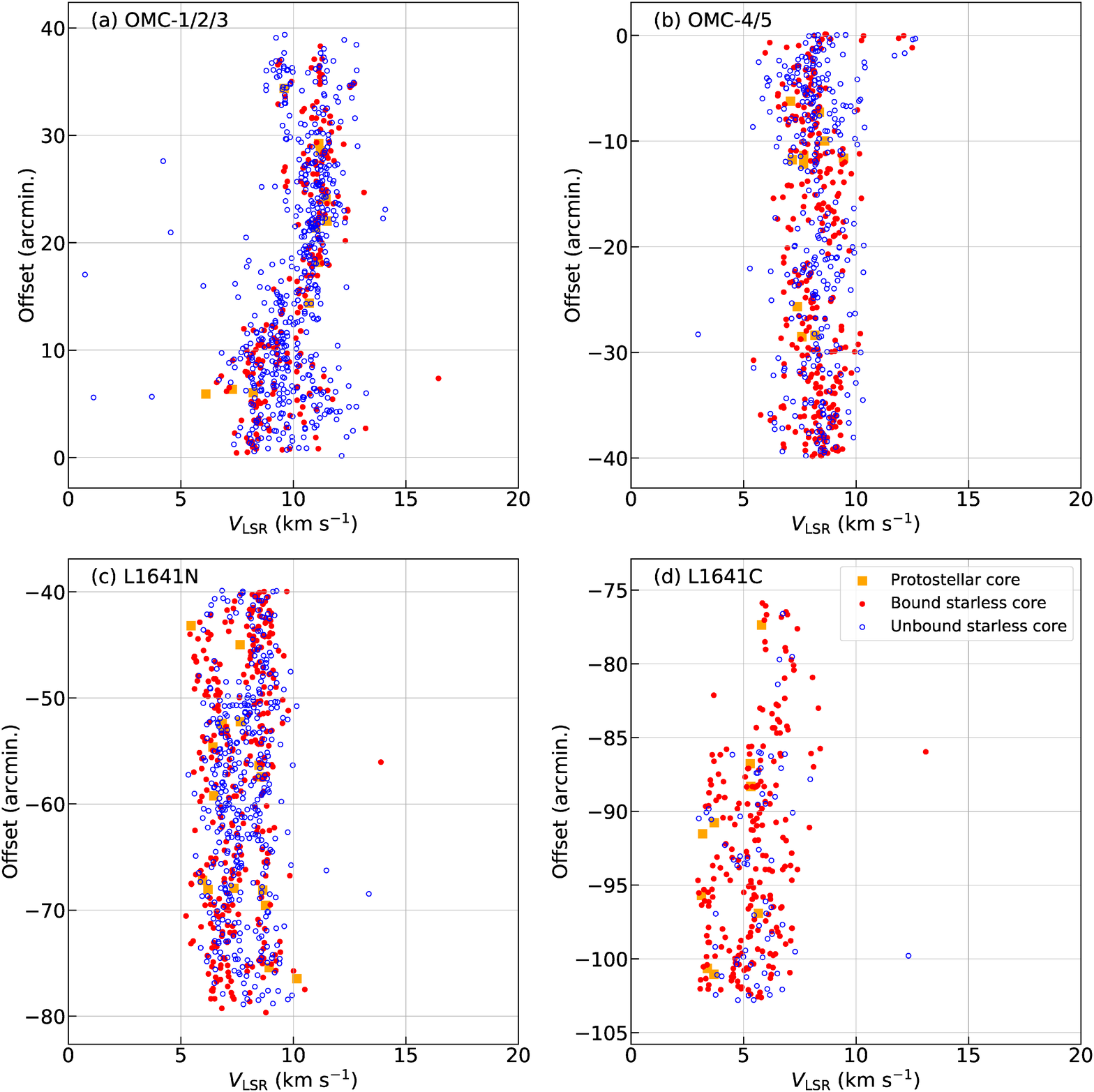}
    \caption{The core declination v.s. the central velocity relation for 4 areas.
    The four panels correspond to the same area of Figure \ref{fig:histo_size}.
    The vertical axis in each panel is the declination in the offset from $-5^\circ30'$ in the unit of arc minute.
    The bound starless cores, unbound starless cores, and protostellar cores are represented as filled red circles, open blue circles, and orange squares, respectively.}
    \label{fig:decl_vcen}
\end{figure}

\begin{figure}[htbp]
    \centering
    \includegraphics[width=16cm]{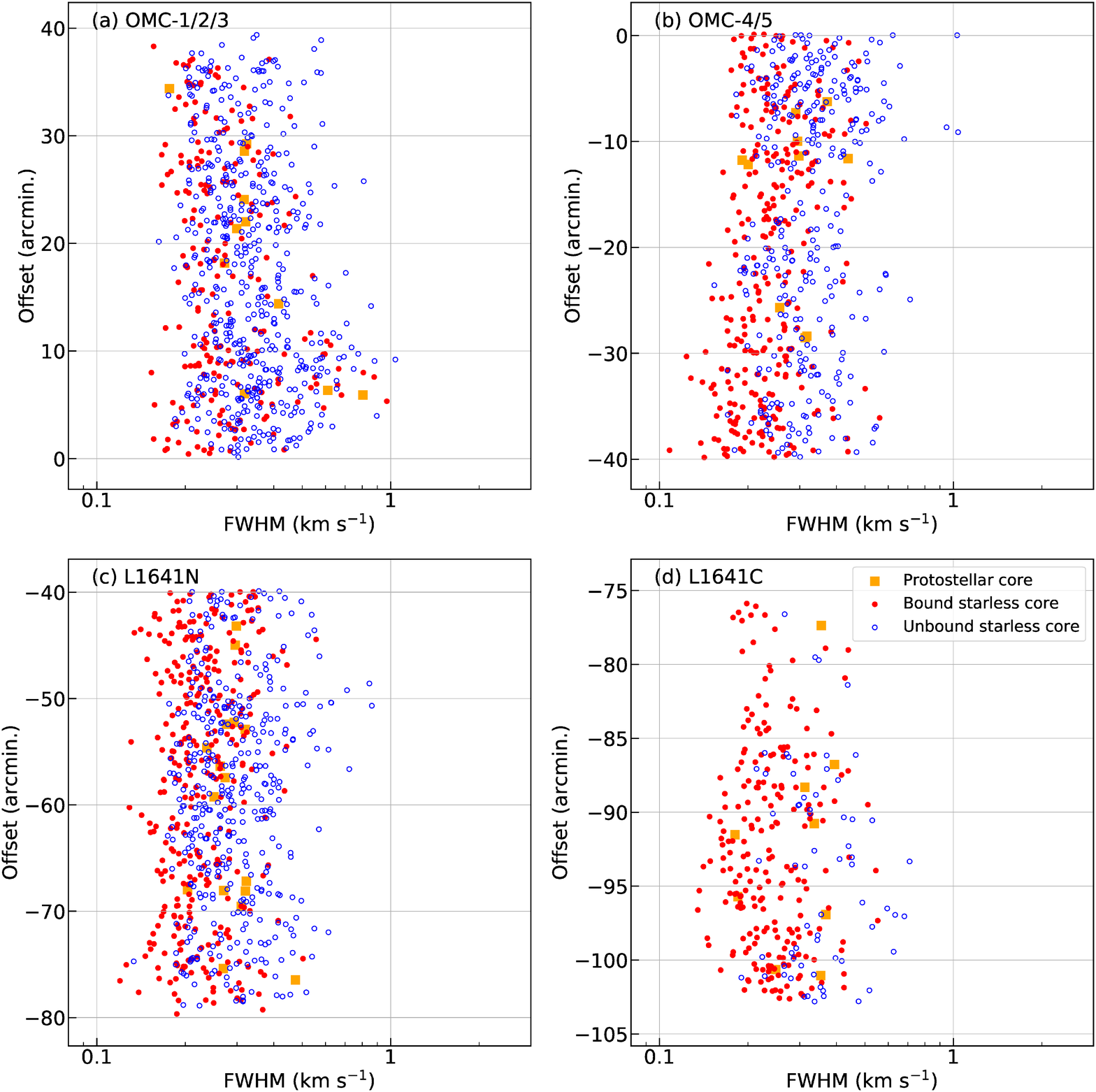}
    \caption{The same figure as Figure \ref{fig:decl_vcen} but for the core declination v.s. the velocity width in FWHM relation.}
    \label{fig:decl_fwhm}
\end{figure}

\begin{figure}[htbp]
    \centering
    \includegraphics[width=16cm]{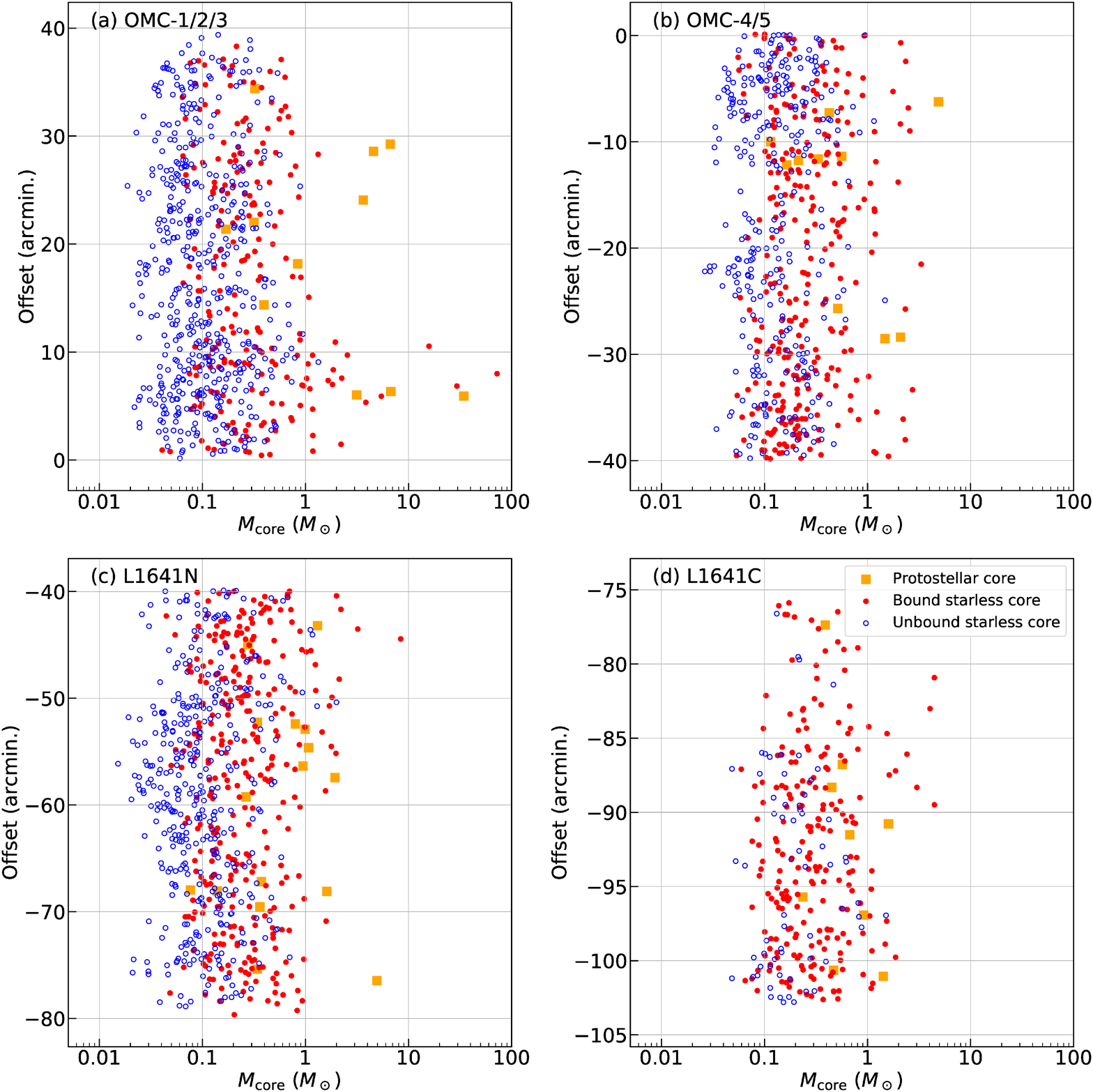}
    \caption{The same figure as Figure \ref{fig:decl_vcen} but for the core declination v.s. the core mass relation.}
    \label{fig:decl_mass}
\end{figure}

\begin{figure}[htbp]
    \centering
    \includegraphics[width=16cm]{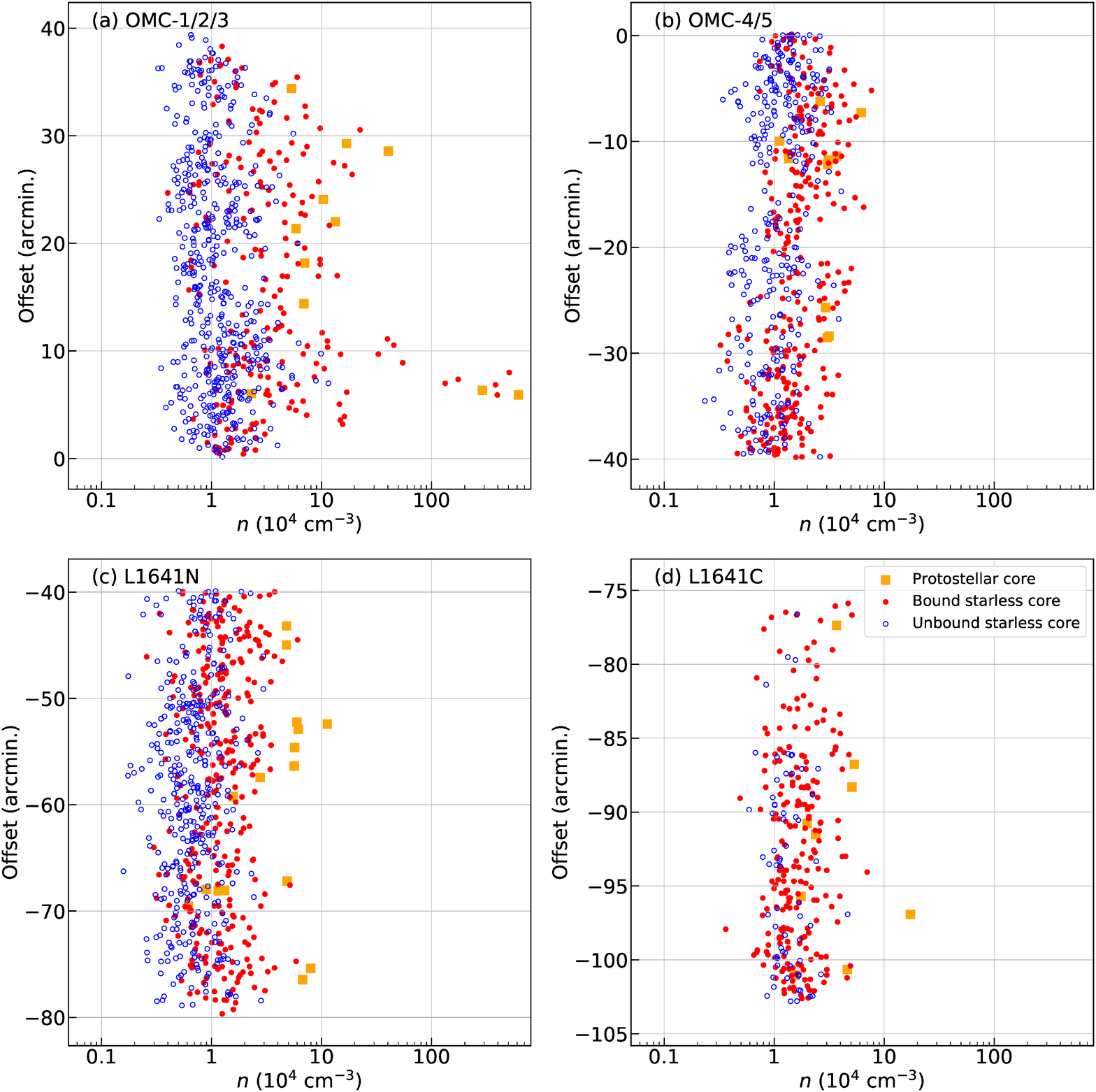}
    \caption{The same figure as Figure \ref{fig:decl_vcen} but for the core declination v.s. the density relation.}
    \label{fig:decl_density}
\end{figure}

In this section, we investigate the relationship between each core property and declination (i.e., position in the filamentary cloud) for the four areas shown in Figure \ref{fig:core} and \ref{fig:oriona_core_bound}.
We focus on the four core properties: the central velocity in Figure \ref{fig:decl_vcen}, the velocity width in Figure \ref{fig:decl_fwhm}, the core mass in Figure \ref{fig:decl_mass}, and core density in Figure \ref{fig:decl_density}.
The core declination is shown as the offset from $-5^\circ30'$ in units of arc minute.

From Figure \ref{fig:decl_vcen}, we see the velocity gradient from the northern region to the southern region; the core central velocity is $\sim$ 10 km s$^{-1}$ in panel (a) and it is $\sim$ 5 km s$^{-1}$ in panel (d).
In the declination -- velocity width relation of Figure \ref{fig:decl_fwhm} (a), there are several bound starless cores with large velocity FWHM of $\sim$ 1 km s$^{-1}$ around the declination of OMC-1.
In addition, higher-mass cores and denser cores are concentrated in OMC-1 as shown in the declination -- core mass relation (Figure \ref{fig:decl_mass}) and declination -- core density relation (Figure \ref{fig:decl_density}).
The properties of cores in OMC-1 are significantly different from typical properties of other cores in OMC-2/3 and Orion A. As shown in Section \ref{subsec:core_property}, core properties of OMC-1/2/3 are dissimilar to those in the other subregions. 
One possible origin of the irregular core properties in OMC-1 is that this region experienced global compression as we discuss in Section \ref{subsec:CMF_different}.

\section{CMFs in Orion A}
\label{sec:cmf}

As discussed in other studies, the CMF can evolve with time \citep[][Paper I]{motte18}.
However, our cores show no clear dependence of the core free-fall time on the core mass except for several cores located mainly in OMC-1/2/3 (see Figure \ref{fig:freefall}).
Therefore, we assume that the core lifetimes are constant for all core masses and the CMF's shape and its slope at the high-mass end do not change with time.
We only allow CMF to move parallel to the core mass axis with time.

We considered the completeness of core identification as follows.
First, we made a three-dimensional artificial gaussian core in a PPV space for each CMF's mass bin.
Each artificial core mass was fixed to the central mass of each mass bin on a log scale.
With the core mass, we calculated the dispersion along the position and velocity axes as sizes of the gaussian core from radius -- mass and velocity FWHM -- mass relations of starless cores in Orion A (Figure \ref{fig:oriona_correlation} (b) and (c)).
To convert the core mass to the core flux, we assumed optically thin C$^{18}$O ($J$=1--0) under LTE conditions.
Here, the temperature and abundance ratio of C$^{18}$O to H$_2$ are fixed to 20 K and 6.5 $\times$ 10$^{-7}$ as in Paper II.
Second, we inserted the core into the {\it trunks} of the observed C$^{18}$O ($J$=1--0) PPV data cube to avoid overlapping with the true cores.
Third, we applied astrodendro to the data with the same parameters as in Section \ref{subsec:core_id} and checked whether the artificial core is identified as an individual {\it leaf}.
Then, we repeated the above steps 1000 times for each mass bin (see also Paper I and Paper II).
The completeness limit where the detection probability becomes 90\% or less is shown in the following CMFs.
We also define the mass detection limit as a mass of an ideal minimum core to pass the core identification procedure in Section \ref{subsec:core_id}.
The total flux of the ideal core is calculated as 4$\sigma\times$ 1 pixel + 3$\sigma\times$ (60-1) pixels.
See Paper I for the derivation of this detection limit.

\subsection{CMFs in the whole Orion A}
\label{sec:cmf_oriona}

Figure \ref{fig:oriona_cmf} (a) shows the observed CMFs for all identified cores,
starless cores and bound starless cores in Orion A.
For comparison, the stellar IMF in the ONC region and CMF for bound starless cores in Orion A are indicated in Figure \ref{fig:oriona_cmf} (b).
The definition of the mass bin is slightly different from Paper I and this causes a slight difference in the mass function of the power-law index.

Figure \ref{fig:oriona_cmf} (a), the CMFs have turnovers at $\sim 0.07\ M_\odot$ for all starless cores
and $\sim 0.12\ M_\odot$ for bound cores, respectively.
The CMFs have power-law shapes at the high-mass ends and the best-fit power-law indices
and their uncertainties are -2.25 $\pm$ 0.1 for all starless cores and -2.18 $\pm$ 0.11 for bound starless cores.
We refer to the peak nearest to the power-law shape of the CMF as a turnover.
We derived a best-fit single power-law function between the mass bin higher than the turnover by two bins and the high-mass end of each CMF.
The properties of the CMFs such as turnover masses and parameters of the best-fit single power-law functions
are summarized in Table \ref{tab:cmf_property}.
There are no clear differences in the power-law indices among the three CMFs.
The detection limit is $\sim 0.016\ M_\odot$ and this is much smaller than the completeness limit.

As seen in Figure \ref{fig:oriona_cmf} (b), CMF and IMF have similar turnover masses of $\sim$ 0.2 $M_\odot$. The slopes are -2.18 $\pm$ 0.11 for Orion A CMF and -2.44 $\pm$ 0.18 for ONC IMF and are not distinguishable when we take into account the uncertainties.
Then, considering the protostellar feedback which carries core mass outward, mass accretion from the surrounding cloud is expected to explain the observed CMF--IMF relation as discussed in Paper I.

\subsection{CMFs in the individual areas}
\label{subsec:cmf_onc}

The CMFs for the individual areas are presented in Figure \ref{fig:cmfs}.
The power-law slopes of the high mass ends for bound starless cores in OMC-1/2/3 and OMC-4/5 are shallower than Salpater-like slope of -2.35.
Especially, the CMF of OMC-1/2/3 has a slope of -1.77 $\pm$ 0.11.
In contrast, CMFs for bound starless cores in L1641N/V 380 Ori and OMC-4/5 have Salpater-like IMF slope.
The turnover mass for bound starless cores is close to $\sim$ 0.1 $M_\odot$ for all areas.
In OMC-1, the CMF for the identified cores has a slightly smaller turnover mass compared to other areas.
In other words, the fraction of lower mass cores is larger.
In L1641C, unbound cores are deficient and therefore, the CMF of starless cores almost coincides with that of bound starless cores.

The most important difference is that only the CMF in OMC-1/2/3 extends to a higher mass of $\gtrsim 10 M_\odot$.
For other areas, the CMFs have a cut-off at around
$ 5-10 M_\odot$. Particularly, in OMC-4/5 and L1641C, the maximum core mass is as small as $\sim 5 M_\odot$.
Even in L1641N/V 380 Ori, there is only one core with a mass of $\sim 10 M_\odot$.
This suggests that the upper mass limit of CMFs depends on the cloud environment, and thus the IMFs are expected to have similar cutoffs at high-mass ends if
the one-to-one correspondence between stellar mass and core mass exits as suggested in \citet{hsu13}.

\section{Discussion}
\label{sec:discussion}

\subsection{CMFs in different environments}
\label{subsec:CMF_different}

As we showed in the previous section, the slope of CMFs
for the mass range of 1 -- 10 $M_\odot$ is similar among different regions. The free-fall times of the bound cores do not have a clear dependence on the core mass in the mass range of $< 10\ M_\odot$.
As for the massive cores in OMC-1/2/3 ($> 10\ M_\odot$), the core free-fall time becomes shorter for denser cores.
Therefore, the slope at the high mass end ($> 10\ M_\odot$) may be influenced by the time evolution.
On the other hand, the shape of CMF below 10 $M_\odot$ is expected to be less affected by time due to the longer free-fall time.
In that case, the high-mass part of CMF is expected to get steeper in time because star formation may proceed faster at the high-mass end of the CMF, whereas the lower-mass part may not change significantly.
Although we cannot rule out the possibility of the time evolution that steepens the slope of the CMF, a steeper CMF than Salpeter IMF is seen neither in this study nor in observations of high-mass star-forming regions. Instead of that, what we observed in this study is CMF with a slope at the high-mass end similar to or shallower than Salpeter-like IMF.
The later CMF is only seen in the OMC-1/2/3 and the power-law index of -1.77 $\pm$ 0.11 is consistent with the CMFs in high-mass star-forming regions \citep[e.g.][]{motte18}. Then, we suggest environmental effects peculiar to OMC-1/2/3 and high-mass star-forming regions make the CMF shallow. The possible physical processes are shown later in this section.
Whether the shallow CMF evolves into or evolves from a CMF with Salpeter-like slope cannot be constrained from this observation.

The CMFs in the areas south of OMC-1, a relatively quiescent portion of the cloud, have an upper mass cutoff of around 5--10 $M_\odot$.
The massive bound starless cores existing only in OMC-1/2/3 tend to have larger densities and larger line widths for similar sizes. These cores also deviate significantly from the relation of $R_\mathrm{core} \propto M_\mathrm{core} ^{1/3}$.
This might imply that massive cores are formed by some global compressional processes. Such processes can be related to global gravitational collapse, global colliding flows, cloud-cloud collisions, or stellar feedback from supernovae and stellar winds from massive stars.

It is difficult to identify which process is most important in Orion A.
For example, a cloud-cloud collision can create a filamentary ridge with a denser northern part through an off-center collision of two clouds \citep{lim21,wu17,fukui21}.
\citet{hacar17} proposed gravitational contraction along the main ridge toward OMC-1.
Interestingly, L1641N/V380 Ori contains several intermediate-mass cores, which extends the CMF to slightly larger masses, compared to those in OMC-4/5 and L1641C, although the properties of these intermediate-mass cores look similar to those of lower-mass cores. In this area, the possibility of a cloud-cloud collision is discussed by \citet{nakamura12a}.
\citet{kounkel21} suggested that the northern part (OMC-1/2/3) was compressed by supernovae and star formation has been triggered \citep[see also][]{bally10}. \citet{ntormousi11} and \citet{gomez14} showed that the global colliding flows can also create similar filamentary structures.
The global gravitational collapse of a sheet-like cloud also creates colliding flows which form the structure similar to Orion A \citep{hartmann07}.

In summary, the CMFs and core properties in Orion A imply that star formation in OMC-1/2/3 is influenced by the global compression due to gravitational contraction, cloud-cloud collision, stellar feedback from massive stars, or global colliding flows.

\subsection{A Possibility of Core Growth in Orion A}

As shown in the previous sections, protostellar cores tend to have larger masses than starless cores.
Such a characteristic is pointed out by \citet{kong21} for the Dragon infrared dark cloud (G28.37+0.07 or G28.34+0.06).
\citet{kong21} discussed that cores grow in mass during star formation. Theoretically, several dynamical scenarios of star formation suggest the importance of such core growth through the accretion of inter-core gas.
Paper I also found a similar core characteristic in the ONC region. By using our complete core sample of Orion A, we found that the same trend is observed for all four subregions, as shown in Figure \ref{fig:histo_mass}.
In addition, the core lifetime is estimated to be several times the free-fall time in Figures \ref{fig:lifetime} and \ref{fig:lifetime_4blocks}.

\begin{figure}[htbp]
    \centering
    \includegraphics[height=16cm]{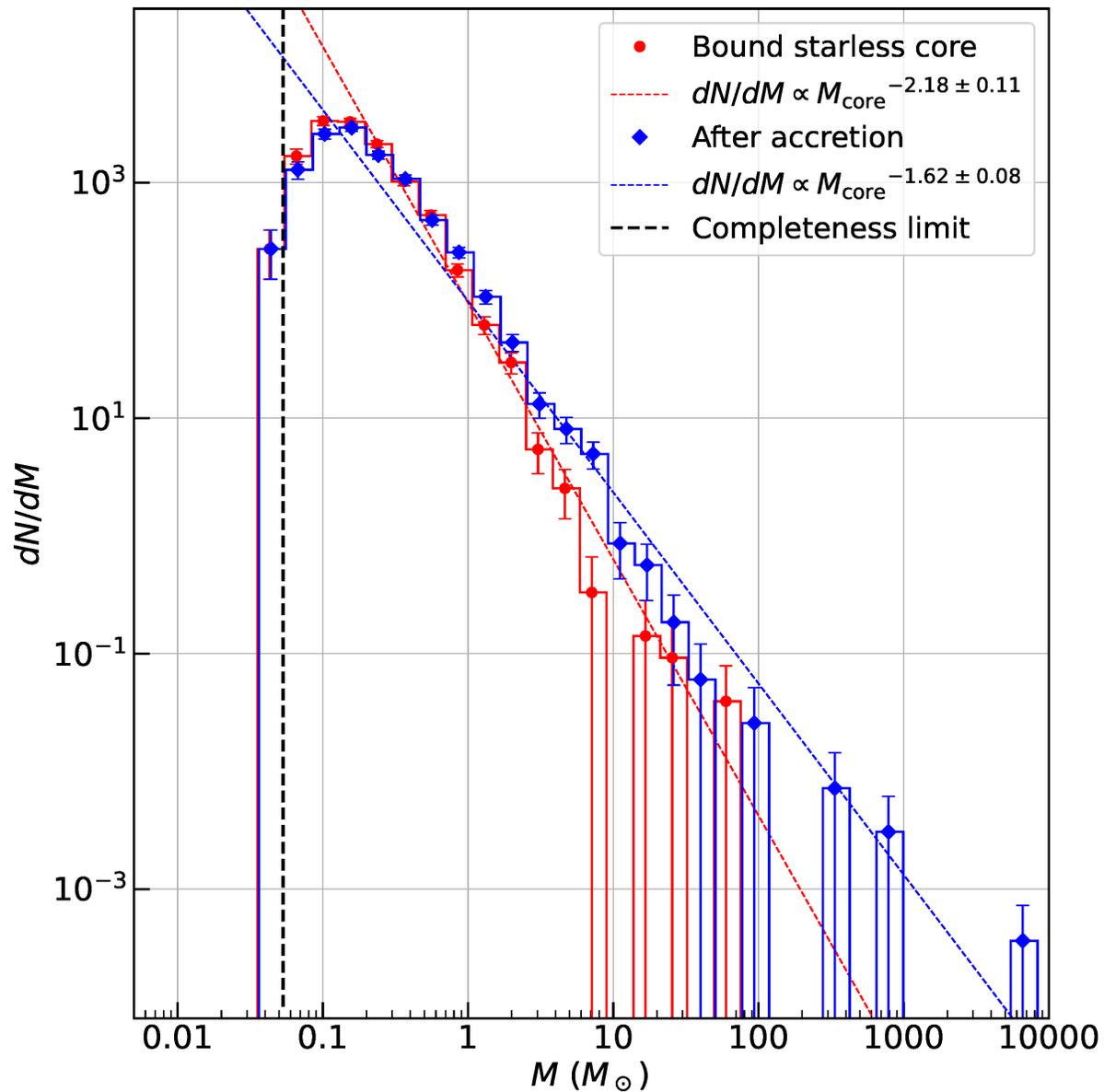}
    \caption{CMF for bound starless cores (red) and expected CMF with core growth by mass accretion (blue). We assumed that $1\ M_\mathrm{core}$ gas accretes on the $1\ M_\mathrm{core}$ core within a certain time and calculated mass accretion rate for each core.}
    \label{fig:cmf_accretion}
\end{figure}

The fact that protostellar cores tend to have larger masses than the starless cores
may indicate that the starless cores gain additional mass from the surroundings by accretion.
To investigate the accretion process, we estimate the accretion rate in two ways as follows.
First, we calculate the accretion rate which needs to double the core mass within the typical core lifetime of 10 $t_{\rm ff}$ and describe it as a required accretion rate.
Here, we use the mean density of bound starless cores in Orion A of $4\times 10^4$ cm$^{-3}$ to derive the free-fall time $t_{\rm ff}$.
Paper I showed that CMF resembles IMF in the ONC region and the relation implies that stars formed have similar masses as current cores.
When the core mass doubles due to mass accretion and stellar mass equal the initial core mass, the mass ratio of a star and final core mass (initial core mass + accretion mass) is 0.5.
We note that this is just a representative case to estimate the accretion rate from ambient gas to a dense core.
We define the required accretion rate of this model as
\begin{equation}
\begin{split}
    \dot{M}_\mathrm{required}
    &= \frac{M_\mathrm{core}}{10t_\mathrm{ff}}\\
    &\sim 6.5 \times 10^{-7} \left(\frac{M_\mathrm{core}}{1\ M_\odot}\right)\\
    &\times \left(\frac{n_\mathrm{core}}{4\times10^4\ \mathrm{cm}^{-3}}\right)^{1/2} M_\odot\ \mathrm{yr}^{-1}
\end{split}
\label{eq:required}
\end{equation}
Since the accretion rate is proportional to core mass, the shape of the CMF such as the slope at the high-mass end does not change with time. Then, CMFs only shift to the right without changing shape.
Second, we calculate the accretion rate of Bondi-Hoyle-Lyttleton accretion as
\begin{equation}
\begin{split}
    \dot{M}_\mathrm{Bondi}
    &= \frac{\pi \rho G^2 M^2}{\sigma^3}\\
    &\sim 4.1 \times 10^{-9} \left(\frac{n_\mathrm{clump}}{10^3\ \mathrm{cm}^{-3}}\right)\\
    &\times \left(\frac{M_\mathrm{core}}{1\ M_\odot}\right)^2
    \left(\frac{\sigma_\mathrm{clump}}{1\ \mathrm{km\ s^{-1}}}\right)^{-3} M_\odot\ \mathrm{yr}^{-1}
\end{split}
\label{eq:bondi}
\end{equation}
for the clump average density of a few $ \times 10^3$ cm$^{-3}$ and velocity dispersion of $\sim$ 1 km s$^{-1}$ \citep[e.g.,][]{krumholz08, kong18}.
This rate is significantly smaller than the required accretion rate derived above, in the representative case.
In a typical bound starless core lifetime of $\sim$ a few Myr,
only a tiny fraction of mass would increase for this low accretion rate.
In other words, a much larger accretion rate than the Bondi-Hoyle-Lyttleton accretion is needed for core growth to take place.
Such large accretion can be achieved by the gravitational contraction of gravitationally unstable clumps, such as in competitive accretion or global gravitational collapse models \citep[e.g.,][]{bonnell06, vazquez-semadeni19}, or by mass converging along dense filaments, such as in the inertial-iinflow model \citep{padoan20, pelkonen21}. However, in the case of the competitive accretion model, the gravitational potential of the clump environment should be much deeper than it is in the observed areas, except for OMC-1.
Mass accretion rate through filaments may possibly achieve such high gas accretion rates. Observations of the gas velocity structures along filaments will give us clues to investigate this possibility.
Another concern is that Bondi-Hoyle-Lyttleton accretion rate is not proportional to $M_\mathrm{core}$ but to $M_\mathrm{core}^2$.
If this is the dominant accretion process, the shape of the evolved CMF will differ from that of the initial CMF: the slope at the high-mass end will become shallower.
We constructed a CMF when mass accretion rate is proportional to $M_\mathrm{core}^2$ in Figure \ref{fig:cmf_accretion} using identified bound starless cores in Orion A.
The masses accreted to cores are calculated by assuming that surrounding gas accretes onto the cores until $1\ M_\mathrm{core}$ obtains another $1\ M_\mathrm{core}$.
We also assumed that there is a sufficient amount of gas for all cores to grow and star formation does not start during mass accretion.
The power-law index of CMF evolves from -2.18 $\pm$ 0.11 to -1.62 $\pm$ 0.08 due to mass accretion. Even if we exclude massive cores with $>100M_\odot$, the power-law index becomes -2.02 $\pm$ 0.05. Thus, when the mass accretion rate is proportional to $M_\mathrm{core}^2$, the CMF gets shallower as accretion proceeds.

Another possibility to explain the larger mass of protostellar cores is that cores grow in mass through merging with other cores and/or only massive cores form stars. The core merging is expected to happen frequently along the filamentary structures if the core lifetime is longer than the core free-fall time at least by a factor of a few. However, there is no clear observational evidence for frequent core merging.
Since the relationship between CMF and IMF depends on the merging properties, various shapes of the mass functions are expected to be observed so far.

From an analytical point of view, \citet{inutsuka01} showed that CMFs have IMF-like slope when the power-law index of a power spectrum of initial line mass of filaments is -1.5. 
\textit{Herschel} observations of \citet{roy15} revealed a power-law index of -1.6 $\pm$ 0.3, which is consistent with the analytical prediction.
In addition, the model includes the mass accretion through filament and the CMF's slope does not evolve in time. This seems to contradict the shallow CMF observed in OMC-1/2/3 in this study and IRDCs \citep[e.g.,][]{motte18}. More detail observations toward various star-forming regions in different star formation stages will give clues to reveal the properties of core growth.

\section{Conclusions}
\label{sec:conclusions}

We conducted a dense core survey toward Orion A molecular cloud with the CARMA+NRO45-m combined C$^{18}$O ($J$=1--0) data and summarise the core properties in Table \ref{tab:catalog}.
We list the main results and discussions as follows.
\begin{enumerate}
\item We applied astrodendro to the observed PPV data and identified 2341 cores after applying several additional selection criteria as described in Section \ref{subsec:core_id}.
Using the Class 0/I objects in the protostar catalog from a Vienna survey \citep{meingast16}, we identified 2295 starless cores and 46 protostellar cores.
Our study is one of the widest-field (1$\times$3 degree) unbiased surveys of dense cores in Orion A among high spatial resolution (8\arcsec\ $\sim$ 3000 AU) surveys.
\item Protostellar cores tend to have higher masses than starless cores. They are also denser and more virialised compared to the starless cores. The core mass difference between starless cores and protostellar cores probably indicates the core growth by mass accretion from surrounding material.
More than half of starless cores are classified as unbound cores based on the virial analysis.
\item In the OMC-1 area, core physical properties are different from those in the other areas in Orion A.
The velocity width distribution has a shallow tail toward large velocity FWHM and the distribution of the core density has larger dispersion compared to the other areas.
In addition, there are massive and dense bound starless cores with large velocity widths in OMC-1.
Such cores do not exist in the other areas.
\item We have expanded the target region from Paper I in this paper and confirm that CMF in Orion A has a similar turnover mass and a power-law index at the high-mass end as that in the ONC region. The slope at the high-mass end also resembles IMF in the ONC region and Salpeter-like IMF.
Although there are no clear differences in the turnover mass of CMF among the four subregions, the CMF in OMC-1/2/3 obviously has a shallower slope at the high-mass end than the other subregions.
However, massive cores ($\geq 10 M_\odot$) are only located in OMC-1/2/3.
\item Except for several massive cores in OMC-1, the free-fall time is almost independent of the core mass.
Most of the cores seem to evolve to stars during the interval of a few times of the free-fall time.
This suggests that the current CMF is also thought to evolve into the stellar mass function with the current stellar MF in the core lifetime resembling the past CMF.
\item The possibility of core growth is suggested by several previous studies and we discussed the process from the point of view of the mass accretion rate.
When we assume the Bondi-Hoyle-Lyttleton accretion from a typical clump has an accretion rate much smaller than that needed to double the core mass over the core lifetime.
Thus, a more effective mass accretion process is expected to grow core masses during the realistic core lifetime.
Since the density of the filamentary structure is much higher compared to the mean density of a whole cloud, the Bondi-Hoyle accretion will be much more effective than that estimated in this paper.
\end{enumerate}

\appendix

\section{KS test in four subregions}
\label{appsec:ks_test}

We show the results of the KS test of each core physical property of each subregion in Table \ref{tab:ks_test_omc123}, \ref{tab:ks_test_omc45}, \ref{tab:ks_test_l1641n}, and \ref{tab:ks_test_l1641c}. Here, we derive p-values between three pairs of core categories in each subregion: starless cores and protostellar cores, bound starless cores and unbound starless cores, and bound starless cores and protostellar cores.
We note that the number of protostellar cores in each subregion is small, the results of the KS test between starless cores and protostellar cores and bound starless cores and protostellar cores seem to have large uncertainties.
For bound starless cores and unbound starless cores the p-values of velocity width in FWHM, mass, and number density are much smaller than 0.05 in all subregions. As explained in Section \ref{subsec:core_property} of the main text, it was statistically confirmed that bound and unbound starless cores have different distributions as well.

Next, we conducted the KS test among core physical properties in four subregions and show the result in Table \ref{tab:ks_test_diamter}, \ref{tab:ks_test_a_ratio}, \ref{tab:ks_test_fwhm}, \ref{tab:ks_test_mass}, \ref{tab:ks_test_density}, and \ref{tab:ks_test_virial_ratio}. We selected cores in the same category such as bound starless cores or unbound starless cores from different subregions.
Interestingly, from Table \ref{tab:ks_test_mass} the null hypothesis that core masses in OMC-1/2/3 and the other region have the same distributions is not rejected with the significance level of 5\%.

\section{Virial analysis}
\label{appsec:virial_ratio}

In Section \ref{subsec:derivation} of the main text, we adopted $\alpha_\mathrm{vir}<2$ as a definition of bound cores. This definition may not be always appropriate since we assumed a spherically symmetric core with no magnetic field.
The virial ratio includes only the gravitational and internal turbulent kinetic energies of a core. Therefore, in this Appendix, we consider how the adopted threshold value of the virial ratio affects our main conclusion.
Below, we adopt $\alpha_\mathrm{vir}<1$ for the definition of bound cores. As expected, the smaller virial ratio reduces the number of bound cores as shown in Table \ref{tab:N_bound_core} and thus affects the detailed shape of CMFs and estimation of core lifetimes.

\begin{figure}[htbp]
    \centering
    \includegraphics[height=14cm]{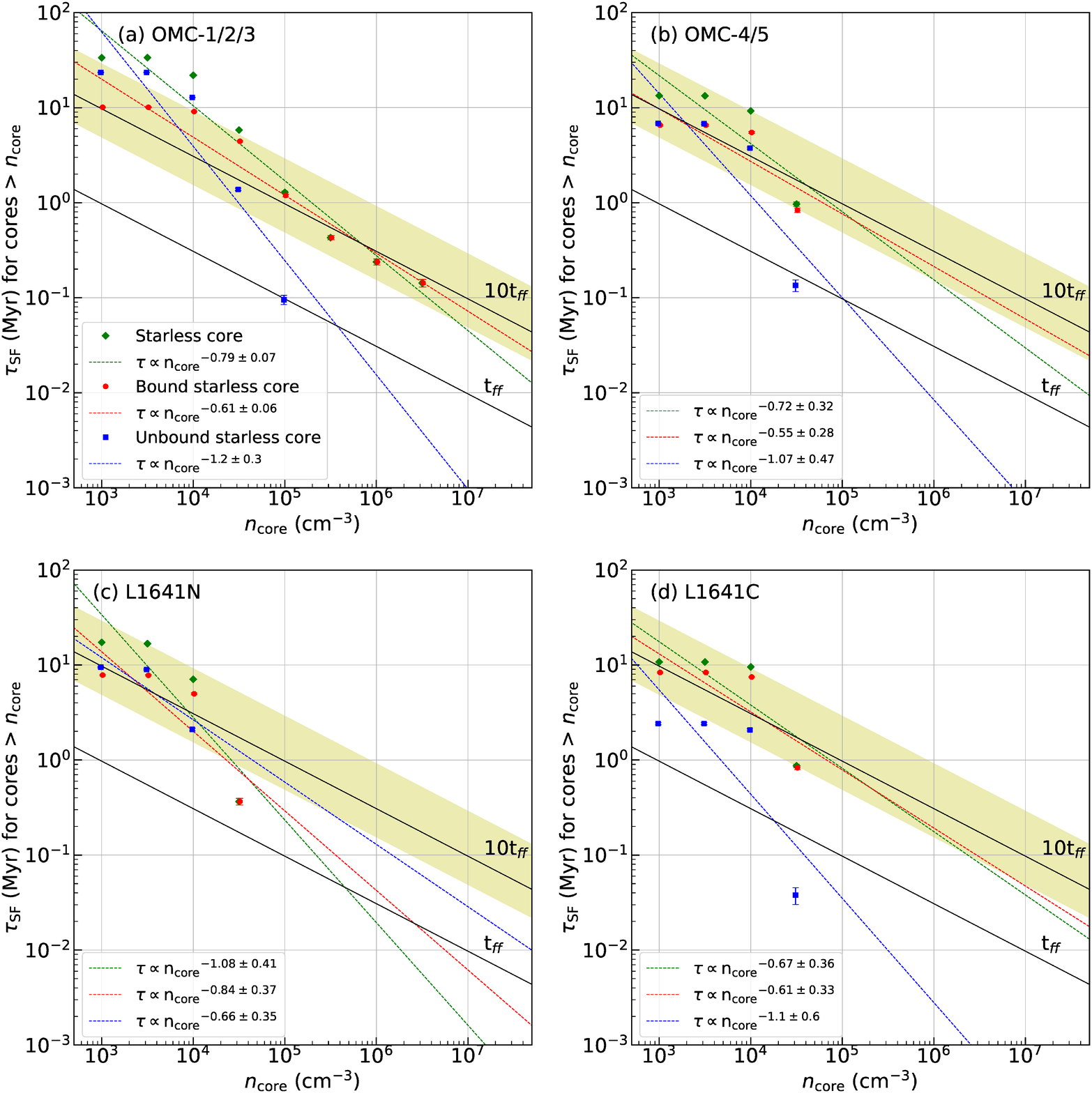}
    \caption{The core lifetime vs. core density for four areas. The solid lines show one $t_{\rm ff}$ and 10 $t_{\rm ff}$.
    The dashed lines are best-fit functions from the second-lowest density points ($n_\mathrm{core}\sim$ 10$^3$ cm$^{-3}$) to the ends in the form of a power-law function.
    The shaded area in each panel indicates the area with the lifetime is between 5 and 30 $t_{\rm ff}$.
    The error bar in each panel just represents the statistical uncertainty calculated with the square root of the
    number of cores of each plot, $\sqrt N$, and other uncertainties such as the number of class II objects are not included.}
    \label{fig:lifetime_4blocks}
\end{figure}

\begin{figure}[htbp]
    \centering
    \includegraphics[height=14cm]{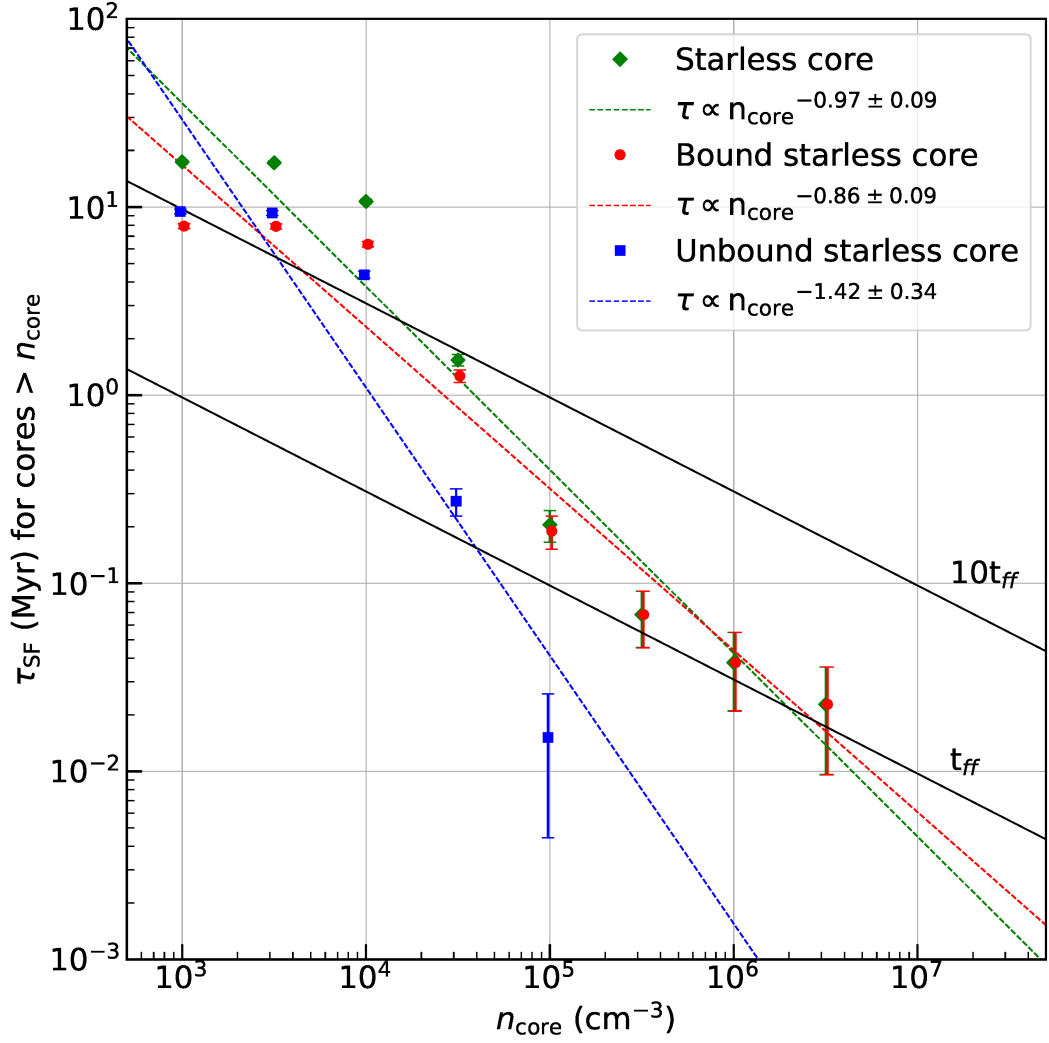}
    \caption{The core lifetime vs. core density in Orion A. Details are the same as for Figure \ref{fig:lifetime_4blocks}.
    }
    \label{fig:lifetime}
\end{figure}

\begin{figure}[htbp]
    \centering
    \includegraphics[height=14cm]{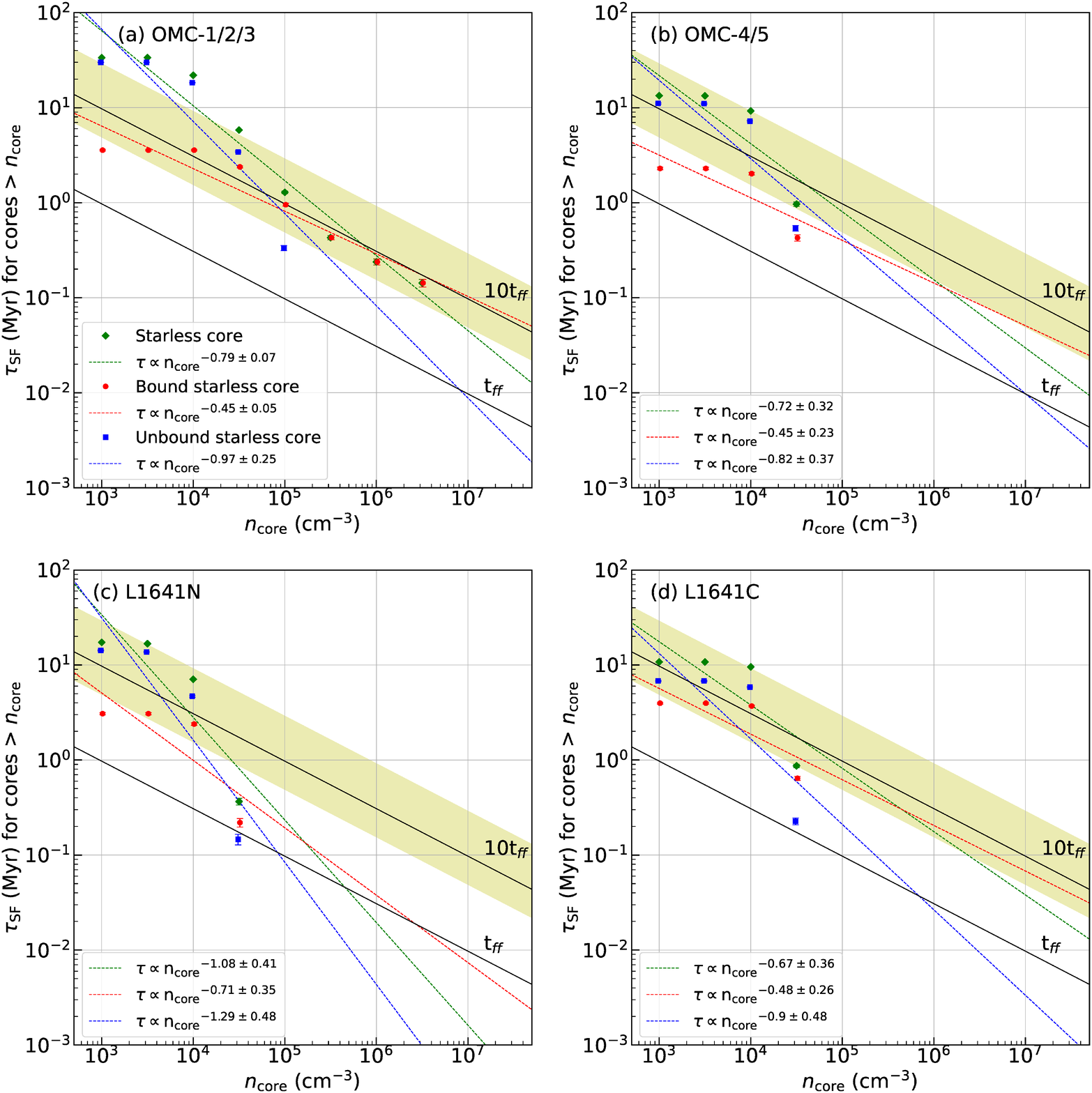}
    \caption{The same as Figure \ref{fig:lifetime_4blocks} but starless cores with virial ratios of unity or less are classified as bound cores.
    Details are the same as for Figure \ref{fig:lifetime}.}
    \label{fig:lifetime_4blocks_a1}
\end{figure}

First, Figure \ref{fig:lifetime_4blocks_a1} shows the relationship between core lifetime and core density for four regions which corresponds to Figure \ref{fig:lifetime_4blocks} in Section \ref{sec:correlations}.
When we focus on the bound cores, the lifetimes of less dense cores ($\leq 10^{5}$ cm$^{-3}$) become shorter but those of denser cores ($\geq 10^{5}$ cm$^{-3}$) are not changed from Figure \ref{fig:lifetime_4blocks} in the main text.
The trend that the core lifetime decreases with core density is also the same as the main text.
In addition, most of the core lifetimes are in between 5 and 30 free-fall times as well.
Therefore, we think the discussion based on the core lifetime in the main text is less affected by the definition of bound cores unless much smaller virial ratios are needed to bound dense cores.

\begin{figure}[htbp]
    \centering
    \includegraphics[width=16cm]{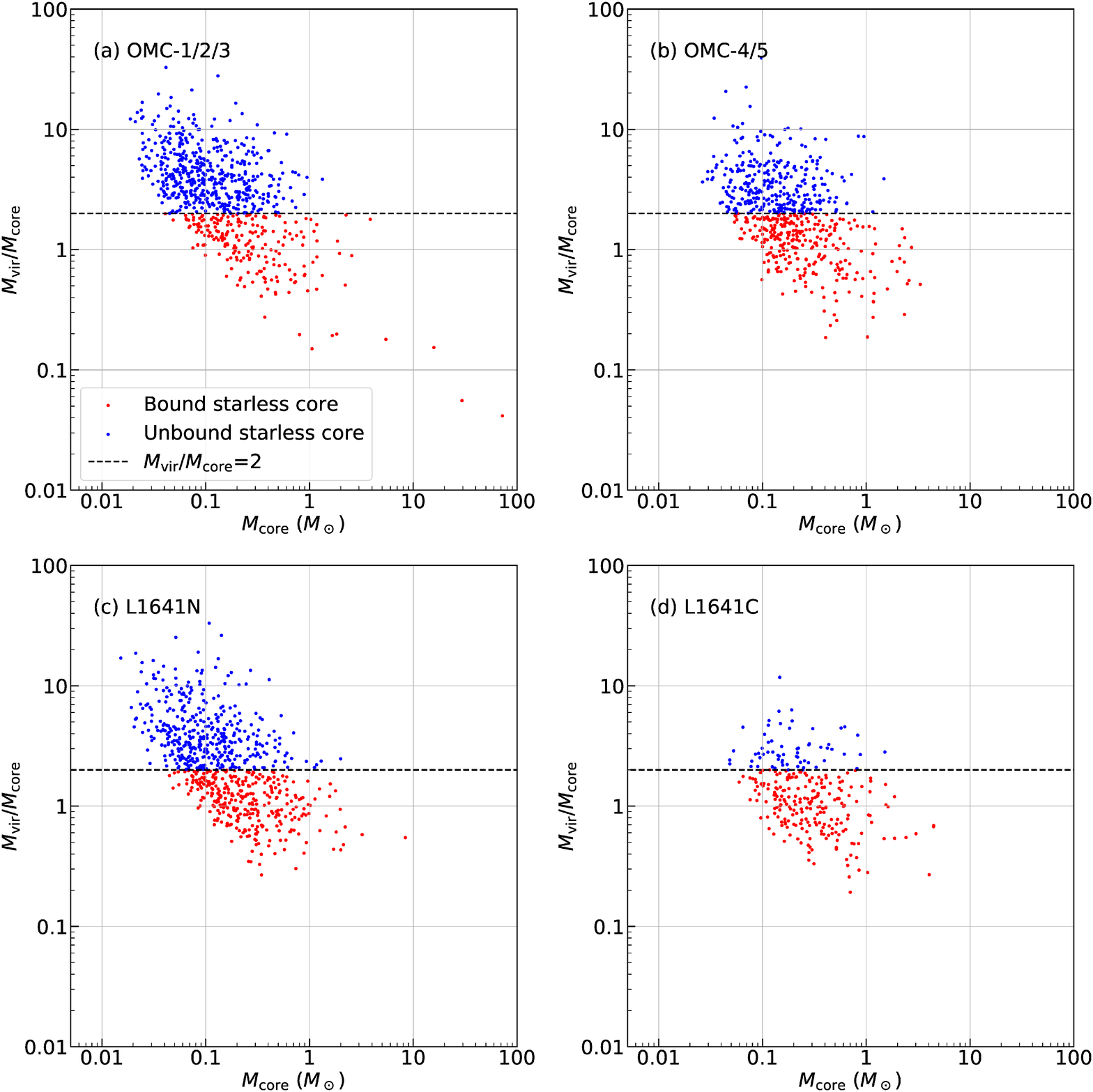}
    \caption{The virial ratio -- mass relation in four subregions. The horizontal dashed lines indicate $\alpha_\mathrm{vir}$=2.}
    \label{fig:oriona_correlation_m-alpha}
\end{figure}

\begin{figure}[htbp]
    \centering
   \includegraphics[height=7cm]{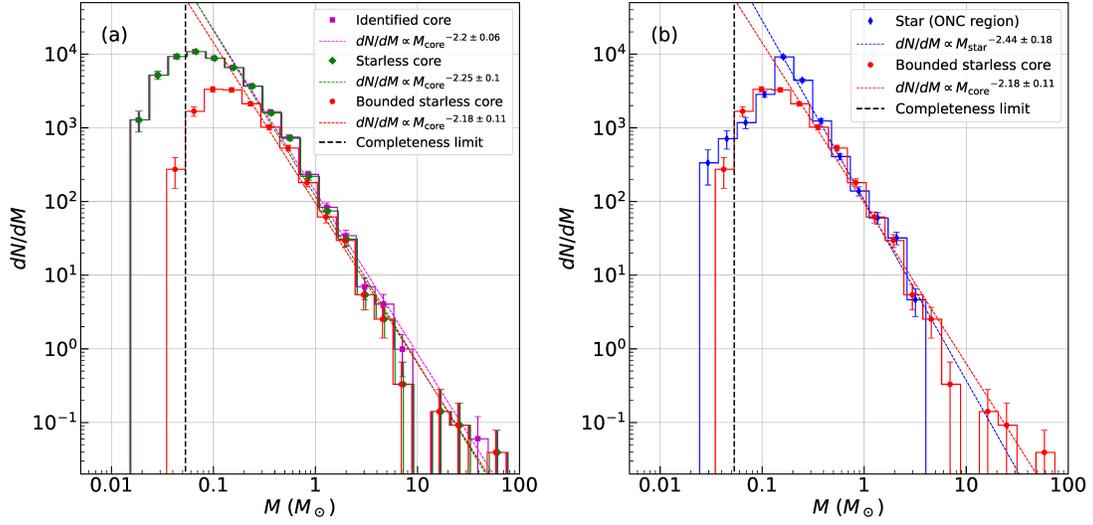}
    \caption{(a) The observed CMFs for all cores (magenta), all starless cores (green)
    and bound starless cores (red) in the whole Orion A area.
    The error bars show the statistical uncertainty calculated as the square root of the
    number of cores in each mass bin, $\sqrt N$.
    The dotted lines show the best-fit power-law functions for each CMF
    between two mass bins above the turnover and the high-mass end.
    The detection mass limit is estimated to be $0.016\ M_\odot$.
    (b) Comparison between the CMF of bounded starless cores (red) and the stellar IMF in the ONC region (blue). Their distributions look very similar, and the peak mass is comparable to each other.
    The vertical dashed lines in both panels are the completeness limits.}
    \label{fig:oriona_cmf}
\end{figure}

\begin{figure}[htbp]
    \centering
    \includegraphics[height=16cm]{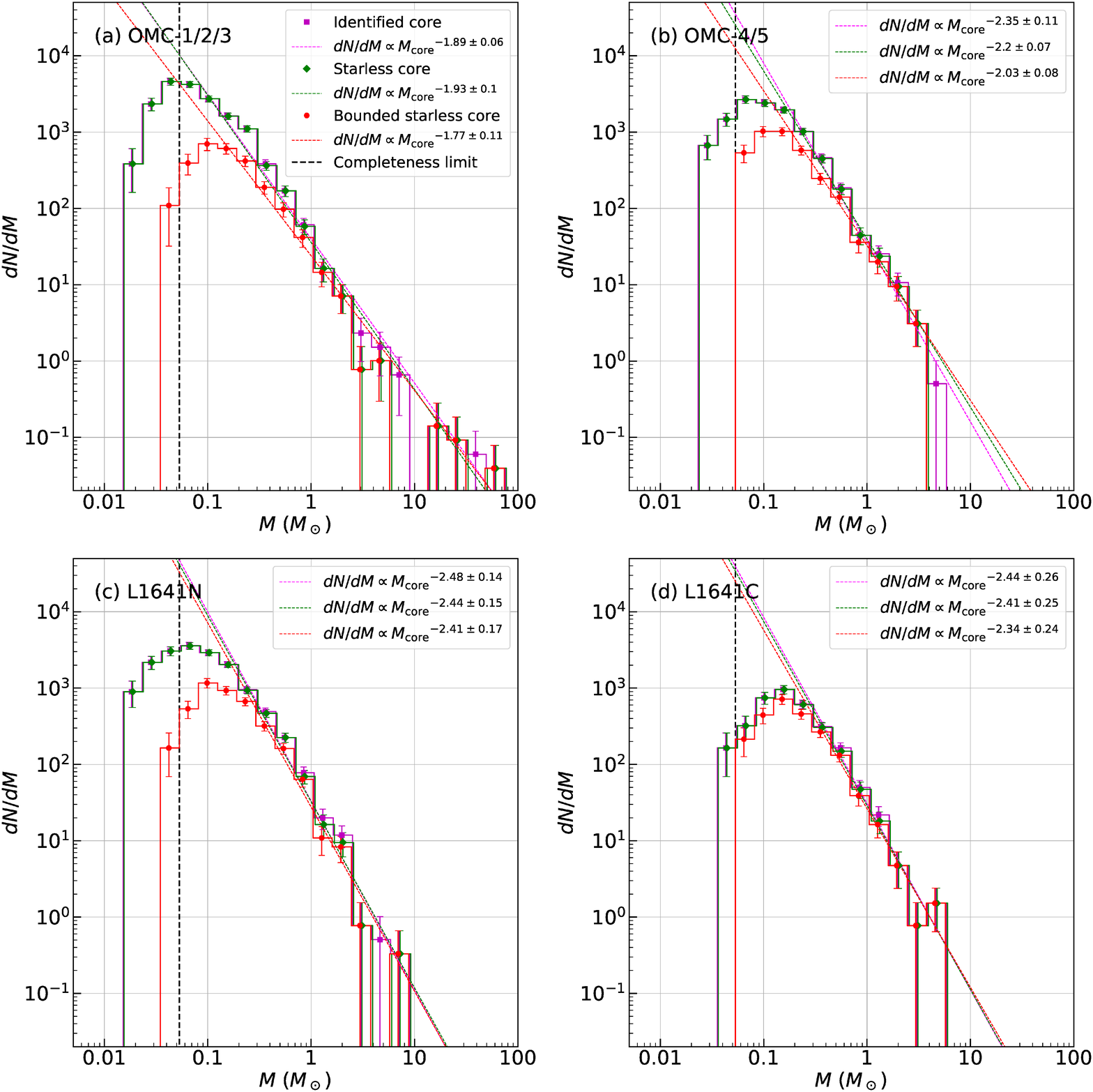}
    \caption{The same as Figure \ref{fig:oriona_cmf}
    {\it left}. Panels (a) though (d) show the CMFs in areas (a) through (d), respectively, delineated in Figure \ref{fig:obsarea}.
    Details are the same as for Figure \ref{fig:oriona_cmf}.
    }
    \label{fig:cmfs}
\end{figure}

\begin{figure}[htbp]
    \centering
    \includegraphics[height=16cm]{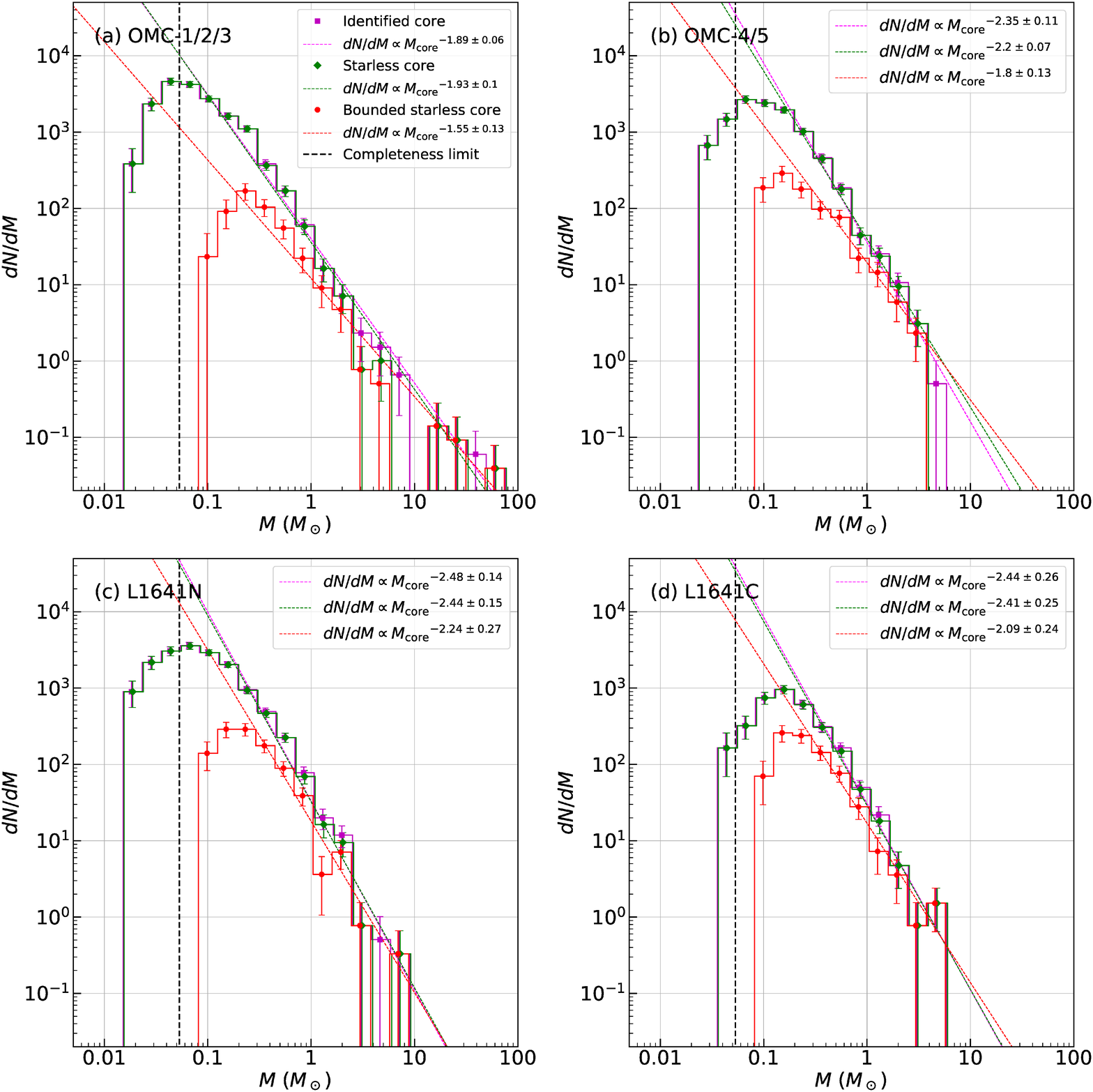}
    \caption{The same as Figure \ref{fig:cmfs} but starless cores with virial ratios of unity or less are classified as bound cores.
    Details are the same as for Figure \ref{fig:oriona_cmf}.}
    \label{fig:cmfs_a1}
\end{figure}

Second, CMFs of four regions are shown in Figure \ref{fig:cmfs_a1} which is equivalent to Figure \ref{fig:cmfs} in Section \ref{subsec:cmf_onc}.
The MFs of starless cores, bound cores, and unbound cores of four regions are also shown and CMFs other than that of bound cores are same as them in Figure \ref{fig:cmfs}.
As shown in Figure \ref{fig:oriona_correlation_m-alpha}, low-mass cores tend to have larger virial ratios.
Thus, such low-mass bound cores in Section \ref{subsec:cmf_onc} are reclassified as unbound cores when we set a small virial ratio as the boundary between a bound core and an unbound core.
In contrast, high-mass cores are reclassified as bound cores again.
This affects a slope at a high-mass and a turnover mass of CMF: the slope becomes shallower and the trunover mass shifts to high-mass direction by one or two mass bins.
The turnover masses of CMFs in 4 regions are $\sim$ 0.2 $M_\odot$ and this is also similar to the turnover mass of IMF in the ONC region (Figure \ref{fig:oriona_cmf} (b)) as we described in Section \ref{sec:cmf_oriona}.
Then, we conclude that there is no sever influences on our discussion based on CMFs even if we set $\alpha_\mathrm{vir}=1$ as a boundary of bound and unbound cores.

\begin{acknowledgments}
This work was supported in part by The Graduate University for Advanced Studies, SOKENDAI and JSPS KAKENHI Grant Number of 22J13587. Data analysis was carried out on the Multi-wavelength Data Analysis System operated by the Astronomy Data Center (ADC), National Astronomical Observatory of Japan. This work was carried out in part at the Jet Propulsion Laboratory, California Institute of Technology, under a contract with NASA (80NM0018D0004). N.S. acknowledges support from the project GENESIS (ANR-16-CE92-0035-01/DFG1591/2-1), the project FEEDBACK (DLR, 50 OR 1916), and the SFB 956 (DFG). P.S. was partially supported by a Grant-in-Aid for Scientific Research (KAKENHI Number 18H01259) of the Japan Society for the Promotion of Science (JSPS). P.P. acknowledges support by the grant PID2020-115892GB-I00 funded by MCIN/AEI/10.13039/501100011033. We thank the anonymous referee for many useful comments that have improved the presentation.

\facility{CARMA, No:45m, Herschel}
\software{The data analysis in this paper uses python packages Astropy \citep{astropy_collaboration13}, SciPy \citep{jones01}, Numpy \citep{van_der_walt11}, Matplotlib \citep{hunter07} and Astrodendro \citep{rosolowsky08}.}
\end{acknowledgments}

\clearpage

\bibliographystyle{aasjournal}
\bibliography{corecatalog_OrionA_ms.bib}

\clearpage

\begin{table}[htbp]
 \centering
  \caption{The Results of Core Identification in Orion A}
  \small
  \begin{tabular}{ccccc} 
\hline \hline 
Region & Category & Total & Bound Cores & Unbound Cores \\ 
\hline  
 & Identified Core & 2341 & 1087 & 1254  \\ 
Entire region & Starless Core & 2295 & 1045 & 1250  \\ 
 & Protostellar Core & 46 & 42 & 4  \\ 
\hline 
 & Identified Core & 716 & 223 & 493  \\ 
(a) OMC-1/2/3 area & Starless Core & 705 & 212 & 493  \\ 
 & Protostellar Core & 11 & 11 & 0  \\ 
\hline 
 & Identified Core & 605 & 299 & 306  \\ 
(b) OMC-4/5 area & Starless Core & 595 & 291 & 304  \\ 
 & Protostellar Core & 10 & 8 & 2  \\ 
\hline 
 & Identified Core & 726 & 335 & 391  \\ 
(c) L1641N area & Starless Core & 710 & 321 & 389  \\ 
 & Protostellar Core & 16 & 14 & 2  \\ 
\hline 
 & Identified Core & 294 & 230 & 64  \\ 
(d) L1641C area & Starless Core & 285 & 221 & 64  \\ 
 & Protostellar Core & 9 & 9 & 0  \\ 
\hline 
\end{tabular} 

  \label{tab:oriona_core_id}
\end{table}

\begin{table}[htbp]
 \centering
  \caption{The Summary of Physical Properties of Dense Cores in Orion A}
  \small
  \begin{tabular}{ccccc} 
\hline \hline 
Property & Category & Minimum & Maximum & Mean $\pm$ Std. \\ 
\hline  
  & Identified Core & 0.030 & 0.260 & 0.076 $\pm$ 0.027  \\ 
Diameter (pc) & Starless Core & 0.030 & 0.260 & 0.076 $\pm$ 0.027  \\ 
 & Protostellar Core & 0.040 & 0.173 & 0.081 $\pm$ 0.033  \\ 
\hline 
  & Identified Core & 1.01 & 5.45 & 1.89 $\pm$ 0.62  \\ 
Aspect Ratio & Starless Core & 1.01 & 5.45 & 1.89 $\pm$ 0.62  \\ 
 & Protostellar Core & 1.11 & 3.39 & 1.80 $\pm$ 0.50  \\ 
\hline 
  & Identified Core & 0.11 & 1.04 & 0.30 $\pm$ 0.11  \\ 
FWHM (km s$^{-1}$) & Starless Core & 0.11 & 1.04 & 0.30 $\pm$ 0.11  \\ 
 & Protostellar Core & 0.18 & 0.80 & 0.32 $\pm$ 0.11  \\ 
\hline 
  & Identified Core & 0.02 & 72.21 & 0.35 $\pm$ 1.85  \\ 
Mass ($M_\odot$) & Starless Core & 0.02 & 72.21 & 0.32 $\pm$ 1.70  \\ 
 & Protostellar Core & 0.08 & 34.36 & 2.05 $\pm$ 5.10  \\ 
\hline 
  & Identified Core & 0.16 & 615.46 & 2.79 $\pm$ 21.43  \\ 
Number Density (10$^4$ cm$^{-3}$) & Starless Core & 0.16 & 508.13 & 2.34 $\pm$ 16.37  \\ 
 & Protostellar Core & 0.62 & 615.46 & 25.21 $\pm$ 97.44  \\ 
\hline 
  & Identified Core & 0.04 & 39.24 & 3.03 $\pm$ 3.06  \\ 
Virial Ratio & Starless Core & 0.04 & 39.24 & 3.07 $\pm$ 3.08  \\ 
 & Protostellar Core & 0.06 & 3.21 & 0.81 $\pm$ 0.71  \\ 
\hline 
\end{tabular} 

  \label{tab:oriona_property}
\end{table}

\begin{table}[htbp]
 \centering
  \caption{The Results of KS Test of Core Properties in Orion A}
  \small
  \begin{tabular}{cccc} 
\hline \hline 
Core Property & Category & p-value \\ 
\hline  
 & Starless Core \& Protostellar Core & 4.36$\times 10^{-1}$  \\ 
Diameter (pc) & Bound Starless Core \& Unbound Starless Core & 1.61$\times 10^{-10}$  \\ 
 & Bound Starless Core \& Protostellar Core & 5.37$\times 10^{-1}$  \\ 
\hline 
 & Starless Core \& Protostellar Core & 6.94$\times 10^{-1}$  \\ 
Aspect Ratio & Bound Starless Core \& Unbound Starless Core & 1.96$\times 10^{-2}$  \\ 
 & Bound Starless Core \& Protostellar Core & 4.34$\times 10^{-1}$  \\ 
\hline 
 & Starless Core \& Protostellar Core & 3.97$\times 10^{-3}$  \\ 
FWHM (km s$^{-1}$) & Bound Starless Core \& Unbound Starless Core  & 3.77$\times 10^{-15}$  \\ 
 & Bound Starless Core \& Protostellar Core & 5.19$\times 10^{-9}$  \\ 
\hline 
 & Starless Core \& Protostellar Core & 7.33$\times 10^{-15}$  \\ 
Mass ($M_\odot$) & Bound Starless Core \& Unbound Starless Core & 3.77$\times 10^{-15}$  \\ 
 & Bound Starless Core \& Protostellar Core & 4.73$\times 10^{-8}$  \\ 
\hline 
 & Starless Core \& Protostellar Core & 4.18$\times 10^{-17}$  \\ 
Number Density (10$^4$ cm$^{-3}$) & Bound Starless Core \& Unbound Starless Core & 3.77$\times 10^{-15}$  \\ 
 & Bound Starless Core \& Protostellar Core & 9.42$\times 10^{-12}$  \\ 
\hline 
 & Starless Core \& Protostellar Core & 3.73$\times 10^{-16}$  \\ 
Virial Ratio & Bound Starless Core \& Unbound Starless Core & -  \\ 
 & Bound Starless Core \& Protostellar Core & 3.20$\times 10^{-8}$  \\ 
\hline 
\end{tabular} 

  \label{tab:ks_test_oriona}
\end{table}

\begin{table}[htbp]
 \centering
  \caption{The Results of Core Identification in the OMC-1/2/3 Area}
  \small
  \begin{tabular}{ccccc} 
\hline \hline 
Property & Category & Minimum & Maximum & Mean $\pm$ Std. \\ 
\hline  
  & Identified Core & 0.030 & 0.159 & 0.066 $\pm$ 0.020  \\ 
Diameter (pc) & Starless Core & 0.030 & 0.159 & 0.065 $\pm$ 0.020  \\ 
 & Protostellar Core & 0.040 & 0.156 & 0.071 $\pm$ 0.034  \\ 
\hline 
  & Identified Core & 1.02 & 5.45 & 1.91 $\pm$ 0.64  \\ 
Aspect Ratio & Starless Core & 1.02 & 5.45 & 1.91 $\pm$ 0.64  \\ 
 & Protostellar Core & 1.38 & 3.39 & 1.97 $\pm$ 0.54  \\ 
\hline 
  & Identified Core & 0.15 & 1.04 & 0.34 $\pm$ 0.13  \\ 
FWHM (km s$^{-1}$) & Starless Core & 0.15 & 1.04 & 0.33 $\pm$ 0.13  \\ 
 & Protostellar Core & 0.18 & 0.80 & 0.38 $\pm$ 0.17  \\ 
\hline 
  & Identified Core & 0.02 & 72.21 & 0.47 $\pm$ 3.26  \\ 
Mass ($M_\odot$) & Starless Core & 0.02 & 72.21 & 0.39 $\pm$ 3.00  \\ 
 & Protostellar Core & 0.17 & 34.36 & 5.55 $\pm$ 9.42  \\ 
\hline 
  & Identified Core & 0.33 & 615.46 & 5.94 $\pm$ 38.54  \\ 
Number Density (10$^4$ cm$^{-3}$) & Starless Core & 0.33 & 508.13 & 4.59 $\pm$ 29.38  \\ 
 & Protostellar Core & 2.29 & 615.46 & 92.13 $\pm$ 183.81  \\ 
\hline 
  & Identified Core & 0.04 & 32.67 & 3.82 $\pm$ 3.32  \\ 
Virial Ratio & Starless Core & 0.04 & 32.67 & 3.87 $\pm$ 3.32  \\ 
 & Protostellar Core & 0.06 & 1.48 & 0.49 $\pm$ 0.50  \\ 
\hline 
\end{tabular} 

  \label{tab:omc123_property}
\end{table}

\begin{table}[htbp]
 \centering
  \caption{The Results of Core Identification in the OMC-4/5 Area}
  \small
  \begin{tabular}{ccccc} 
\hline \hline 
Property & Category & Minimum & Maximum & Mean $\pm$ Std. \\ 
\hline  
  & Identified Core & 0.034 & 0.206 & 0.077 $\pm$ 0.027  \\ 
Diameter (pc) & Starless Core & 0.034 & 0.206 & 0.077 $\pm$ 0.026  \\ 
 & Protostellar Core & 0.053 & 0.173 & 0.088 $\pm$ 0.036  \\ 
\hline 
  & Identified Core & 1.03 & 4.97 & 1.90 $\pm$ 0.62  \\ 
Aspect Ratio & Starless Core & 1.03 & 4.97 & 1.90 $\pm$ 0.63  \\ 
 & Protostellar Core & 1.13 & 2.21 & 1.71 $\pm$ 0.36  \\ 
\hline 
  & Identified Core & 0.11 & 1.04 & 0.30 $\pm$ 0.11  \\ 
FWHM (km s$^{-1}$) & Starless Core & 0.11 & 1.04 & 0.30 $\pm$ 0.11  \\ 
 & Protostellar Core & 0.19 & 0.44 & 0.30 $\pm$ 0.07  \\ 
\hline 
  & Identified Core & 0.03 & 4.88 & 0.29 $\pm$ 0.43  \\ 
Mass ($M_\odot$) & Starless Core & 0.03 & 3.31 & 0.28 $\pm$ 0.38  \\ 
 & Protostellar Core & 0.11 & 4.88 & 1.08 $\pm$ 1.40  \\ 
\hline 
  & Identified Core & 0.23 & 7.66 & 1.53 $\pm$ 0.98  \\ 
Number Density (10$^4$ cm$^{-3}$) & Starless Core & 0.23 & 7.66 & 1.51 $\pm$ 0.96  \\ 
 & Protostellar Core & 1.12 & 6.18 & 3.03 $\pm$ 1.31  \\ 
\hline 
  & Identified Core & 0.19 & 39.24 & 2.67 $\pm$ 2.73  \\ 
Virial Ratio & Starless Core & 0.19 & 39.24 & 2.69 $\pm$ 2.74  \\ 
 & Protostellar Core & 0.31 & 3.21 & 1.10 $\pm$ 1.05  \\ 
\hline 
\end{tabular} 

  \label{tab:omc45_property}
\end{table}

\begin{table}[htbp]
 \centering
  \caption{The Results of Core Identification in the L1641N Area}
  \small
  \begin{tabular}{ccccc} 
\hline \hline 
Property & Category & Minimum & Maximum & Mean $\pm$ Std. \\ 
\hline  
  & Identified Core & 0.037 & 0.234 & 0.082 $\pm$ 0.029  \\ 
Diameter (pc) & Starless Core & 0.037 & 0.234 & 0.082 $\pm$ 0.029  \\ 
 & Protostellar Core & 0.049 & 0.157 & 0.083 $\pm$ 0.031  \\ 
\hline 
  & Identified Core & 1.01 & 5.02 & 1.87 $\pm$ 0.60  \\ 
Aspect Ratio & Starless Core & 1.01 & 5.02 & 1.87 $\pm$ 0.60  \\ 
 & Protostellar Core & 1.11 & 2.88 & 1.80 $\pm$ 0.51  \\ 
\hline 
  & Identified Core & 0.12 & 0.86 & 0.28 $\pm$ 0.10  \\ 
FWHM (km s$^{-1}$) & Starless Core & 0.12 & 0.86 & 0.28 $\pm$ 0.10  \\ 
 & Protostellar Core & 0.20 & 0.47 & 0.29 $\pm$ 0.06  \\ 
\hline 
  & Identified Core & 0.02 & 8.39 & 0.27 $\pm$ 0.48  \\ 
Mass ($M_\odot$) & Starless Core & 0.02 & 8.39 & 0.25 $\pm$ 0.44  \\ 
 & Protostellar Core & 0.08 & 4.93 & 0.98 $\pm$ 1.15  \\ 
\hline 
  & Identified Core & 0.16 & 11.25 & 1.13 $\pm$ 0.98  \\ 
Number Density (10$^4$ cm$^{-3}$) & Starless Core & 0.16 & 6.05 & 1.05 $\pm$ 0.73  \\ 
 & Protostellar Core & 0.62 & 11.25 & 4.51 $\pm$ 2.87  \\ 
\hline 
  & Identified Core & 0.26 & 33.09 & 3.16 $\pm$ 3.30  \\ 
Virial Ratio & Starless Core & 0.27 & 33.09 & 3.21 $\pm$ 3.32  \\ 
 & Protostellar Core & 0.26 & 2.18 & 0.89 $\pm$ 0.64  \\ 
\hline 
\end{tabular} 

  \label{tab:l1641n_property}
\end{table}

\begin{table}[htbp]
 \centering
  \caption{The Results of Core Identification in the L1641C Area}
  \small
  \begin{tabular}{ccccc} 
\hline \hline 
Property & Category & Minimum & Maximum & Mean $\pm$ Std. \\ 
\hline  
  & Identified Core & 0.036 & 0.260 & 0.082 $\pm$ 0.031  \\ 
Diameter (pc) & Starless Core & 0.036 & 0.260 & 0.082 $\pm$ 0.031  \\ 
 & Protostellar Core & 0.053 & 0.136 & 0.083 $\pm$ 0.029  \\ 
\hline 
  & Identified Core & 1.03 & 4.37 & 1.88 $\pm$ 0.60  \\ 
Aspect Ratio & Starless Core & 1.03 & 4.37 & 1.88 $\pm$ 0.61  \\ 
 & Protostellar Core & 1.12 & 2.55 & 1.67 $\pm$ 0.48  \\ 
\hline 
  & Identified Core & 0.14 & 0.71 & 0.28 $\pm$ 0.10  \\ 
FWHM (km s$^{-1}$) & Starless Core & 0.14 & 0.71 & 0.28 $\pm$ 0.10  \\ 
 & Protostellar Core & 0.18 & 0.39 & 0.30 $\pm$ 0.08  \\ 
\hline 
  & Identified Core & 0.05 & 4.46 & 0.42 $\pm$ 0.55  \\ 
Mass ($M_\odot$) & Starless Core & 0.05 & 4.46 & 0.41 $\pm$ 0.54  \\ 
 & Protostellar Core & 0.24 & 1.59 & 0.75 $\pm$ 0.45  \\ 
\hline 
  & Identified Core & 0.36 & 17.35 & 1.85 $\pm$ 1.33  \\ 
Number Density (10$^4$ cm$^{-3}$) & Starless Core & 0.36 & 7.00 & 1.75 $\pm$ 0.92  \\ 
 & Protostellar Core & 1.55 & 17.35 & 4.86 $\pm$ 4.63  \\ 
\hline 
  & Identified Core & 0.19 & 11.75 & 1.52 $\pm$ 1.19  \\ 
Virial Ratio & Starless Core & 0.19 & 11.75 & 1.55 $\pm$ 1.19  \\ 
 & Protostellar Core & 0.28 & 1.36 & 0.74 $\pm$ 0.32  \\ 
\hline 
\end{tabular} 

  \label{tab:l1641c_property}
\end{table}

\begin{threeparttable}[htbp]
 \centering
  \caption{The Summary of CMF Parameters for Figures \ref{fig:oriona_cmf} and \ref{fig:cmfs}}
  \small
  \begin{tabular}{ccccc} 
\hline \hline 
Region & Category & Turnover Mass \tnote{a} & High-mass Slope & Highest Mass \\ 
 &  & ($M_\odot$) & Power-low index $\pm$ Error & ($M_\odot$) \\ 
\hline  
 & Identified Core & 0.07 & -2.20 $\pm$ 0.06 & 72.21  \\ 
Orion A & Starless Core & 0.07 & -2.25 $\pm$ 0.10 & 72.21  \\ 
 & Bound Starless Core & 0.10 & -2.18 $\pm$ 0.11 & 72.21  \\ 
\hline 
 & Identified Core & 0.04 & -1.89 $\pm$ 0.06 & 72.21  \\ 
(a) OMC-1/2/3 area & Starless Core & 0.04 & -1.93 $\pm$ 0.10 & 72.21  \\ 
 & Bound Starless Core & 0.10 & -1.77 $\pm$ 0.11 & 72.21  \\ 
\hline 
 & Identified Core & 0.07 & -2.35 $\pm$ 0.11 & 4.88  \\ 
(b) OMC-4/5 area & Starless Core & 0.07 & -2.20 $\pm$ 0.07 & 3.31  \\ 
 & Bound Starless Core & 0.10 & -2.03 $\pm$ 0.08 & 3.31  \\ 
\hline 
 & Identified Core & 0.07 & -2.48 $\pm$ 0.14 & 8.39  \\ 
(c) L1641N area & Starless Core & 0.07 & -2.44 $\pm$ 0.15 & 8.39  \\ 
 & Bound Starless Core & 0.10 & -2.41 $\pm$ 0.17 & 8.39  \\ 
\hline 
 & Identified Core & 0.15 & -2.44 $\pm$ 0.26 & 4.46  \\ 
(d) L1641C area & Starless Core & 0.15 & -2.41 $\pm$ 0.25 & 4.46  \\ 
 & Bound Starless Core & 0.15 & -2.34 $\pm$ 0.24 & 4.46  \\ 
\hline 
\end{tabular} 

  \label{tab:cmf_property}
 \begin{tablenotes}
 \item[a] This column shows the central mass of the turnover mass bin in a log-scale of each CMF.
 \end{tablenotes}
\end{threeparttable}

\begin{sidewaystable}[htbp]
\centering
\caption{Physical Properties of the Starless C$^{18}$O Cores in Orion A (The full version is online only.)}
\tiny
\begin{tabular}{cccd{-2}d{-1}d{-3}d{-2}d{-3}d{-3}d{-2}d{-2}d{-2}d{-1}d{-2}d{-2}d{-2}cc} 
\hline \hline    
\rm{ID} & \rm{R.A.} & \rm{Dec.} & \multicolumn{1}{c}{$V_\mathrm{lsr}$} & \multicolumn{1}{c}{$A$} & \multicolumn{1}{c}{2$R_\mathrm{core}$} & \multicolumn{1}{c}{\rm{FWHM}} & 
\multicolumn{1}{c}{$\sigma_\mathrm{major}$} & \multicolumn{1}{c}{$\sigma_\mathrm{minor}$} & \multicolumn{1}{c}{\rm{P.A.}} & \multicolumn{1}{c}{$T_\mathrm{mb, peak}$} & 
\multicolumn{1}{c}{$T_\mathrm{mb, total}$} & \multicolumn{1}{c}{$M_\mathrm{core}$} & \multicolumn{1}{c}{$M_{\mathrm{vir}}$} & \multicolumn{1}{c}{$\alpha_\mathrm{vir}$} & 
\multicolumn{1}{c}{$n_\mathrm{core}$} & \multicolumn{1}{c}{\rm{Starless}} & \multicolumn{1}{c}{\rm{Region}}  \\ 
 & \rm{(J2000)} & \rm{(J2000)} & \multicolumn{1}{c}{(\rm{km s}$^{-1}$)} & \multicolumn{1}{c}{(\rm{arcsec}$^2$)} & \multicolumn{1}{c}{\rm{(pc)}} & 
\multicolumn{1}{c}{(\rm{km s}$^{-1}$)} & \multicolumn{1}{c}{\rm{(pc)}} & \multicolumn{1}{c}{\rm{(pc)}} & \multicolumn{1}{c}{\rm{(degree)}} & \multicolumn{1}{c}{\rm{(K)}} & \multicolumn{1}{c}{\rm{(K)}} & 
\multicolumn{1}{c}{($M_\odot$)} & \multicolumn{1}{c}{($M_\odot$)} &   & \multicolumn{1}{c}{(10$^4$\ \rm{cm}$^{-3}$)} &   &   \\ 
\hline   \\   
 1 & 5 37 02 & -4 55 39 & 8.27 & 772.0 & 0.063 & 0.35 & 0.027 & 0.011 & 112.85 & 3.50 & 681.9 & 0.10 & 0.49 & 5.05 & 1.07 & Y & OMC-1/2/3  \\ 
 2 & 5 37 04 & -4 50 29 & 9.48 & 712.0 & 0.060 & 0.28 & 0.016 & 0.015 & -158.03 & 3.78 & 795.3 & 0.08 & 0.29 & 3.82 & 0.95 & Y & OMC-1/2/3  \\ 
 3 & 5 37 04 & -4 57 08 & 8.19 & 684.0 & 0.059 & 0.16 & 0.021 & 0.010 & 55.77 & 4.89 & 820.6 & 0.08 & 0.09 & 1.10 & 1.11 & Y & OMC-1/2/3  \\ 
 4 & 5 37 05 & -4 55 26 & 8.25 & 364.0 & 0.043 & 0.36 & 0.013 & 0.007 & 152.50 & 4.34 & 579.2 & 0.04 & 0.35 & 8.55 & 1.41 & Y & OMC-1/2/3  \\ 
 5 & 5 37 05 & -5 07 57 & 8.22 & 1620.0 & 0.091 & 0.40 & 0.033 & 0.023 & 83.63 & 2.86 & 1083.6 & 0.36 & 0.92 & 2.57 & 1.30 & Y & OMC-4/5  \\ 
 6 & 5 37 06 & -4 23 03 & 9.23 & 1072.0 & 0.074 & 0.34 & 0.029 & 0.012 & 81.89 & 4.94 & 1156.0 & 0.10 & 0.54 & 5.55 & 0.66 & Y & OMC-1/2/3  \\ 
 7 & 5 37 06 & -4 39 06 & 10.67 & 368.0 & 0.043 & 0.23 & 0.011 & 0.010 & 118.03 & 3.30 & 362.0 & 0.05 & 0.14 & 2.92 & 1.62 & Y & OMC-1/2/3  \\ 
 8 & 5 37 06 & -4 40 11 & 10.70 & 828.0 & 0.065 & 0.25 & 0.021 & 0.013 & -156.42 & 3.81 & 853.5 & 0.09 & 0.26 & 2.84 & 0.92 & Y & OMC-1/2/3  \\ 
 9 & 5 37 07 & -4 37 11 & 10.55 & 904.0 & 0.068 & 0.24 & 0.025 & 0.013 & 63.83 & 4.06 & 1015.4 & 0.07 & 0.26 & 3.53 & 0.63 & Y & OMC-1/2/3  \\ 
 10 & 5 37 07 & -4 40 27 & 10.21 & 1292.0 & 0.081 & 0.23 & 0.025 & 0.021 & -159.61 & 3.91 & 1013.0 & 0.14 & 0.26 & 1.82 & 0.73 & Y & OMC-1/2/3  \\ 
\multicolumn{1}{c}{$\cdot$} & \multicolumn{1}{c}{$\cdot$} & \multicolumn{1}{c}{$\cdot$} & \multicolumn{1}{c}{$\cdot$} & \multicolumn{1}{c}{$\cdot$} & \multicolumn{1}{c}{$\cdot$} & 
\multicolumn{1}{c}{$\cdot$} & \multicolumn{1}{c}{$\cdot$} & \multicolumn{1}{c}{$\cdot$} & \multicolumn{1}{c}{$\cdot$} & \multicolumn{1}{c}{$\cdot$} & \multicolumn{1}{c}{$\cdot$} & 
\multicolumn{1}{c}{$\cdot$} & \multicolumn{1}{c}{$\cdot$} & \multicolumn{1}{c}{$\cdot$} & \multicolumn{1}{c}{$\cdot$} & \multicolumn{1}{c}{$\cdot$} & \multicolumn{1}{c}{$\cdot$}   \\   
\multicolumn{1}{c}{$\cdot$} & \multicolumn{1}{c}{$\cdot$} & \multicolumn{1}{c}{$\cdot$} & \multicolumn{1}{c}{$\cdot$} & \multicolumn{1}{c}{$\cdot$} & \multicolumn{1}{c}{$\cdot$} & 
\multicolumn{1}{c}{$\cdot$} & \multicolumn{1}{c}{$\cdot$} & \multicolumn{1}{c}{$\cdot$} & \multicolumn{1}{c}{$\cdot$} & \multicolumn{1}{c}{$\cdot$} & \multicolumn{1}{c}{$\cdot$} & 
\multicolumn{1}{c}{$\cdot$} & \multicolumn{1}{c}{$\cdot$} & \multicolumn{1}{c}{$\cdot$} & \multicolumn{1}{c}{$\cdot$} & \multicolumn{1}{c}{$\cdot$} & \multicolumn{1}{c}{$\cdot$}   \\   
\multicolumn{1}{c}{$\cdot$} & \multicolumn{1}{c}{$\cdot$} & \multicolumn{1}{c}{$\cdot$} & \multicolumn{1}{c}{$\cdot$} & \multicolumn{1}{c}{$\cdot$} & \multicolumn{1}{c}{$\cdot$} & 
\multicolumn{1}{c}{$\cdot$} & \multicolumn{1}{c}{$\cdot$} & \multicolumn{1}{c}{$\cdot$} & \multicolumn{1}{c}{$\cdot$} & \multicolumn{1}{c}{$\cdot$} & \multicolumn{1}{c}{$\cdot$} & 
\multicolumn{1}{c}{$\cdot$} & \multicolumn{1}{c}{$\cdot$} & \multicolumn{1}{c}{$\cdot$} & \multicolumn{1}{c}{$\cdot$} & \multicolumn{1}{c}{$\cdot$} & \multicolumn{1}{c}{$\cdot$}   \\   
\hline   \\   
\end{tabular} 

\label{tab:catalog}
\end{sidewaystable}
\normalsize

\begin{table}[htbp]
 \centering
  \caption{The Results of KS KS Test of Core Properties in OMC-1/2/3 Area}
  \small
  \begin{tabular}{cccc} 
\hline \hline 
Core Property & Category & p-value \\ 
\hline  
 & Starless Core \& Protostellar Core & 6.43$\times 10^{-1}$  \\ 
Diameter (pc) & Bound Starless Core \& Unbound Starless Core & 9.12$\times 10^{-1}$  \\ 
 & Bound Starless Core \& Protostellar Core  & 7.39$\times 10^{-1}$  \\ 
\hline 
 & Starless Core \& Protostellar Core & 6.95$\times 10^{-1}$  \\ 
Aspect Ratio & Bound Starless Core \& Unbound Starless Core & 1.30$\times 10^{-2}$  \\ 
 & Bound Starless Core \& Protostellar Core  & 9.42$\times 10^{-1}$  \\ 
\hline 
 & Starless Core \& Protostellar Core & 2.35$\times 10^{-1}$  \\ 
FWHM (km s$^{-1}$) & Bound Starless Core \& Unbound Starless Core & 1.13$\times 10^{-14}$  \\ 
 & Bound Starless Core \& Protostellar Core  & 4.69$\times 10^{-3}$  \\ 
\hline 
 & Starless Core \& Protostellar Core & 2.32$\times 10^{-6}$  \\ 
Mass ($M_\odot$) & Bound Starless Core \& Unbound Starless Core & 6.81$\times 10^{-25}$  \\ 
 & Bound Starless Core \& Protostellar Core  & 3.59$\times 10^{-3}$  \\ 
\hline 
 & Starless Core \& Protostellar Core & 1.75$\times 10^{-8}$  \\ 
Number Density (10$^4$ cm$^{-3}$) & Bound Starless Core \& Unbound Starless Core & 7.05$\times 10^{-31}$  \\ 
 & Bound Starless Core \& Protostellar Core  & 6.04$\times 10^{-5}$  \\ 
\hline 
 & Starless Core \& Protostellar Core & 6.04$\times 10^{-5}$  \\ 
Virial Ratio & Bound Starless Core \& Unbound Starless Core  & -  \\ 
 & Bound Starless Core \& Protostellar Core  & 2.22$\times 10^{-5}$  \\ 
\hline 
\end{tabular} 

  \label{tab:ks_test_omc123}
\end{table}

\begin{table}[htbp]
 \centering
  \caption{The Results of KS KS Test of Core Properties in OMC-4/5 Area}
  \small
  \begin{tabular}{cccc} 
\hline \hline 
Core Property & Category & p-value \\ 
\hline  
 & Starless Core \& Protostellar Core & 7.53$\times 10^{-1}$  \\ 
Diameter (pc) & Bound Starless Core \& Unbound Starless Core & 5.85$\times 10^{-3}$  \\ 
 & Bound Starless Core \& Protostellar Core  & 9.01$\times 10^{-1}$  \\ 
\hline 
 & Starless Core \& Protostellar Core & 4.76$\times 10^{-1}$  \\ 
Aspect Ratio & Bound Starless Core \& Unbound Starless Core & 7.27$\times 10^{-1}$  \\ 
 & Bound Starless Core \& Protostellar Core  & 3.83$\times 10^{-1}$  \\ 
\hline 
 & Starless Core \& Protostellar Core & 2.91$\times 10^{-1}$  \\ 
FWHM (km s$^{-1}$) & Bound Starless Core \& Unbound Starless Core & 0  \\ 
 & Bound Starless Core \& Protostellar Core  & 8.68$\times 10^{-3}$  \\ 
\hline 
 & Starless Core \& Protostellar Core & 1.16$\times 10^{-2}$  \\ 
Mass ($M_\odot$) & Bound Starless Core \& Unbound Starless Core & 6.76$\times 10^{-14}$  \\ 
 & Bound Starless Core \& Protostellar Core  & 9.16$\times 10^{-2}$  \\ 
\hline 
 & Starless Core \& Protostellar Core & 3.91$\times 10^{-5}$  \\ 
Number Density (10$^4$ cm$^{-3}$) & Bound Starless Core \& Unbound Starless Core & 2.84$\times 10^{-14}$  \\ 
 & Bound Starless Core \& Protostellar Core  & 6.26$\times 10^{-4}$  \\ 
\hline 
 & Starless Core \& Protostellar Core & 6.26$\times 10^{-4}$  \\ 
Virial Ratio & Bound Starless Core \& Unbound Starless Core  & -  \\ 
 & Bound Starless Core \& Protostellar Core  & 3.85$\times 10^{-3}$  \\ 
\hline 
\end{tabular} 

  \label{tab:ks_test_omc45}
\end{table}

\begin{table}[htbp]
 \centering
  \caption{The Results of KS KS Test of Core Properties in L1641N Area}
  \small
  \begin{tabular}{cccc} 
\hline \hline 
Core Property & Category & p-value \\ 
\hline  
 & Starless Core \& Protostellar Core & 8.72$\times 10^{-1}$  \\ 
Diameter (pc) & Bound Starless Core \& Unbound Starless Core & 1.83$\times 10^{-5}$  \\ 
 & Bound Starless Core \& Protostellar Core  & 3.58$\times 10^{-1}$  \\ 
\hline 
 & Starless Core \& Protostellar Core & 7.71$\times 10^{-1}$  \\ 
Aspect Ratio & Bound Starless Core \& Unbound Starless Core & 1.11$\times 10^{-1}$  \\ 
 & Bound Starless Core \& Protostellar Core  & 4.85$\times 10^{-1}$  \\ 
\hline 
 & Starless Core \& Protostellar Core & 1.75$\times 10^{-2}$  \\ 
FWHM (km s$^{-1}$) & Bound Starless Core \& Unbound Starless Core & 9.99$\times 10^{-16}$  \\ 
 & Bound Starless Core \& Protostellar Core  & 7.57$\times 10^{-5}$  \\ 
\hline 
 & Starless Core \& Protostellar Core & 4.73$\times 10^{-6}$  \\ 
Mass ($M_\odot$) & Bound Starless Core \& Unbound Starless Core & 9.99$\times 10^{-16}$  \\ 
 & Bound Starless Core \& Protostellar Core  & 3.75$\times 10^{-3}$  \\ 
\hline 
 & Starless Core \& Protostellar Core & 5.86$\times 10^{-7}$  \\ 
Number Density (10$^4$ cm$^{-3}$) & Bound Starless Core \& Unbound Starless Core & 9.99$\times 10^{-16}$  \\ 
 & Bound Starless Core \& Protostellar Core  & 4.73$\times 10^{-6}$  \\ 
\hline 
 & Starless Core \& Protostellar Core & 4.73$\times 10^{-6}$  \\ 
Virial Ratio & Bound Starless Core \& Unbound Starless Core  & -  \\ 
 & Bound Starless Core \& Protostellar Core  & 7.41$\times 10^{-3}$  \\ 
\hline 
\end{tabular} 

  \label{tab:ks_test_l1641n}
\end{table}

\begin{table}[htbp]
 \centering
  \caption{The Results of KS KS Test of Core Properties in L1641C Area}
  \small
  \begin{tabular}{cccc} 
\hline \hline 
Core Property & Category & p-value \\ 
\hline  
 & Starless Core \& Protostellar Core & 5.98$\times 10^{-1}$  \\ 
Diameter (pc) & Bound Starless Core \& Unbound Starless Core & 1.26$\times 10^{-1}$  \\ 
 & Bound Starless Core \& Protostellar Core  & 5.57$\times 10^{-1}$  \\ 
\hline 
 & Starless Core \& Protostellar Core & 5.53$\times 10^{-1}$  \\ 
Aspect Ratio & Bound Starless Core \& Unbound Starless Core & 8.73$\times 10^{-1}$  \\ 
 & Bound Starless Core \& Protostellar Core  & 5.60$\times 10^{-1}$  \\ 
\hline 
 & Starless Core \& Protostellar Core & 8.59$\times 10^{-2}$  \\ 
FWHM (km s$^{-1}$) & Bound Starless Core \& Unbound Starless Core & 2.78$\times 10^{-15}$  \\ 
 & Bound Starless Core \& Protostellar Core  & 1.50$\times 10^{-2}$  \\ 
\hline 
 & Starless Core \& Protostellar Core & 2.03$\times 10^{-3}$  \\ 
Mass ($M_\odot$) & Bound Starless Core \& Unbound Starless Core & 3.26$\times 10^{-4}$  \\ 
 & Bound Starless Core \& Protostellar Core  & 5.15$\times 10^{-3}$  \\ 
\hline 
 & Starless Core \& Protostellar Core & 1.05$\times 10^{-2}$  \\ 
Number Density (10$^4$ cm$^{-3}$) & Bound Starless Core \& Unbound Starless Core & 1.18$\times 10^{-4}$  \\ 
 & Bound Starless Core \& Protostellar Core  & 2.05$\times 10^{-2}$  \\ 
\hline 
 & Starless Core \& Protostellar Core & 2.05$\times 10^{-2}$  \\ 
Virial Ratio & Bound Starless Core \& Unbound Starless Core  & -  \\ 
 & Bound Starless Core \& Protostellar Core  & 9.60$\times 10^{-2}$  \\ 
\hline 
\end{tabular} 

  \label{tab:ks_test_l1641c}
\end{table}

\begin{table}[htbp]
 \centering
  \caption{The Results of KS Test of Core Diameter Among Four Subregions}
  \small
  \begin{tabular}{cccccc} 
\hline \hline 
 &  & OMC-1/2/3 & OMC-4/5 & L1641N & L1641C \\ 
\hline  
OMC-1/2/3 & Bound Starless Core & - &  3.75$\times 10^{-10}$ &  6.66$\times 10^{-16}$ &  2.34$\times 10^{-11}$ \\ 
  & Unbound Starless Core & - &  1.67$\times 10^{-5}$ &  2.26$\times 10^{-9}$ &  1.37$\times 10^{-4}$ \\ 
 \hline  
OMC-4/5 & Bound Starless Core & - & - &  3.17$\times 10^{-5}$ &  1.20$\times 10^{-1}$ \\ 
  & Unbound Starless Core & - & - &  9.72$\times 10^{-2}$ &  2.93$\times 10^{-1}$ \\ 
 \hline 
L1641N & Bound Starless Core & - & - & - &  4.02$\times 10^{-2}$ \\ 
  & Unbound Starless Core & - & - & - &  7.80$\times 10^{-1}$ \\ 
 \hline 
\end{tabular} 

  \label{tab:ks_test_diamter}
\end{table}

\begin{table}[htbp]
 \centering
  \caption{The Results of KS Test of Core Aspect Ratio Among Four Subregions}
  \small
  \begin{tabular}{cccccc} 
\hline \hline 
 &  & OMC-1/2/3 & OMC-4/5 & L1641N & L1641C \\ 
\hline  
OMC-1/2/3 & Bound Starless Core & - &  2.22$\times 10^{-1}$ &  1.84$\times 10^{-1}$ &  4.74$\times 10^{-2}$ \\ 
  & Unbound Starless Core & - &  4.53$\times 10^{-1}$ &  5.89$\times 10^{-1}$ &  3.65$\times 10^{-1}$ \\ 
 \hline  
OMC-4/5 & Bound Starless Core & - & - &  7.34$\times 10^{-1}$ &  8.49$\times 10^{-1}$ \\ 
  & Unbound Starless Core & - & - &  8.76$\times 10^{-1}$ &  8.20$\times 10^{-1}$ \\ 
 \hline 
L1641N & Bound Starless Core & - & - & - &  3.39$\times 10^{-1}$ \\ 
  & Unbound Starless Core & - & - & - &  6.40$\times 10^{-1}$ \\ 
 \hline 
\end{tabular} 

  \label{tab:ks_test_a_ratio}
\end{table}

\begin{table}[htbp]
 \centering
  \caption{The Results of KS Test of Core Velocity Width Among Four Subregions}
  \small
  \begin{tabular}{cccccc} 
\hline \hline 
 &  & OMC-1/2/3 & OMC-4/5 & L1641N & L1641C \\ 
\hline  
OMC-1/2/3 & Bound Starless Core & - &  3.22$\times 10^{-3}$ &  4.46$\times 10^{-8}$ &  1.69$\times 10^{-2}$ \\ 
  & Unbound Starless Core & - &  3.75$\times 10^{-1}$ &  1.13$\times 10^{-6}$ &  2.63$\times 10^{-1}$ \\ 
 \hline  
OMC-4/5 & Bound Starless Core & - & - &  1.50$\times 10^{-2}$ &  8.26$\times 10^{-1}$ \\ 
  & Unbound Starless Core & - & - &  5.21$\times 10^{-4}$ &  1.15$\times 10^{-1}$ \\ 
 \hline 
L1641N & Bound Starless Core & - & - & - &  1.17$\times 10^{-2}$ \\ 
  & Unbound Starless Core & - & - & - &  7.38$\times 10^{-5}$ \\ 
 \hline 
\end{tabular} 

  \label{tab:ks_test_fwhm}
\end{table}

\begin{table}[htbp]
 \centering
  \caption{The Results of KS test of Core Mass Among Four Subregions}
  \small
  \begin{tabular}{cccccc} 
\hline \hline 
 &  & OMC-1/2/3 & OMC-4/5 & L1641N & L1641C \\ 
\hline  
OMC-1/2/3 & Bound Starless Core & - &  3.87$\times 10^{-1}$ &  7.25$\times 10^{-1}$ &  1.67$\times 10^{-1}$ \\ 
  & Unbound Starless Core & - &  2.99$\times 10^{-4}$ &  2.83$\times 10^{-1}$ &  2.61$\times 10^{-8}$ \\ 
 \hline  
OMC-4/5 & Bound Starless Core & - & - &  5.91$\times 10^{-1}$ &  6.04$\times 10^{-3}$ \\ 
  & Unbound Starless Core & - & - &  1.07$\times 10^{-4}$ &  1.88$\times 10^{-3}$ \\ 
 \hline 
L1641N & Bound Starless Core & - & - & - &  4.39$\times 10^{-2}$ \\ 
  & Unbound Starless Core & - & - & - &  1.73$\times 10^{-8}$ \\ 
 \hline 
\end{tabular} 

  \label{tab:ks_test_mass}
\end{table}

\begin{table}[htbp]
 \centering
  \caption{The Results of KS Test of Core Density Among Four Subregions}
  \small
  \begin{tabular}{cccccc} 
\hline \hline 
 &  & OMC-1/2/3 & OMC-4/5 & L1641N & L1641C \\ 
\hline  
OMC-1/2/3 & Bound Starless Core & - &  9.55$\times 10^{-13}$ &  6.66$\times 10^{-16}$ &  4.26$\times 10^{-14}$ \\ 
  & Unbound Starless Core & - &  4.12$\times 10^{-1}$ &  3.12$\times 10^{-26}$ &  8.19$\times 10^{-6}$ \\ 
 \hline  
OMC-4/5 & Bound Starless Core & - & - &  1.49$\times 10^{-7}$ &  1.14$\times 10^{-1}$ \\ 
  & Unbound Starless Core & - & - &  2.00$\times 10^{-15}$ &  1.17$\times 10^{-5}$ \\ 
 \hline 
L1641N & Bound Starless Core & - & - & - &  1.25$\times 10^{-9}$ \\ 
  & Unbound Starless Core & - & - & - &  2.06$\times 10^{-24}$ \\ 
 \hline 
\end{tabular} 

  \label{tab:ks_test_density}
\end{table}

\begin{table}[htbp]
 \centering
  \caption{The Results of KS Test of Virial Ratio Among Four Subregions}
  \small
  \begin{tabular}{cccccc} 
\hline \hline 
 &  & OMC-1/2/3 & OMC-4/5 & L1641N & L1641C \\ 
\hline  
OMC-1/2/3 & Bound Starless Core & - &  9.78$\times 10^{-1}$ &  3.45$\times 10^{-1}$ &  8.44$\times 10^{-3}$ \\ 
  & Unbound Starless Core & - &  2.83$\times 10^{-6}$ &  2.19$\times 10^{-2}$ &  4.24$\times 10^{-9}$ \\ 
 \hline  
OMC-4/5 & Bound Starless Core & - & - &  6.96$\times 10^{-1}$ &  4.16$\times 10^{-3}$ \\ 
  & Unbound Starless Core & - & - &  2.38$\times 10^{-2}$ &  2.58$\times 10^{-3}$ \\ 
 \hline 
L1641N & Bound Starless Core & - & - & - &  2.84$\times 10^{-2}$ \\ 
  & Unbound Starless Core & - & - & - &  5.20$\times 10^{-6}$ \\ 
 \hline 
\end{tabular} 

  \label{tab:ks_test_virial_ratio}
\end{table}

\begin{table}[htbp]
 \centering
  \caption{The Summary of Number of Bound Starless Cores}
  \small
  \begin{tabular}{ccc} 
\hline \hline 
 & \multicolumn{2}{c}{Condition of bound core}  \\ 
Region & $\alpha_\mathrm{vir}<2$ & $\alpha_\mathrm{vir}<1$  \\ 
\hline  
Orion A & 1045 & 408  \\ 
(a) OMC-1/2/3 area & 212 & 75  \\ 
(b) OMC-4/5 area & 291 & 102  \\ 
(c) L1641N area & 321 & 126  \\ 
(d) L1641C area & 221 & 105  \\ 
\hline 
\end{tabular} 

  \label{tab:N_bound_core}
\end{table}

\begin{table}[htbp]
 \centering
  \caption{The Summary of CMF Parameters for Figure \ref{fig:cmfs_a1} and Orion A}
  \small
  \begin{tabular}{ccccc} 
\hline \hline 
Region & Category & Turnover Mass \tnote{a} & High-mass Slope & Highest Mass \\ 
 &  & ($M_\odot$) & Power-low index $\pm$ Error & ($M_\odot$) \\ 
\hline  
 & Identified Core & 0.15 & -2.44 $\pm$ 0.26 & 72.21  \\ 
Orion A & Starless Core & 0.15 & -2.41 $\pm$ 0.25 & 72.21  \\ 
 & Bound Starless Core & 0.15 & -2.34 $\pm$ 0.24 & 72.21  \\ 
\hline 
 & Identified Core & 0.04 & -1.89 $\pm$ 0.06 & 72.21  \\ 
(a) OMC-1/2/3 area & Starless Core & 0.04 & -1.93 $\pm$ 0.10 & 72.21  \\ 
 & Bound Starless Core & 0.24 & -1.55 $\pm$ 0.13 & 72.21  \\ 
\hline 
 & Identified Core & 0.07 & -2.35 $\pm$ 0.11 & 4.88  \\ 
(b) OMC-4/5 area & Starless Core & 0.07 & -2.20 $\pm$ 0.07 & 3.31  \\ 
 & Bound Starless Core & 0.15 & -1.80 $\pm$ 0.13 & 3.31  \\ 
\hline 
 & Identified Core & 0.07 & -2.48 $\pm$ 0.14 & 8.39  \\ 
(c) L1641N area & Starless Core & 0.07 & -2.44 $\pm$ 0.15 & 8.39  \\ 
 & Bound Starless Core & 0.15 & -2.24 $\pm$ 0.27 & 8.39  \\ 
\hline 
 & Identified Core & 0.15 & -2.44 $\pm$ 0.26 & 4.46  \\ 
(d) L1641C area & Starless Core & 0.15 & -2.41 $\pm$ 0.25 & 4.46  \\ 
 & Bound Starless Core & 0.15 & -2.09 $\pm$ 0.24 & 4.46  \\ 
\hline 
\end{tabular} 

  \label{tab:cmf_property_a1}
\end{table}


\clearpage


\end{document}